\documentclass{article}
\usepackage{graphics,epsfig}
\title {Foundations of quantum physics: a general \\ realistic and
operational approach\footnote{Published as: Aerts, D., 1999, ``Foundations of quantum physics: a general realistic and
operational approach", {\it International Journal of Theoretical Physics}, {\bf 38}, 289.}}
\author {Diederik Aerts}
\date {}
\font\smallroman=cmr10 at 8pt

\font\num=msbm10

\newtheorem{basicnotion}{Basic Notion}
\newtheorem{theorem}{Theorem}
\newtheorem{definition}{Definition}

\newtheorem{proposition}{Proposition}

\begin{document}
\maketitle
\centerline{FUND and CLEA,}
\centerline {Brussels Free University, Krijgskundestraat 33,}
\centerline {1160 Brussels, Belgium,}
\centerline {e-mail: diraerts@vub.ac.be}
\bigskip

\bigskip
\begin{abstract}
\noindent We present a general formalism with the aim of
describing the situation of an entity, how it is, how it reacts to
experiments, how we can make statistics with it, and how it
`changes' under the influence of the rest of the universe.
Therefore we base our formalism on the following basic notions: (1)
the states of the entity; they describe the modes of being of the
entity, (2) the experiments that can be performed on the entity;
they describe how we act upon and collect knowledge about the
entity, (3) the probabilities; they describe our repeated
experiments and the statistics of these repeated experiments, (4)
the symmetries; they describe the interactions of the entity with
the external world without being experimented upon. Starting from
these basic notions we formulate the necessary derived notions: 
mixed states, mixed experiments and events, an eigen closure
structure describing the properties of the entity, an ortho closure
structure introducing an orthocomplementation, outcome
determination, experiment determination, state determination and
atomicity giving rise to some of the topological separation axioms
for the closures.  We define the notion of sub entity in a general
way and identify the morphisms of our structure. We study specific
examples in detail in the light of this formalism: a classical
deterministic entity and a quantum entity described by the standard
quantum mechanical formalism. We present a possible solution to the
problem of the description of sub entities within the standard quantum
mechanical procedure using the tensor product of the Hilbert spaces,
by introducing a new completed quantum mechanics in Hilbert space,
were new `pure' states are introduced, not represented by rays of the
Hilbert space.
\end{abstract}
\section{Introduction}
Several scientists have worked in the past on the
elaboration of axiomatic approaches to quantum mechanics and it would
lead us to far to present in this paper an overview of all these
approaches. It is however possible to indicate two specific lines that
have inspired most of the `traditional' ones of these approaches.

\subsection{An axiomatics for standard quantum mechanics}

The first line of inspiration was the recovery of standard
quantum mechanics in an axiomatic way. In the standard quantum
formalism a state $p_{\bar c}$ of an entity
$S$ is represented by the one dimensional subspace or the ray
$\bar c$ of a separable complex Hilbert space ${\cal H}$.
An experiment $e_H$ testing an observable is represented by a self
adjoint operator $H$ on ${\cal H}$, and the set of outcomes of this
experiment $e_H$ is the spectrum $spec(H) \subset {\num R}$.
Measurable subsets $A
\subset spec(H)$ represent the events (in the sense of
probability theory) of outcomes. The interaction of the experiment
$e_H$ with the physical entity being in state $p_{\bar c}$
is described in the following way: (1) the probability for a
specific event $A \subset spec(H)$ to occur if the entity is in a
specific state $p_{\bar c}$ is given by
$<c, P_A(c)>$, where $P_A$ is the spectral projection
corresponding to $A$,  $c$ is the unit vector in state ${\bar c}$
and $<\ ,\ >$ is the inproduct in the Hilbert space ${\cal H}$ ; (2)
if the outcome is contained in
$A$, the state
$p_{\bar c}$ is changed to $p_{\bar d}$ where ${\bar d}$ is the ray
generated by
$P_A(c)$. 

This standard quantum mechanical formalism was the
inspiration for most of the axiomatic approaches.
In it, however, the structure of the set of states and of
the experiments is derived from the structure of a complex separable
Hilbert space. The presence of this Hilbert space is ad hoc, in the
sense that there are no physically obvious and plausible reasons why
the Hilbert space structure should be at the origin of both the
structure of the state space, as well as the structure of the
experiments. This initiated the search for an axiomatic theory for
quantum mechanics where the Hilbert space structure would be derived
from more general and physically more plausible axioms (Birkhoff and
Von Neumann 1936, Zierler 1961, Mackey 1963, Piron 1964, Jauch 1968,
Varadarajan 1968, Beltrametti and Cassinelli 1981).
 Due to the original focus (Birkhoff and Von Neumann 1936) on
the collection of `experimental propositions' of a physical entity
- with the conviction that such an `experimental proposition' would
be a good basic concept - most of the later axiomatics were
constructed taking as their basic concept the set
${\cal L}$ of experimental propositions concerning an entity
$S$. The first real breakthrough (Piron 1964) came with a theorem of
Constantin Piron, who proved that if ${\cal L}$ is a complete
[axiom 1], orthocomplemented [axiom 2] atomic [axiom 3] lattice,
which is weakly modular [axiom 4] and satisfies the covering law
[axiom 5], then each irreducible component of the lattice ${\cal
L}$ can be represented as the lattice of all `biorthogonal'
subspaces of a vector space $V$ over a division ring $K$ (with some
other properties satisfied that we shall not explicit here). Such a
vector space is called an `orthomodular space' and also sometimes a
`generalized Hilbert space'. It can be shown that an infinite
dimensional orthomodular space over a division ring which is the
real or complex numbers, or the quaternions, is a Hilbert space.
For a long time there did not even exist any other example of an
infinite dimensional orthomodular space. The search for a further
characterization of the real, complex or quaternionic Hilbert space
started (Wilbur 1977). Then Keller constructed a non classical
orthomodular space (Keller 1980), and recently Sol{\`e}r could
prove that any orthomodular space that contains an infinite
orthonormal sequence is a real, complex or quaternionic Hilbert
space (Sol{\`e}r 1995, Holland 1995). It is under investigation in
which way this result of Sol{\`e}r can be used to formulate new
physically plausible axioms (Pulmannova 1994, 1996, Holland 1995,
Aerts and Van Steirteghem 1998).

\subsection{An operational axiomatic approach}

A second line of inspiration could be called `operationality'.
Going along with the search for `good' axioms was also the idea of
founding the basic notions for this axiomatics in a physically
clear and operational way. `Operationality' means that the axioms
should be introduced in such a way that they can be related   to
`real physical operations' that can be performed in the laboratory.
We have to say some words about this philosophical preoccupation
with operationality. A first triumph for the `operational method'
was certainly the well known analysis of the concept of
simultaneity in physics by Albert Einstein that was also at the
origin of the Einsteinian interpretation of relativity theory.
Standard quantum mechanics is an example of a very non-operational
theory: the basic concept, the wave function, is in principle a
mathematical object with no clear physical interpretation. The
three approaches that have tried to formulate quantum mechanics
operationally are, the Geneva-Brussels approach (Jauch 1968, Piron
1964, 1976, 1989, 1990, Aerts 1981, 1982, 1983a,b), the Amherst
approach (Foulis and Randall 1978, 1981, Foulis, Piron and Randall
1981, Randall and Foulis 1976, 1978, 1981, 1983), and the Marburg
approach (Ludwig 1983, 1985). In all three approaches different
concepts have been used as basic notions and different aspects of
the possibility of operational foundation have been investigated.
The approach that we present in this paper has `learned' from these
three and puts forward a new scheme that takes into account
important results of the earlier approaches, but also gives new
insights that have meanwhile grown out of the theoretical and
experimental progress of the last decades (e.g non locality is an
experimental fact now and not a theoretical hypothesis any
longer).

We also want
to be explicitly critical against a general attitude that we would
classify as `naive operationalism'. As `naive realism' believes
that reality is just like it appears to us and in this way ignores
the problem related to the way we gather knowledge about this
reality, `naive operationalism' believes that it is only our
laboratory experiments that are `real' and the rest is a
construction out of the data and structure that we gather from
these laboratory experiments. The extreme weight that naive
operationalism puts on the laboratory experiments as the only
candidates for foundational concepts is somewhat similar to the
positivist and empirist attitude in philosophy. Meanwhile it is
known that to make experiments we need a theory and that as a
consequence there is no nice hierarchy in the way naive
operationalism proposes. We agree with the naive operationalists
that our contact with reality is our experience and hence our
experiments. In this sense it is good to make the effort and try to
introduce as many possible basic concepts that are directly linked
to these experiences and/or experiments.  On the other hand we are
convinced of the fact that the overall structure of reality,
although it comes to us partially and in a fragmented way through
our immediate experience with it, is revealed to us much more by
the combination of a great many different experiences and by the
way these different experiences form coherent wholes and are
interrelated and also by the way they structure our long term
interaction with reality. In this sense we are also convinced of
the fact that this overall contact with reality - which our
immediate sense experience and hence also our concrete laboratory
experiments are only one aspect of - reveals to us the global
ontological structure of reality: `the way things are' and `what is
the calculus of being'. It is by taking explicitly this fact into
account that we will construct our foundational approach and in
this sense we do not want to call it an `operational' approach -
because operationalism is often interpreted as what we have called
naive operationalism - but a realistic and operational approach.

There is another aspect of our approach that we have to point
out. As we have mentioned briefly in 1.1., most of quantum
axiomatics have been influenced by the original article of Birkhoff
and Von Neumann, and as a consequence have chosen the concept of
`operational proposition' as their basic concept (called `property'
in the Geneva-Brussels approach). In the Amherst approach the
concept of `operation' is primary, but here one also tries to
derive `operational propositions' from this concept. We think that
it is more fruitful to have more basic concepts than just the one
of `experimental proposition' or `operation'. Therefore we will
found our approach on 5 basic concepts and/or structures: the
states, the experiments, the outcomes, the probabilities and the
symmetries. These basic concepts express the naive operationalist
foundational aspects, the laboratory experiments, but are also used
to derive a `calculus of being', structuring the global reality as
it is revealed to us from the overall structure of our experiences
with it.

\subsection{A possible solution of the problem of the description sub
entities in standard quantum mechanics}

In standard quantum mechanics a sub entity of a big entity is described
within the tensor product procedure of the corresponding Hilbert
spaces. As a consequence of the tensor product procedure there exists
pure states (the so called non-product states) of the big entity that
are such that if the big entity is in one of these pure states, the
sub entity is not in a pure state. This is a deep problem in standard
quantum mechanics that has not been solved in a satisfactory way. In
this paper we present a possible solution to this problem that comes
to the definition of a new `completed' quantum mechanics in Hilbert
space, were new `pure' states are introduced that cannot be
represented by rays of the corresponding Hilbert spaces. We show how
this solution follows naturally from the general approach that we
have introduced and how it also is linked with earlier findings. We
also want to mention that for those readers that are only interested
in this newly introduced version of a `completed' quantum mechanics,
but do not want to study the new formalism in detail, that we have
written section~\ref{comquant} in a self contained way. These
readers might immediately switch to section~\ref{comquant}.  

\bigskip
\noindent
The object of our description is the situation of a physical entity
$S$ in its most general way. The archetypical notions that
we consider are the following: 

\medskip
\noindent {\bf The states:} The physical entity $S$ `is' at each
moment in a certain state $p$. In our approach the states
describe the reality of the entity and the structure of the set of
states expresses the main part of the `calculus of being'.

\medskip
\noindent {\bf The experiments:} We gather knowledge about
the entity by means of experiments $e, f, g, ...$ that we can
perform on it. The structure of the set of these experiments
expresses the main part of the way we investigate the reality of
the entity. 

\medskip
\noindent {\bf The outcomes:} The structure of the possible
outcomes, i.e. the ways that the entity and the experiments
performed on it can `be' and `behave' together, is at the root of
our formalism. 

\medskip
\noindent {\bf The probabilities:} For many entities these
possibilities for certain outcomes can be structured in a
probabilistic theory, probability being the representation of the
relative frequencies of repeated experiments. 

\medskip
\noindent {\bf The symmetries: } The entity changes also when we do
not disturb it by a measurement and these changes are governed by
symmetry principles on the reality of the entity, expressing its
relation with the rest of the world.

\bigskip
\noindent These are the basic notions that we want to formalize in
our approach. Derived concepts will be introduced step by step. 

As we will see, an entity will be determined by a well defined set
of relevant states, a well defined set of relevant experiments, a
well defined set of relevant outcomes, and the way in which these
experiments interact with the entity in a state to give rise to an
outcome. This entity corresponds to a physical phenomenon of the
real physical world. In this way it is clear that what we often
will classify, in our intuitive classification of phenomena of the
real world, by the same phenomenon, may correspond to different
entities. Similarly, one entity may also correspond to different
phenomena. In the traditional philosophical scheme it could be said
that entities are `models' of the phenomenon. However, we do not
want to fix this traditional interpretation a priori, since we
believe that a rigorous approach where an entity is defined by well
defined sets of the basic ontological notions of phenomena (states,
experiments, outcomes, probabilities and symmetries), may well
lead, also philosophically, to a better `ontological'
classification. 
\section{Basic notions}
At a certain moment an entity $S$ is in a certain state $p$. This
state represents the reality of the entity at that moment. In this
way we connect a well defined set of states
$\Sigma$ to the entity $S$. 

\begin{basicnotion}[states] Let $S$ be an entity, then $\Sigma$
is the set of states of this entity
$S$. At each moment the entity $S$ `is' in a state $p \in
\Sigma$, that will be referred to as the entity's `actual' state.
This state $p$ represents the reality of the entity $S$ at that
moment. We shall denote states by symbols $p, q, r, ..$. 
\end{basicnotion}
We gather our knowledge about the entity $S$ and we act upon
the entity by means of experiments that can be performed on
$S$. A well defined set of relevant experiments that are
connected to a given entity $S$ is denoted by $\cal E$ and we will
denote experiments by $e,f,g,...$.

\begin{basicnotion}[experiments]
Let $S$ be an entity with a set of states $\Sigma$. The set of
experiments that we use to gather knowledge about $S$ and to act on
$S$ is denoted by $\cal E$. If an entity is in a certain state $p
\in
\Sigma$ and we perform an experiment $e \in {\cal E}$, then
an outcome $x(e,p)$ occurs.
\end{basicnotion}
Different outcomes can possibly occur for an experiment $e$ on an
entity $S$ in state $p$. The set of possible outcomes for $e$ if
$S$ is in $p$ is characteristic for the way in which the experiment
and the entity interact, and will play a major role in our
formalism. We denote this set of possible outcomes by
$O(e,p)$.

\begin{basicnotion}[outcomes]
We denote by the non-empty set $O(e,p)$ the set of
possible outcomes for experiment $e$ if $S$ is in the state $p$. We
denote the set of all non empty sets of possible outcomes for $S$
being in state $p \in \Sigma$ and performing the experiment $e \in
{\cal E}$ by ${\cal O} = \{O(e,p) \ \vert\ e \in {\cal E}, p \in
\Sigma\}$. The set of possible outcomes of the experiment
$e$ will we denoted by
$O(e) =
\cup_{p
\in
\Sigma}O(e,p)$. The set of possible outcomes for all
experiments, the entity $S$ being in state $p$ will be denoted by
$O(p) = \cup_{e \in {\cal E}}O(e,p)$, and the set of all possible
outcomes is denoted by $X = \cup_{p \in
\Sigma, e \in {\cal E}}O(e,p)$. 
\end{basicnotion}
In principle we could consider situations where $O(e,p)
= \emptyset$, but in  certain sense this would mean that the
experiment $e$ in question is not really applicable to the entity
in this state $p$. Since this is a non physical situation, we make
the hypothesis that for $p \in \Sigma, e \in {\cal E}$ we have
$O(e,p) \not= \emptyset$.

We represent mathematically the entity $S$ by a set of experiments
${\cal E}$, a set of states
$\Sigma$, a set of outcomes $X$, and set of non-empty sets of
outcomes ${\cal O} =
\{O(e,p) \ \vert\ e \in {\cal E}, p \in \Sigma\}$. We denote the
entity $S$ by $S({\cal E}, \Sigma, X, {\cal O})$ and will call it a
`experiment state outcome entity', to indicate that the basic
notions that we use to describe the entity are the experiments, the
states and the outcomes. Since we do not want to repeat each time
the characterization `experiment state outcome' we will just write
`the entity $S({\cal E}, \Sigma, X, {\cal O})$' in those cases that
it does not lead to confusion.
\section{Pre-order and orthogonality}

The archetypical situation that we
consider is that of an entity
$S({\cal E}, \Sigma, X, {\cal
O})$ that `is' in a state $p \in \Sigma$ and whereon an experiment
$e \in {\cal E}$ can be performed, that gives rise an outcome
$x(e,p) \in O(e,p)$. There are natural structures on ${\cal E}
\times \Sigma$, on $\Sigma$, on
$\cal E$, and on $X$. Our method to formalize these structures is
the following: first we introduce the physical ideas and then we
define the mathematical structure expressing these physical ideas.
We do this in such a way that the mathematical structure is
independent of the physical interpretation, but that, if
interpreted, it gives rise to the original physical ideas.

Consider an entity $S({\cal E}, \Sigma, X, {\cal O})$ and two
states $p, q \in \Sigma$. If it is such that for all experiments $
e \in {\cal E}$ whenever
$S$ is in state $p$, the set of outcomes that can occur for an
experiment
$e$ is contained in the set of outcomes that can occur for the
experiment $e$ if $S$ is in the state $q$ , we say that
$p$ `implies'
$q$ and denote $p < q$. We call this implication the `state
implication'. This is the first example of a physical idea that we
want to formalize. Let us first introduce a mathematical definition.
\begin{definition} {\bf (pre-order, equivalence)}
Consider a set $Z$ and $a,b,c \in Z$. The relation $<$ is
a pre-order relation iff :  
\begin{equation}
\begin{array}{c}
a < a \\
a < b, b < c \Rightarrow a < c
\end{array}
\end{equation}
We say that two elements $a, b \in Z$ are equivalent, and we denote
$a \approx b$, iff $a < b$ and $b < a$.
\end{definition}
\begin{definition} \label{defprestat} {\bf (state implication)}
For an entity $S({\cal E}, \Sigma, X, {\cal
O})$, for $e \in {\cal E}$ and $p,q \in \Sigma$ we define:
\begin{eqnarray}
p <_e q &\Leftrightarrow& O(e,p) \subset O(e,q) \\
p < q &\Leftrightarrow& \forall f \in {\cal E}, p <_f q
\end{eqnarray}
and we say respectively that $p$ `$e$-implies' $q$ and that $p$
`implies' $q$, and call $<_e$ the `$e$-state implication' and $<$
the state implication.
\end{definition}
\begin{theorem}
For an entity $S({\cal E}, \Sigma, X, {\cal
O})$, the state implications $<_e$ and $<$
introduced on $\Sigma$ in definition~\ref{defprestat} are pre-order
relations.
\end{theorem}
Proof: Clearly for $p \in \Sigma$ we have $p < p$. Consider $p,q,r
\in \Sigma$ such that $p < q$ and $q < r$. Then
$\forall \ e \in {\cal E}$ we have $O(e,p) \subset O(e,q)$ and
$O(e,q) \subset O(e,r)$. From this follows that
$\forall \ e \in {\cal E}$ we have $O(e,p) \subset O(e,q)$, which
shows that $p < r$. 

\bigskip
\noindent
In a similar way we introduce natural implications on ${\cal E}
\times
\Sigma$ and on ${\cal E}$ that we call the `central implication'
and the `experiment implication'.
\begin{definition} {\bf (central implication, experiment
implication)} \label{defpreexp} For an entity $S({\cal E}, \Sigma,
X, {\cal O})$, for $(e,p), (f,q) \in {\cal E} \times \Sigma$, $e, f
\in {\cal E}$ and $p \in \Sigma$ we define:
\begin{eqnarray}
(e,p) < (f,q)  &\Leftrightarrow& O(e,p) \subset O(f,q) \\
e <_p f &\Leftrightarrow&  O(e,p) \subset O(f,p) \\
e < f &\Leftrightarrow& \forall\ q \in \Sigma, e <_q f
\end{eqnarray}
and we respectively say $(e,p)$ `implies' $(f,q)$, $e$
`$p$-implies' $f$, and $e$ `implies' $f$, and call these
implications respectively the `central implication' and the
`$p$-experiment implication' and the `experiment implication'.
\end{definition}
\begin{theorem}
For an entity $S({\cal E}, \Sigma, X, {\cal O})$, the implication
relations $<$ and $<_p$ defined on ${\cal E} \times \Sigma$ and on
$\cal E$ in definition~\ref{defpreexp} are pre-order relations.
\end{theorem}
Consider an entity $S({\cal E}, \Sigma, X, {\cal O})$ and two states $p, q \in \Sigma$. If it is such
that state $p$ and state $q$ can be `distinguished' for the entity
$S$, then we say that $p$ and $q$ are `orthogonal', and we denote
$p \perp q$. Before we formalize this physical concept of
`distinguished states' in our approach, let us introduce the
mathematical concept of an orthogonality relation.
\begin{definition} {\bf (orthogonality)}
Consider a set $Z$ and $a, b \in Z$. The
relation $\perp$ is an orthogonality relation iff:
\begin{equation}
\begin{array}{c}
a \not\perp a \\
a \perp b \Rightarrow b \perp a
\end{array}
\end{equation}
\end{definition}
\begin{definition} {\bf (state orthogonality)} \label{deforthstat}
For an entity $S({\cal E}, \Sigma, X, {\cal
O})$ and for $p,q \in \Sigma$ we define:
\begin{eqnarray}
p \perp_e q &\Leftrightarrow& O(e,p) \cap O(e,q) = \emptyset \\
p \perp q &\Leftrightarrow& \exists \ e \in {\cal E}, p \perp_e q
\end{eqnarray}
we say that $p$ is `$e$-orthogonal' to $q$ if $p
\perp_e q$, and $p$ is `orthogonal' to $q$ if $p \perp q$. We call
$\perp_e$ the `$e$-state orthogonality' and $\perp$ the `state
orthogonality'.
\end{definition}
\begin{theorem}
For an entity $S({\cal E}, \Sigma, X, {\cal
O})$, the $e$-state orthogonality $\perp_e$ and the state
orthogonality $\perp$ introduced on $\Sigma$ in
definition~\ref{deforthstat} is an orthogonality relation.
\end{theorem}
Proof: Clearly for $p \in \Sigma$ we have $p \not\perp_e p$ and $p
\not\perp p$. Consider $p,q \in \Sigma$ such that $p
\perp_e q$. Then $O(e,p) \cap O(e,q) = \emptyset$ and hence $q
\perp_e p$. In an analogous way we show that $p
\perp q$ implies $q \perp p$. 

\bigskip
\noindent
In a similar way we introduce natural orthogonality relations on
${\cal E} \times
\Sigma$ and on ${\cal E}$ that we call the `central
orthogonality' and the `experiment orthogonality'.
\begin{definition} {\bf (central orthogonality,
experiment orthogonality)} \label{deforthcen} For an entity
$S({\cal E}, \Sigma, X, {\cal O})$, for $(e,p), (f,q) \in {\cal E}
\times \Sigma$ and $e, f \in {\cal E}$ we define:
\begin{eqnarray}
(e,p) \perp (f,q) &\Leftrightarrow& O(e,p) \cap O(f,q) = \emptyset
\\ e \perp_p f &\Leftrightarrow& O(e,p) \cap O(f,p) = \emptyset \\
e \perp f &\Leftrightarrow&  \exists \ p \in \Sigma, e \perp_p f
\end{eqnarray}
we say that $(e,p)$ is `orthogonal' to $(f,q)$, $e$
is `$p$-orthogonal' to $f$ if $e \perp_p f$ and $e$ is `orthogonal'
to $f$ if $e \perp f$. We call the orthogonality relations
respectively the `central orthogonality', the `$p$-experiment
orthogonality' and the `experiment orthogonality'.
\end{definition}
There exists a natural orthogonality relation on the set of
outcomes.
\begin{definition} {\bf (outcome orthogonality)}
\label{deforthout} For an entity $S({\cal E}, \Sigma, X, {\cal O})$
and $x, y \in X$ we define:
\begin{eqnarray}
x \perp_{e,p} y &\Leftrightarrow&  x,y \in O(e,p), x \not= y \\
x \perp y &\Leftrightarrow& \ \exists \ e \in {\cal E}, p \in
\Sigma, x \perp_{e,p} y
\end{eqnarray}
we say that $x$ is $(e,p)$-orthogonal to $y$ if $x \perp_{e,p}
y$ and $x$ is orthogonal to $y$ if $x \perp y$, and we call these
relations respectively the `$(e,p)$-outcome orthogonality' and the
`outcome orthogonality'.
\end{definition}
\begin{theorem}
Consider an entity $S({\cal E}, \Sigma, X, {\cal
O})$. The central orthogonality, the $p$-experiment orthogonality
and the experiment orthogonality as introduced in
definition~\ref{deforthcen} and the outcome orthogonality as
introduced in definition~\ref{deforthout} are orthogonality
relations.
\end{theorem}
\begin{proposition}
For an entity $S({\cal E}, \Sigma, X, {\cal
O})$ and $(e,p), (f,q) \in {\cal E} \times \Sigma$, $p,q \in
\Sigma$ and $e,f \in {\cal E}$ we have :
\begin{equation}
(e,p) < (f,q) \Rightarrow (e,p) \not\perp (f,q)
\end{equation}
\begin{equation}
p < q \Rightarrow p \not\perp q 
\end{equation}
\begin{equation}
e < f \Rightarrow e \not\perp f 
\end{equation}
Moreover, the orthogonalities defined on ${\cal E} \times \Sigma,
{\cal E}, \Sigma$, have the following property:
\begin{equation}
a \perp b, c < a, d < b \Rightarrow c \perp d
\end{equation}
\end{proposition}
We remark that a couple $(e,p)$ is equivalent with a couple
$(f,q)$, and we denote $(e,p) \approx (f,q)$, iff
$(e,p) < (f,q)$ and $(f,q) < (e,p)$, that two states
$p, q \in \Sigma$ are equivalent, and we denote $p \approx q$, iff
$p < q$, and
$q < p$, and that two experiments $e, f \in {\cal E}$ are
equivalent, and we denote $e \approx f$, iff $e < f$ and $f < e$.
\begin{definition} {\bf (eigen state, eigen couple)}
Suppose that we have an entity $S({\cal E}, \Sigma, X, {\cal
O})$. We say that a state $p
\in \Sigma$ is an `eigenstate' for the experiment $e \in {\cal E}$
with `eigen-outcome' x(e,p) iff
$O(e,p)$ is a singleton, and hence $O(e,p) = \{x(e,p)\}$. We also
say in this case that $(e,p)$ is an eigen couple with eigen outcome
$x(e,p)$.
\end{definition}
If the state $p \in \Sigma$ is an eigenstate of the experiment $e
\in {\cal E}$ with eigen outcome
$x(e,p)$, this means that the experiment $e$ has a `determined'
outcome for $S$ being in state $p$.
\section{Mixed states, mixed experiments and events}
\label{mixedstat} Often we are in a position that we `lack
knowledge' about the state $p$ in which the entity $S$ `is' or
about the experiment $e$ that will be performed on the entity, or
about the outcome that will occur. We should include a description
of this possible lack of knowledge in our formalism. Suppose that
we have an entity $S({\cal E}, \Sigma, X, {\cal O})$. Consider non
empty subsets $P \subset \Sigma$, $E \subset {\cal E}$ and $A
\subset X$. If we know that the entity is in one of the states of
$P$, but we do not know in which one exactly, we are in a situation
of `lack of knowledge' about the state of the entity, and we will
indicate this situation by the mixed state $p(P)$. If we know that
an experiment of $E$ will be performed, but we do not know exactly
which one, we will indicate this situation by the mixed experiment
$e(E)$. If one of the outcomes of $A$  occurs, but we do not know
which one exactly, we shall say that the event $x(A)$ connected to
$A$ occurs. 

At first sight we would think that to one subset $P \subset
\Sigma$ can correspond different situations of `lack of knowledge'
and hence different mixed states. Similarly one subset $E
\subset {\cal E}$ can give rise to different mixed experiments and
one subset $A \subset X$ to different events. This is in fact true,
but we will choose to distinguish these different situations of
lack of knowledge by means of the probability structure that we
shall introduce later. At this stage of the formalism, we mean with
mixed state (mixed experiment, event) the specification of a
situation of lack of knowledge where we do not know its nature. We
lack the knowledge and also lack the knowledge about the nature of
this lack of knowledge. This is again a unique situation and it
allows us to introduce mixed states, mixed experiments and events
in the following way.
\begin{definition} {\bf (mixed experiments, mixed states and
events)} Consider an entity $S({\cal E}, \Sigma, X, {\cal O})$, and
given non empty subsets
$E \subset {\cal E}$, $P \subset \Sigma$ and $A \subset X$. The
mixed experiment $e(E)$ consists of performing one of the
experiments $f \in E$. The entity is in a mixed state $p(P)$ iff it
is in one of the states $q \in P$. An event $x(A)$ occurs iff one
of the outcomes $y \in A$ occurs.
\end{definition}
Obviously we can consider a state $q$ as being the trivial mixed
state on the singleton $\{q\}$ and an experiment $f$ to be the
mixed experiment on the singleton $\{f\}$ and an outcome $y$ to be
the event connected with the singleton $\{y\}$.
\begin{proposition}
Suppose that we have an entity $S({\cal E}, \Sigma, X,
{\cal O})$. For $f \in {\cal E}$, $q \in \Sigma$ and $y \in X$ we
have:
\begin{equation}
\begin{array}{lll}
q = p(\{q\}) &
f = e(\{f\}) &
y = x(\{y\})
\end{array}
\end{equation}
for the mixed state $p(P)$ and the mixed experiment $e(E)$ we have:
\begin{equation}
O(e(E),p) = \cup_{e \in E}O(e,p) \quad O(e,p(P)) = \cup_{p \in
P}O(e,p)
\end{equation}
\begin{equation}
O(e(E),p(P)) = \cup_{e \in E, p \in P}O(e,p)
\end{equation}
\end{proposition}
\begin{definition} {\bf (mixed entity)}
An entity $S({\cal E}, \Sigma, X,
{\cal O})$ is a `mixed entity' iff there is a well defined set
of mixed experiments, mixed states and events associated to the
entity. We denote the set of mixed experiments, mixed states, and
events by
$M({\cal E})$, $M(\Sigma)$ and $M(X)$.
\end{definition}
\begin{definition}
Suppose that we have a mixed entity $S({\cal E}, \Sigma, X,
{\cal O})$ with set of mixed states $M(\Sigma)$, set of mixed
experiments $M({\cal E})$, set of events $M(X)$. We generalize the
pre-order relations and the orthogonality relations that are
defined on ${\cal E} \times \Sigma$,
${\cal E}$, $\Sigma$, and $X$, to pre-order relations and
orthogonality relations defined on $M({\cal E}) \times M(\Sigma)$,
$M({\cal E})$, $M(\Sigma)$, and $M(X)$. All the generalizations are
straightforward, with the exception of the one for the events,
which we will state explicitly: two events $x(A)$ and $x(B)$ are
$(e,p)$-orthogonal iff
$A \subset O(e,p)$ and $B \subset O(e,p)$ and $A \cap B =
\emptyset$: we denote $x(A) \perp_{e,p} x(B)$. Two events
$x(A)$ and $x(B)$ are orthogonal iff there exists $e \in {\cal E}$
and $p \in  \Sigma$ such that $x(A) \perp_{e,p} x(B)$. We introduce
a pre-order relation on the set of events in a straightforward way:
$x(A) < x(B)
\Leftrightarrow A \subset B$.
\end{definition}
We have to verify whether the pre-order relation and
the orthogonality relation that we generalize on
$M({\cal E}) \times M(\Sigma)$, on $M({\cal E})$, on $M(\Sigma)$,
and on $M(X)$ coincides with the old pre-order relation and
orthogonality relation on ${\cal E} \times \Sigma$, ${\cal E}$,
$\Sigma$ and $X$. Since we have
$e(\{f\}) = f$ for all $f \in {\cal E}$ and $p(\{q\}) = q$ for all
$q \in \Sigma$, this is easily checked for the pre-order relation
and orthogonality relation. For the relations on $M({\cal E})$,
$M(\Sigma)$ and $M(X)$ we have to be more careful.
\begin{proposition}
Suppose that we have a mixed entity $S({\cal E}, \Sigma, X,
{\cal O})$ with set of mixed states $M(\Sigma)$, set of mixed
experiments $M({\cal E})$, set of events $M(X)$. For states $q, r
\in \Sigma$, experiments
$f, g \in {\cal E}$ and outcomes $y, z \in X$, we have :
\begin{equation}
\begin{array}{ll}
p(\{q\}) < p(\{r\}) \Leftrightarrow q < r \label{eqpreord} &
p(\{q\}) \perp_{e(\{f\})} p(\{r\}) \Leftrightarrow q \perp_f r \\
p(\{q\}) \perp p(\{r\}) \Leftrightarrow q \perp r \label{eqorth} &
e(\{f\}) < e(\{g\}) \Leftrightarrow f < g \\
e(\{f\}) \perp_{p(\{q\})} e(\{g\}) \Leftrightarrow f \perp_q g &
e(\{f\}) \perp e(\{g\}) \Leftrightarrow f \perp g \\
x(\{y\}) \perp_{e,p} x(\{z\}) \Leftrightarrow y \perp_{e,p} z &
x(\{y\}) \perp x(\{z\}) \Leftrightarrow y \perp z
\end{array}
\end{equation}
\end{proposition}
Proof: Let us prove for example~(\ref{eqpreord}). We have
$p(\{q\}) < p(\{r\}) \Leftrightarrow \forall f \in {\cal E}:
O(e(\{f\}),p(\{q\})) \subset O(e(\{f\}), p(\{q\})) \Leftrightarrow
\forall f \in {\cal E}: O(f,q) \subset O(f,r)
\Leftrightarrow q < r$.  Let us also prove
for example~(\ref{eqorth}). We have $p(\{q\}) \perp p(\{r\})
\Leftrightarrow \exists e(E) \in M({\cal E})$ such that $O(e(E),q)
\cap O(e(E),r) = \emptyset$. But this is equivalent to the fact
that $O(e,q) \cap O(e,r) = \emptyset \ \forall e \in E$, which
shows that $q \perp r$. 
\begin{proposition}
Suppose that we have a mixed entity $S({\cal E}, \Sigma, X,
{\cal O})$ with set of mixed states $M(\Sigma)$, set of mixed
experiments $M({\cal E})$, and set of events $M(X)$. For mixed
states $p(P), p(Q)$, mixed experiments $e(E), e(F)$, and events
$x(A), x(B)$ we have:
\begin{equation}
\begin{array}{ll}
E \subset F \Rightarrow e(E) < e(F) &
P \subset Q \Rightarrow p(P) < p(Q)
\end{array}
\end{equation}
\begin{equation}
\begin{array}{l}
(e(E), p(P)) < (e(F), p(Q)) \Leftrightarrow (e,p) < (e(F),p(Q)) \
\forall e \in E, p \in P \\
(e(E), p(P)) \perp (e(F), p(Q)) \Leftrightarrow (e,p) \perp (f,q) \
\forall e \in E, f \in F, p \in P, q \in Q 
\end{array}
\end{equation}
\begin{equation}
\begin{array}{l}
p(P) < p(Q) \Leftrightarrow p < p(Q) \ \forall p \in P \\
e(E) < e(F) \Leftrightarrow e < e(F) \ \forall e \in E \\
p(P) \perp_{e(E)} p(Q) \Leftrightarrow p \perp_{e(E)} q \ \forall
p \in P, q \in Q \\
 e(E) \perp_{p(P)} e(F) \Leftrightarrow e
\perp_{p(P)} f \ \forall e \in E, f \in F \\ 
x(A) \perp_{e,p} x(B)
\Leftrightarrow x \perp_{e,p} y \ \forall x \in A, y \in B \\
p(P) \perp p(Q) \Rightarrow p \perp q \ \forall p \in P, q \in Q
\\ 
e(E) \perp e(F) \Rightarrow e \perp f \ \forall e \in E, f \in
F \\ x(A) \perp x(B) \Rightarrow x \perp y \ \forall x \in A, y
\in B
\end{array}
\end{equation}
\end{proposition}
Proof:
\noindent We have: $(e(E), p(P)) <
(e(F), p(Q))$ $\Leftrightarrow$ $O(e(E),p(P)) \subset O(e(F),p(Q))$
$\Leftrightarrow$
$\cup_{e \in E, p \in P}O(e,p) \subset O(e(F),p(Q))$
$\Leftrightarrow$ $O(e,p) \subset O(e(F),p(Q)) \ \forall e \in E, p
\in P$. We also have: $(e(E), p(P)) \perp (e(F),
p(Q))$ $\Leftrightarrow$ $O(e(E),p(P)) \cap O(e(F),p(Q)) =
\emptyset$
$\Leftrightarrow$ $O(e(E),p(P)) \subset O(e(F),p(Q))^C$
$\Leftrightarrow$ $\cup_{e \in E, p \in P}O(e,p) \subset \cap_{f
\in F, q \in Q} O(f,q)^C$ $\Leftrightarrow$ $O(e,p) \subset
O(f,q)^C \ \forall e \in E,  p \in P,  f \in F,  q \in Q$
$\Leftrightarrow$ $(e,p) \perp (f,q) \ \forall e \in E,  p \in P, 
f \in F,  q \in Q$. The other implications are proved in an
analogous way.
\begin{definition} {\bf (supremum and infimum)}
Suppose that $Z$ is a set with a pre-order relation $<$. Consider a
set $\{a_j, j \in J\}$ of elements of $Z$. We say that $\vee_{j
\in J}a_j$ is a supremum and $\wedge_{j \in J}a_j$ is an infimum
iff for $b \in Z$ we have:  
\begin{eqnarray}
a_j < b \ \forall j \in J &\Leftrightarrow& \vee_{j \in J}a_j <
b \\ b < a_j \ \forall j \in J &\Leftrightarrow& b < \wedge_{j
\in J}a_j
\end{eqnarray}
\end{definition}
\begin{theorem}
Suppose that we have a mixed entity $S({\cal E}, \Sigma, X,
{\cal O})$ with set of mixed states $M(\Sigma)$, set of mixed
experiments $M({\cal E})$, set of events $M(X)$. The mixed
experiment $e(E) \in M({\cal E})$ is a supremum of the set of
experiments $E$ for the pre-order relation on $M({\cal E})$, the
mixed state $p(P)$ is a supremum for the set of states $P$ for the
pre-order relation on $M(\Sigma)$ and the event $x(A)$ is a
supremum for the set of outcomes $A$ for the pre-order relation on
$M(X)$.
\end{theorem}
Proof: We have that $f < e(E)$ for $f \in E$. Suppose now that $f <
g$ for all $f \in E$. This means that
$O(f,p) \subset O(g,p)$ for all $p \in M(\Sigma)$ and $f \in E$.
But then $\cap_{f \in E}O(f,p) \subset O(g,p)$ for all $p \in
M(\Sigma)$. Hence $O(e(E),p) \subset O(g,p)$ for all $p \in
M(\Sigma)$. 
\begin{definition}
Suppose that we have a mixed entity $S({\cal E}, \Sigma, X,
{\cal O})$ with set of mixed states $M(\Sigma)$, set of
mixed experiments $M({\cal E})$, set of events $M(X)$. Because of
the foregoing proposition we shall also denote $e(E) = \vee_{f
\in E}f$, $p(P) = \vee_{q \in P}q$ and $x(A) = \vee_{y \in
x(A)}y$.
\end{definition}
We have to remark that although $e(E) = \vee_{f \in E}f$ is well
defined, it is not necessarily a unique supremum of the set $E$.
The same remark holds for $P$. 

Suppose that we consider a set of mixed states $P
\subset M(\Sigma)$ of an entity $S$. Then we can again consider the
situation of `lack of knowledge' where we know that the entity is
in one of the mixed states $q \in P$, but we do not know in which
one: let us denote this mixed state (of mixed states) by $p(P)$.
This is again a mixed state, but at first sight it is a type of
mixed state that we did not yet consider explicitly in our
formalism, namely a mixed state of mixed states. If this would be
really a new type of mixed state, we would arrive in a regressum ad
infinitum, and this would be a problem. Luckily this is not the
case. The new type $p(P)$ of mixed state is of the type that we
have already introduced. Indeed, suppose that we denote an element
$q \in P$ by $p(Q_q)$ where $Q_q
\subset \Sigma$ is the set of states which $q$ is a mixed state
on. With the state
$p(P)$ of lack of knowledge about the set of mixed states
$q \in P$ corresponds the state of lack of knowledge about the set
$\cup_{q \in P}Q_q$, i.e.
$\vee_{q \in P}p(Q_q)$. And since the mixed state $p(P)$ exists,
$\vee_{q \in P}p(Q_q) \in M(\Sigma)$. 
\begin{proposition}
Suppose that we have a mixed entity $S({\cal E}, \Sigma, X,
{\cal O})$ with set of mixed states $M(\Sigma)$, set of mixed
experiments $M({\cal E})$, set of events $M(X)$. We have:
\begin{equation}
\begin{array}{lll}
M(M(\Sigma)) \subset M(\Sigma) &
M(M({\cal E})) \subset M({\cal E}) &
M(M(X) \subset M(X)
\end{array}
\end{equation}
\end{proposition}
\begin{definition}
Suppose that we have a mixed entity $S({\cal E}, \Sigma, X,
{\cal O})$ with set of mixed states $M(\Sigma)$, set of mixed
experiments $M({\cal E})$, set of events $M(X)$. We will say that
the entity is `full' of mixed states iff there exists a mixed state
$p(P)$ connected to each subset $P \subset \Sigma$. We will say
that the entity is `full' of mixed experiments iff there exists a
mixed experiment $e(E)$ connected to each subset $E \subset {\cal
E}$. We will say that an entity is `full' of events iff there
exists an event $x(A)$ for each $A \subset X$. 
\end{definition}
\begin{definition} {\bf (complete pre-order set)}
\label{complpreord}
 Consider a set $Z$ with a pre-order relation
$<$, then $Z$ is a `complete' pre-order set iff for each subset of
$Z$ there exists a supremum and an infimum.
\end{definition}
\begin{theorem}
Suppose that we have a mixed entity $S({\cal E}, \Sigma, X,
{\cal O})$ with set of mixed states $M(\Sigma)$, set of mixed
experiments $M({\cal E})$, set of events $M(X)$. If the entity is
full of mixed states, then the pre-order relation on $M(\Sigma)$
gives rise to a complete pre-order set $M(\Sigma)$. If
the entity is full of mixed measurements, the pre-order relation on
$M({\cal E})$ gives rise to a complete pre-order set
$M({\cal E})$. If the entity is full of events, the pre-order
relation on
$M(X)$ gives rise to a complete pre-order set $M(X)$. 
More concretely for $P_i \subset \Sigma$ and $P =
\cup_iP_i$, for $E_j \subset {\cal E}$ and $E = \cup_jE_j$ and for
$A_k \subset X$ and $A = \cup_kA_k$ we have:
\begin{equation}
\begin{array}{lll}
p(P) = \vee_ip(P_i) &
e(E) = \vee_je(E_j) &
x(A) = \vee_kx(A_j)
\end{array}
\end{equation}
\end{theorem}
\section{Probability}
So far we have always referred to `possible outcomes'. For most of
the entities studied in physics these possibilities will be
structured in such a way that they give rise to probabilities as
limits of relative frequencies of repeated experiments. Indeed, for
an entity
$S$ in state $p$, for an experiment $e$ and for an outcome $x$ we
introduce the probability that, if the entity is in state $p$, the
experiment $e$ gives the outcome $x$, denoted by $\mu(e,p,x)$, as
the limit of the relative frequency of the occurrence of the
outcome $x$.
\begin{definition}
Suppose that we have a mixed entity $S({\cal E}, \Sigma, X,
{\cal O})$ with set of mixed states $M(\Sigma)$, set of mixed
experiments $M({\cal E})$, set of events $M(X)$. Consider a map
$\mu$:
\begin{equation}
\begin{array}{ll}
\mu : M({\cal E}) \times M(\Sigma) \times M(X) \rightarrow [0,1]
& (e,p,x) \mapsto \mu(e,p,x)
\end{array}
\end{equation}
We say that $\mu$ is a generalized probability measure iff for $e_i
\in M({\cal E})$, $p_j \in M(\Sigma)$, $x_k \in M(X)$,
countable sets, such that
$e_i
\perp e_l$ for
$i
\not= l$, $p_j \perp p_m$ for $j
\not= m$, $x_k \perp x_n$ for $k
\not= n$, and such that $\vee_ie_i$ is a mixed
experiment, $\vee_jp_j$ is a mixed state, $\vee_kx_k$ is an
event, we have:
\begin{equation}
\mu(\vee_ie_i, \vee_jp_j, \vee_kx_k) =
\sum_{i,j,k}\mu(e_i,p_j,x_k) 
\end{equation}
we also have that $x(O(e,p))$ is an event and we have:
\begin{equation}
\mu(e,p,x(O(e,p))) = 1
\end{equation}
\end{definition}
We say that the entity is probabilistic iff the different states
of lack of knowledge are described by different generalized
probability measures $\mu$ that correspond to limits of relative
frequencies of outcomes in these states of lack of knowledge. Hence
probability $\mu(e,p,x)$ is the probability that the event $x$
occurs when the entity $S$ is in state $p$ and the experiment $e$
is performed in the state of lack of knowledge described by $\mu$.
This motivates the following definition:
\begin{definition}
Suppose that we have a mixed entity $S({\cal E}, \Sigma, X,
{\cal O})$ with set of mixed states $M(\Sigma)$, set of mixed
experiments $M({\cal E})$, set of events $M(X)$. The entity $S$ is
probabilistic iff it has a associated well defined set
${\cal M}$ of generalized probability measures. We denote a probabilistic entity by $S({\cal E}, \Sigma, X,
{\cal O}, {\cal M})$
\end{definition}
From now on we will only distinguish between states and mixed
states, experiments and mixed experiments and outcomes and events,
when it is explicitly necessary. The results that are valid for a
general entity are of course also valid for a mixed entity,
considered as a special type of entity.
\section{State property entities} \label{statpropent}
In this section we want to introduce the concept of `property' of
an entity. We give a new description that is inspired by the way
that properties are introduced in the Geneva-Brussels approach
(Piron 1976, 1989, 1990, Aerts 1981, 1982, 1983). The main
differences are: i) we distinguish between properties and `testable'
properties, a difference that has not been made in the earlier
approaches, and, ii) we consider a property and a state as
different concepts, while in the earlier approaches a state was
represented by the set of all actual properties.

Let us consider an entity $S$. We remark that in this section $S$
is not necessarily an `experiment state outcome entity'. A property
$a$ of $S$ is an attribute of $S$. The property $a$ can be
`actual', which means that $S$ is in a state such that it `has' the
property $a$ `in acto', or `potential', which means that
$S$ is in a state such that it does not have the property $a$, but
can eventually acquire it. Let us denote the set of properties
corresponding to the entity
$S$ by
${\cal L}$. If the entity $S$ is in a state $p$ we can consider the
set $\xi(p)$ of all properties that are actual. We call $\xi(p)$
the property state connected to $p$. Let us call ${\cal J}$ the set
of property states. 

If it
is such that for the entity being in an arbitrary state $p \in
\Sigma$ we have that if $a \in {\cal L}$ is `actual' then also $b
\in {\cal L}$ is `actual', we say that $a$ `implies' $b$ (or $a$ is
`stronger than'
$b$). This `implication' introduces a `pre-order' relation on the
set of properties ${\cal L}$. There exists also a natural pre-order
relation on the set of states for a state property entity. Indeed,
if for two states
$p, q \in \Sigma$, the set of properties $\xi(p)$ that is actual if
the entity is in state $p$ contains the set of properties $\xi(q)$
that is actual if the entity is in state $q$, then we say that $p$
`property implies' $q$.

We have now
introduced all the necessary physical concepts to give a formal
definition of an entity described by its states and properties.
\begin{definition} [state property entity] \label{defimpprop}
We say that $S$ is a state property entity iff it is
characterized by a set of states $\Sigma$, a set of properties
${\cal L}$, and a function
$\xi$:
\begin{equation}
\begin{array}{ll}
\xi: \Sigma \rightarrow {\cal P}({\cal L}) &
p \mapsto \xi(p)
\end{array}
\end{equation}
where $\xi(p)$ is the set of properties that are `actual' if
the entity $S$ is in state $p$. We call $\xi$ the state property
function. Hence, for a property
$a
\in {\cal L}$ and a state
$p
\in
\Sigma$ we have:
\begin{equation}
a \ {\rm is \ actual\ if}\ S\ {\rm is\ in\ state\ } p
\Leftrightarrow a \in \xi(p)
\end{equation}
We call $\xi(p)$ the property state corresponding to $p$,
and introduce ${\cal J} = \xi(\Sigma)$ the set of all property
states. Further we have that for $p, q \in \Sigma$ and $a, b \in
{\cal L}$:
\begin{eqnarray}
p \prec q &\Leftrightarrow& \xi(q) \subset \xi(p) \\
a \prec b &\Leftrightarrow& {\rm if\ for\ } p \in \Sigma \ {\rm we\
have\ } a \in \xi(p)\ {\rm then\ } b \in
\xi(p)
\end{eqnarray}
and we say that $p$ `property implies' $q$ and $a$ `implies' $b$
and call this implication the `property implication'. We denote a
state property entity $S$ by $S(\Sigma, {\cal L}, \xi)$.
\end{definition} 
\begin{theorem}
Consider a state property entity $S(\Sigma, {\cal L}, \xi)$. The
implications on $\Sigma$ and on ${\cal L}$ that are introduced in
definition~\ref{defimpprop} are pre-order relations.
\end{theorem}
\begin{definition} {\bf (pre-order set with an ordering set)}
Consider a set $Z$ with a pre-order relation $<$ and consider a
set $U \subset {\cal P}(Z)$. We say that $U$ is an ordering set for
$Z$ iff it is so that for $a, b \in Z$ we have $a < b$ iff whenever
$u \in U$ such that $a \in u$ we have $b \in u$.
\end{definition}
\begin{theorem}
Consider a state property entity $S(\Sigma, {\cal L}, \xi)$, then
the set of property states $\xi(\Sigma) = {\cal J}$ is an ordering
set for ${\cal L}, <$.
\end{theorem}
Proof: Consider $a, b \in {\cal L}$ such that $a \prec b$. Consider
$p \in \Sigma$ such that $a \in \xi(p)$, then $b
\in \xi(p)$. On the other hand suppose that for $\xi(p) \in {\cal
J}$ we have $a \in \xi(p)$ implies $b \in \xi(p)$, then $a \prec b$.

\bigskip
\noindent
It makes sense to identify equivalent properties. Indeed,
equivalent properties are always `actual' and potential together
which makes it possible to indicate them as `the same property' for
the entity $S$. This is the reason that we introduce the following
type of entity where such an identification has been made.
\begin{definition} [identified state property entity]
Consider a state property entity $S(\Sigma, {\cal L}, \xi)$. We say
that $S(\Sigma, {\cal L}, \xi)$ is an `identified' state property
entity iff for $a, b \in {\cal L}$ we have $a \approx b \Rightarrow
a = b$.
\end{definition}
\begin{theorem}
For an identified state property entity $S(\Sigma, {\cal L},
\xi)$, the pre-order relation on the set of properties is a partial
order relation.
\end{theorem}
We have formalized the concept of state property entity. This is
an entity for which we only consider the `ontological' notions of
`state' and `property' and how they are related. Properties can
often also be directly tested. We will analyze now how this can be
formalized. Consider an experiment $e$ and a subset $A \subset
O(e)$ of the outcome set of $e$. Suppose that we have a situation
such that we are `certain' that if we would perform $e$ we find an
outcome contained in $A$. Then it is possible to make correspond a
`property' $a(e,A)$ with this situation, $a(e,A)$ being `actual'
iff this situation is present. The property $a(e,A)$ that we have
defined in this way is a `testable' property.
\begin{definition} \label{def:testent}
Consider an entity $S({\cal E}, \Sigma, X, {\cal O})$ and an
experiment $e \in {\cal E}$. For a set of outcomes $A \subset O(e)$
we introduce an $e$-testable property $a(A)$ such that:
\begin{equation}
a(A) {\rm \ is\ actual\ if\ } S {\rm \ is\ in\ state\ } p
\Leftrightarrow O(e,p) \subset A 
\end{equation}
We denote the set of $e$-testable properties of $S$
by ${\cal L}(e)$.
\end{definition}
We will see now that a state property entity for which the set of
properties is ${\cal L}(e)$ for a given experiment
$e$ has more structure than a general state property entity. Let
us investigate this additional structure. Although we need only one
experiment to define a state property entity for which the set of
properties is ${\cal L}(e)$ it will be more interesting - and
we will not loose generality if we do - to investigate the structure
of these entities for the case of an experiment state outcome
entity. Let us first introduce a mathematical definition.
\begin{proposition}
Consider an entity $S({\cal E}, \Sigma, X, {\cal O})$ and for $e
\in {\cal E}$ the set of $e$-testable properties ${\cal L}(e)$.
Consider also the state property entity $S(\Sigma, {\cal L}(e),
\xi_e)$. For $p \in
\Sigma$, $A, B \subset O(e)$, $(A_i)_i, A_i \subset O(e)$ and $q,
r \in \Sigma$ we have:
\begin{eqnarray}
a(A) \in \xi_e(p) &\Leftrightarrow& O(e,p) \subset A
\label{eq:testent} \\ a(A) \prec a(B) &\Leftrightarrow& \forall p
\in \Sigma:  O(e,p) \subset A {\rm \ then\ } O(e,p) \subset B
\label{eq:testent02}
\\ a(A_j) \in \xi_e(p) \ \forall\ j \ &\Leftrightarrow& a(\cap_iA_i) \in \xi_e(p) \label{eq:compl}
\label{eq:inf} \\ 
q \prec r &\Leftrightarrow& q <_e r \label{eq:statimp}
\end{eqnarray} 
\end{proposition}
Proof: The proof of \ref{eq:testent} and \ref{eq:testent02} are
immediate consequences of definition~\ref{defimpprop} and
\ref{def:testent}. Let us prove \ref{eq:compl}. We have
$a(A_j)
\in \xi_e(p) \ \forall\ j \Leftrightarrow O(e,p) \subset A_j \
\forall\ j \Leftrightarrow O(e,p) \subset
\cap_iA_i
\Leftrightarrow a(\cap_iA_i) \in \xi_e(p)$. Let us prove
\ref{eq:statimp}. Suppose that $q \prec r$. This means that
$\xi_e(r) \subset \xi_e(q)$. We have that $a(O(e,r)) \in \xi_e(r)$
and hence $a(O(e,r)) \in
\xi_e(q)$. From this, using \ref{eq:testent}, follows that $O(e,q)
\subset O(e,r)$ and as a consequence we have
$q <_e r$. Suppose now that $q <_e r$ and hence $O(e,q) \subset
O(e,r)$. Consider $a(A) \in \xi_e(r)$. Then we have $O(e,q) \subset
O(e,r) \subset A$ and hence $a(A) \in \xi_e(q)$. This shows that
$\xi_e(r) \subset
\xi_e(q)$ and as a consequence $q \prec r$.
\begin{theorem} 
Consider an entity $S({\cal E}, \Sigma, X, {\cal O})$ and for $e
\in {\cal E}$ the set of $e$-testable properties ${\cal L}(e)$. The
pre-order set of properties of the state property entity
$S(\Sigma, {\cal L}(e), \xi)$ is a complete pre-order set
(see definition \ref{complpreord}) with a maximal element
$I = a(O(e))$ and minimal element $0 = a(\emptyset)$.
\end{theorem}
Proof:  Consider $(a_i)_i, a_i \in {\cal L}(e)$. For each $a_i$
there exists a set of outcomes $A_i \subset O(e)$ such that $a(A_i)
= a_i$. Consider the $e$-testable property $a(\cap_iA_i)$. Let us
show that
$a(\cap_iA_i)$ is an infimum for the set $(a_i)_i$. From
\ref{eq:inf} follows that $a(\cap_iA_i) \prec a(A_j)
\ \forall \ j$. Suppose that $a(A) \prec a(A_j) \ \forall\ j$.
Consider $O(e,p) \subset A$. From this follows that $O(e,p) \subset
A_j\ \forall\ j$ and hence $O(e,p) \subset \cap_iA_i$. This shows,
taking into account \ref{eq:testent02}, that $a(A) \prec
a(\cap_iA_i)$. So $a(\cap_iA_i)$ is an infimum for the set
$(a_i)_i$. There is a natural construction for a supremum that
consists of taking the infimum of all elements that are `implied'
by all elements of the considered set. We remark that however this
supremum depends in principal on all elements of the pre-ordered
set. Let us identify a maximal and minimal element. We have
$O(e,p)
\subset O(e)$ always and hence
$a(A) \prec a(O(e))$ for an arbitrary $A \subset O(e)$. This shows
that $I = a(O(e))$ is a maximal element of
${\cal L}(e)$. On the contrary $O(e,p) \subset \emptyset$ never,
which shows that $a(\emptyset) \prec a(A)$ for an arbitrary $A
\subset O(e)$. Hence $a(\emptyset)$ is a minimal element of ${\cal
L}(e)$.

\bigskip
\noindent
We will introduce now the mathematical concept of a `state
property system' and then show that the state property entity
$S(\Sigma, {\cal L}(e), \xi_e)$ (once properties are identified) is
well described by a state property system.
\begin{definition} [state property system]
We say that $(\Sigma, \prec, {\cal L}, \prec, \wedge, \vee, \xi)$,
or shorter $(\Sigma, {\cal L}, \xi)$, is a state property system
iff $(\Sigma,
\prec)$ is a pre-ordered set, $({\cal L},
\prec, \wedge, \vee)$ is a complete lattice, and $\xi$ is a
function:
\begin{equation}
\begin{array}{ll}
\xi: \Sigma \rightarrow {\cal P}({\cal L}) &
p \mapsto \xi(p)
\end{array}
\end{equation}
For $p \in \Sigma$, $I$ the maximal element and $0$ the minimal
element of ${\cal L}$, and $a_i \in {\cal L}$, we have:
\begin{equation}
\begin{array}{lll}
I \in \xi(p) &
O \not\in \xi(p) &
a_i \in \xi(p) \Leftrightarrow \wedge_ia_i \in \xi(p)
\end{array}
\end{equation}
Further, for $p, q
\in \Sigma$ and for $a, b, a_i \in {\cal L}$, we have:
\begin{eqnarray}
p \prec q &\Leftrightarrow& \xi(q) \subset \xi(p) \\
a \prec b &\Leftrightarrow& a \in \xi(r) {\rm \ then}\  b
\in \xi(r) \ \forall\ r \in \Sigma
\end{eqnarray}
\end{definition}
\begin{theorem} \label{the:statprop}
Consider an entity $S({\cal E}, \Sigma, X, {\cal O})$ and for $e \in
{\cal E}$ we consider the identified state property entity
$S(\Sigma, {\cal L}(e), \xi_e)$, then
$(\Sigma, {\cal L}(e), \xi_e)$ is a state property system.
\end{theorem}
Proof: We only have to remark that for an identified state property
entity, the infimum and supremum that are constructed in
\ref{complpreord} are the infimum and supremum for the partially
ordered set
${\cal L}(e)$. This makes ${\cal L}(e)$ into a complete lattice
with maximal element $I$ and minimal element
$0$.

\bigskip
\noindent
We will now show that the state property systems are naturally
connected to closure structures on the set
of states.
\begin{definition} [Cartan map]
Consider a state property entity $S(\Sigma, {\cal L}, \xi)$.
We introduce the function
\begin{equation}
\begin{array}{ll}
\kappa: {\cal L} \rightarrow {\cal P}(\Sigma) &
a \mapsto \kappa(a) 
\end{array}
\end{equation} 
\begin{equation}
p \in \kappa(a) \Leftrightarrow a \in \xi(p)
\end{equation}
that we call the `Cartan map' (Aerts 1981, 1982, 1983, Piron 1990).
\end{definition}
The meaning of the Cartan map is the following: $\kappa(a)$ is the
set of all states that make $a$ actual. Let us now introduce the
eigen maps.
\begin{definition} {\bf (eigen maps on the states)}
\label{eigenmap} Consider an experiment state outcome entity
$S({\cal E}, \Sigma, X, {\cal O})$. For $e \in {\cal E}$ and $A
\subset O(e)$, we define a map
$eig_e$, that we shall call the eigen map corresponding to the
experiment $e$: 
\begin{equation}
\begin{array}{ll}
eig_e : {\cal P}(O(e)) \rightarrow {\cal P}(\Sigma) &
A \mapsto eig_e(A) 
\end{array}
\end{equation}
\begin{equation}
p \in eig_e(A) \Leftrightarrow O(e,p) \subset A
\end{equation}
\end{definition}
The eigen map $eig_e$ connects a subset $A$ of outcomes of $e$ with
a subset of states $eig_e(A)$ such that if the entity $S$ is in one
of the states of $eig_e(A)$, one of the outcomes of $A$ occurs with
certainty. 
\begin{proposition} \label{testprop}
Consider an entity $S({\cal E}, \Sigma,
X, {\cal O})$ and for $e \in {\cal E}$ consider the state
property entity $S(\Sigma, {\cal L}(e), \xi_e)$. For
$a(A) \in {\cal L}(e)$ we have:
\begin{equation}
\kappa(a) = eig_e(A) \label{eigcartan}
\end{equation}
\end{proposition}
Before we proceed let us point out some of the properties of the
Cartan map and of the eigen maps.
\begin{proposition} \label{strimplic}
Consider a state property entity $S(\Sigma, {\cal L}, \xi)$. For
$a \in {\cal L}$ we have:
\begin{equation}
a \prec b \Leftrightarrow \kappa(a) \subset \kappa(b)
\end{equation}
\end{proposition}
\begin{proposition}
Consider an entity $S({\cal E}, \Sigma, X, {\cal O})$. The map
$eig_e$ introduced in definition~\ref{eigenmap} satisfies the
following properties:
\begin{eqnarray}
eig_e (\emptyset) &=& \emptyset \\
eig_e (O(e)) &=& \Sigma \\
eig_e(\cap_iA_i) &=& \cap_ieig_e(A_i)
\end{eqnarray}
\end{proposition}
Proof:  $p \in eig_e(\cap_iA_i) \Leftrightarrow O(e,p) \subset
\cap_iA_i$ $\Leftrightarrow$ $O(e,p)
\subset A_i \ \forall \ i$ $\Leftrightarrow$ $p \in eig_e(A_i)
\ \forall \ i$ $\Leftrightarrow$ $p \in
\cap_ieig_e(A_i)$.  

\bigskip
\noindent
Let us introduce some definitions.
\begin{definition} {\bf (closure system)}
Consider a set $W$. We say that ${\cal F} \subset {\cal P}(W)$ is a
closure system iff
\begin{eqnarray}
\emptyset &\in& {\cal F} \\
W &\in& {\cal F} \\
F_i \in {\cal F} &\Rightarrow& \cap_iF_i \in {\cal F}
\label{closinter}
\end{eqnarray}
\end{definition}
\begin{definition} {\bf (closure operator)}
Consider a set $W$. We say that $cl$ is a closure operator on $W$
iff, for $K,L \subset W$, we have:
\begin{eqnarray}
K &\subset& cl(K) \\
K \subset L &\Rightarrow& cl(K) \subset cl(L) \\
cl(cl(K)) &=& cl(K) \\
cl(\emptyset) &=& \emptyset
\end{eqnarray}
\end{definition}
\begin{proposition} 
If a set $W$ is equipped with a closure operator $cl$ and we
define a subset $F \subset W$ to be closed iff
$cl(F) = F$, then the set ${\cal F}$ of closed subsets of $W$
forms a closure system on $W$. Suppose on the other hand that we
consider a closure system ${\cal F}$ on $W$. If, for an arbitrary
$K \subset \Sigma$, we define :
\begin{equation}
cl(K) = \cap_{K \subset F, F \in {\cal F}}F \label{closure}
\end{equation}
then $cl$ is a closure operator on $W$ and ${\cal F}$ is the set of
closed subsets of $W$ defined by this closure operator.
\end{proposition}
Proof: First we prove~(\ref{closinter}). We have that $cl(\cap _i
F_i) \subset cl(F_i) \forall i$ implies
 $cl(\cap _i F_i)
\subset
\cap _i cl(F_i) = \cap _i F_i \subset cl (\cap _i F_i)$.
Now we show that (\ref{closure}) defines a closure operator on W.
So consider $K, L \subset W$. Clearly $cl(\emptyset) = \emptyset$,
$K \subset cl(K)$ and if $K \subset L$, then $cl(K)
\subset cl(L)$. If
$F \in {\cal F}$ then $cl(F) = F$, whence $cl(K) \in {\cal F}$
implies $cl(cl(K)) = cl(K)$. This shows that $cl$ is a closure
operator. Consider a set $K$ such that $cl(K) = K$, then $K =
\cap_{K \subset F, F \in {\cal F}}F$, and hence $K \in {\cal F}$.
It follows that $\cal F$ is the set of closed subsets for this $cl$.
\begin{theorem} \label{the:eigenclos}
Consider an entity $S({\cal E}, \Sigma, X, {\cal O})$, and eigen
maps $eig_e$, $e \in {\cal E}$. Let us denote the image of an eigen
map $eig_e$ by ${\cal F}(e)$, in other words,
\begin{equation}
{\cal F}(e) = \{F \subset \Sigma \ \vert\ \exists\ A \subset O(e),
F = eig_e(A)\}
\end{equation}
then ${\cal F}(e)$ is a closure system on $\Sigma$. We will call
the elements of ${\cal F}(e)$ the $e$-eigen closed sets.
\end{theorem} 
Proof: We have $eig_e(\emptyset) = \emptyset$ and $eig_e(O(e)) =
\Sigma$. Consider $F_i \in {\cal F}(e)$. Then
$F_i = eig_e(A_i)$. We have $\cap_iF_i = \cap_ieig_e(A_i)
= eig_e(\cap_iA_i)$. This shows that $\cap_iF_i \in {\cal F}(e)$.
\begin{theorem} \label{the:cartaneig}
Consider an entity $S({\cal E}, \Sigma, X, {\cal O})$ and for $e \in
{\cal E}$ consider the state property entity $S(\Sigma, {\cal
L}(e), \xi_e)$. We have:
\begin{equation}
\kappa({\cal L}(e)) = {\cal F}(e)
\end{equation}
\end{theorem}
Proof: Consider $F \in \kappa({\cal L}(e))$. Then there
exists $a(A) \in {\cal L}(e)$ such that $\kappa(a(A)) = F$. From
\ref{eigcartan} follows that $\kappa(a(A)) = eig_e(A)$ and hence $F
\in {\cal F}(e)$. Suppose now that $F \in {\cal F}(e)$, then we
have $F = eig_e(A)$ for some $A \subset O(e)$. Again from
\ref{eigcartan} follows that $F = \kappa(a(A))$ and hence $F \in 
\kappa({\cal L}(e))$.
\section{State property systems and closure spaces}
From theorem~\ref{the:statprop} follows that an identified state
property entity $S(\Sigma, {\cal L}(e), \xi_e)$ is represented
mathematically by a state property system. From
theorem~\ref{the:eigenclos} and
\ref{the:cartaneig} follows that there is a closure system on the
states connected with the state property entity $S(\Sigma, {\cal
L}(e), \xi_e)$. We will see now that the connection between state
property systems and closure systems is even much more intimate
than we would expect from the foregoing section. Since we will
encounter the mathematical concept of state property system and
closure system again for the description of entities, we want to
make the results of this section independent of the physical
content. Therefore we introduce some concepts again that have been
introduced earlier within a specific physical context.
\begin{proposition}
Suppose that $(\Sigma, {\cal L}, \xi)$ is a state property system.
We introduce the function $\kappa$, the Cartan map:
\begin{equation}
\begin{array}{ll}
\kappa: {\cal L} \rightarrow {\cal P}(\Sigma) &
a \mapsto \kappa(a) = \{p\ \vert\ a \in \xi(p)\}
\end{array}
\end{equation}
For $a, b, a_i \in {\cal L}$ we have:
\begin{eqnarray}
\kappa(I) &=& \Sigma \\
\kappa(0) &=& \emptyset \\
a \prec b &\Leftrightarrow& \kappa(a) \subset \kappa(b) \\
\kappa(\wedge_ia_i) &=& \cap_i\kappa(a_i)
\end{eqnarray}
\end{proposition}
Proof: Since $I \in \xi(p)\ \forall\ p \in \Sigma$, we
have $\kappa(I) = \Sigma$. Since $0 \not\in \xi(p) \ \forall\ p \in
\Sigma$ we have $\kappa(0) = \emptyset$.
 Take $a \prec b$ and consider $p \in \kappa(a)$.
Then $a \in \xi(p)$ and since $a \prec b$ we have $b \in \xi(p)$.
This implies that $p \in \kappa(b)$. Hence we have shown that
$\kappa(a) \subset \kappa(b)$. Take now $\kappa(a) \subset
\kappa(b)$. Consider $p \in \Sigma$ such that $a \in
\xi(p)$. Then $p \in \kappa(a)$ and hence $p \in
\kappa(b)$. From this follows that $b \in \xi(p)$. This means that
$a \prec b$. We have $\wedge_ia_i \prec a_j\
\forall\ j$. This implies that $\kappa(\wedge_ia_i) \subset
\kappa(a_j)\ \forall\ j$. Hence $\kappa(\wedge_ia_i)
\subset \cap_i\kappa(a_i)$. Take now $p \in \cap_i\kappa(a_i)$,
then $p \in \kappa(a_j)\ \forall\ j$. Hence $a_j \in
\xi(p)\ \forall\ j$ which implies that $\wedge_ia_i \in
\xi(p)$. From this follows that $p \in \kappa(\wedge_ia_i)$. As a
consequence we have $\cap_i\kappa(a_i) \subset
\kappa(\wedge_ia_i)$. This shows that $\kappa(\wedge_ia_i) =
\cap_i\kappa(a_i)$.
\begin{theorem} \label{statprop}
Suppose that $(\Sigma, {\cal L}, \xi)$ is a state property system.
Let us introduce
${\cal F} =
\{\kappa(a)\ \vert\ a \in {\cal L}\}$. Then ${\cal F}$ is a
closure system on $\Sigma$.
\end{theorem}
Proof: From the foregoing theorem follows that $\Sigma \in {\cal
F}$ and $\emptyset \in {\cal F}$. Consider $F_i \in {\cal F}$. Then
there exists $a_i \in {\cal L}$ such that $\kappa(a_i) = F_i$. We
have $\kappa(\wedge_ia_i) = \cap_i\kappa(a_i) =
\cap_iF_i$. This shows that $\cap_iF_i \in {\cal F}$.

\bigskip
\noindent
This theorem shows that with a state property system correspond in
a natural way a closure system on the set of states, where the
properties are represented by the closed subsets. We can show that
with each closure system on the set of states corresponds also a
state property system.
\begin{theorem} \label{closure}
Consider a set $\Sigma$ with a closure system ${\cal F}$ on
$\Sigma$. We define ${\cal L}$ in the following way. The elements
of ${\cal L}$ are the elements of ${\cal F}$, hence ${\cal L} =
{\cal F}$ where we identify the maximal element $I$ of ${\cal L}$
with
$\Sigma$ and the minimal element $0$ of ${\cal L}$
with $\emptyset$. For $F, G \in {\cal L}$ we define $F \prec G$ iff
$F \subset G$. For $F_i \in {\cal L}$ we define $\wedge_iF_i =
\cap_iF_i$ and $\vee_iF_i = cl(\cup_iF_i)$. We introduce the
function $\xi$ in the following way:
\begin{equation}
\begin{array}{ll}
\xi: \Sigma \rightarrow {\cal P}({\cal L}) &
p \mapsto \{F\ \vert\ F \in {\cal F}, p \in F\}
\end{array}
\end{equation}
We introduce a pre-order relation on $\Sigma$. For $p,q \in
\Sigma$ we define:
\begin{equation}
p \prec q \Leftrightarrow \xi(q) \subset \xi(p)
\end{equation}
Then $(\Sigma, {\cal L}, \xi)$ is a
state property system.
\end{theorem}
Proof: It is easy to show that ${\cal L}, \prec, \wedge, \vee$ is a
complete lattice. We have $I \in \xi(p)\ \forall\ p \in
\Sigma$ and $0 \not\in \xi(p)\ \forall\ p \in \Sigma$. Suppose that
$F_i \in \xi(p)\ \forall\ i$. This means that $p \in F_i\
\forall\ i$ and hence $p \in \cap_iF_i$. As a consequence we
have $\cap_iF_i \in \xi(p)$. Let us verify that ${\cal J} =
\{\xi(p)\ \vert\ p \in \Sigma\}$ is an ordering set. Suppose that
$F, G \in {\cal L}$ and $F \subset G$. Consider $p$ such that
$F \in \xi(p)$ and hence $p \in F$. This implies that $p \in G$
and hence $G \in \xi(p)$. Suppose now that for $p \in
\Sigma$ we have that $F \in \xi(p)$ implies that $G \in
\xi(p)$. Consider then $p \in F$ and hence $F \in \xi(p)$. Then
$G \in \xi(p)$ and as a consequence $p \in G$. This shows that $F
\subset G$. It is easy to verify that $\prec$ is a pre-order on
$\Sigma$.

\bigskip
\noindent
Theorem~\ref{statprop} and theorem~\ref{closure} show that there is
a natural correspondence between state property systems and closure
systems. Let us introduce the morphisms of these structures.
Consider two state property systems $(\Sigma, {\cal L},  \xi)$ and
$(\Sigma',{\cal L}', \xi')$. As we have explained, both state
property systems describe respectively identified state property
entities
$S(\Sigma, {\cal L}, \xi) $ and $S'(\Sigma', {\cal L}', \xi') $. We
will arrive at the notion of morphism by analyzing the situation
where the entity
$S$ is a sub entity of the entity $S'$. If the entity $S$ is a sub
entity of the entity $S'$ then we have natural requirements that
have to be satisfied.

\smallskip
\noindent i) If the entity $S'$ is in a state $p'$ then also
the entity $S$ is in a state $m(p')$, and all states of
$S$ are of this type. This defines a surjective function $m$ from
the set of states of $S'$ to the set of states of $S$.

\smallskip
\noindent ii) If we consider a property $a$ of the entity $S$,
then there corresponds a property $n(a)$ of the entity $S'$ with
this property $a$. This defines a function $n$ from the set of
properties of $S$ to the set of properties of $S'$.

\bigskip
\noindent
{\it Requirement of covariance connected to the relation of `being
a sub entity' of `an entity'}

\medskip
\noindent
The most important and fundamental
requirement as to the concept of sub entity and the derived concept
of morphism will be put forward now. It is a requirement of
`covariance' on the ontological level. We want to express now that
the reality of the physical phenomenon described by the entity or
by the sub entity, depending of whether we consider a bigger piece
(the entity) or smaller piece (the sub entity) of this reality, is
independent of this choice.

This implies that if the entity $S'$ is in state $p'$ then the sub
entity
$S$ is in state $m(p')$. Suppose that the property $a$ is actual,
then also the property $n(a)$ must be actual. This shows that we
must have:
\begin{equation}
a \in \xi(m(p')) \Leftrightarrow n(a) \in \xi'(p')
\end{equation}
We are ready now to present a formal definition of a morphism.
\begin{definition} \label{mor} \label{morphism}
Consider two state property systems $(\Sigma,{\cal L}, \xi)$ and
$(\Sigma',{\cal L}', \xi')$. We say that a couple of functions
$(m,n)$ is a morphism iff $m$ is a function:
\begin{equation}
\begin{array}{ll}
m: \Sigma' \rightarrow \Sigma &
p' \mapsto m(p')
\end{array}
\end{equation}
and $n$ is a function:
\begin{equation}
\begin{array}{ll}
n: {\cal L} \rightarrow {\cal L}' &
a \mapsto n(a)
\end{array}
\end{equation}
such that
\begin{equation} \label{comm}
a \in \xi(m(p')) \Leftrightarrow n(a) \in \xi'(p')
\end{equation}
\end{definition}
\begin{proposition}
Consider two state property systems $(\Sigma, {\cal L}, \xi)$
and $(\Sigma', {\cal L}', \xi')$. The couple of functions $(m,n)$,
as introduced in definition~\ref{morphism} is a morphism iff we
have:
\begin{equation}
\xi \circ  m = n^{-1} \circ \xi'
\end{equation}
\end{proposition}
\begin{proposition}
Consider two state property systems $(\Sigma, {\cal L}, \xi)$ and
$(\Sigma', {\cal L}',
\xi')$ connected by a morphism $(m,n)$. For $p', q' \in \Sigma'$,
$a, b \in {\cal L}$, $a_i \in {\cal L}$, we have:
\begin{eqnarray}
a \prec b &\Rightarrow& n(a) \prec n(b) \\
n(\wedge_ia_i) &=& \wedge_in(a_i) \\
n(I) &=& I' \\
n(0) &=& 0' \\
p' \prec q ' &\Rightarrow& m(p') \prec m(q')
\end{eqnarray}
\end{proposition}
Proof: Suppose that $a
\prec b$. Consider $\xi'(p')$ such that $n(a) \in \xi'(p')$. Then
we have $a \in \xi(m(p'))$. Since $a \prec b$ we have
$b \in \xi(m(p'))$. From this follows that $n(b) \in \xi'(p')$. So
we have shown that $n(a) \prec n(b)$. We have
$\wedge_ia_i \prec a_j\ \forall\ j$ and hence $n(\wedge_ia_i) \prec
n(a_j)\ \forall\ j$. This shows that $n(\wedge_ia_i)
\prec \wedge_in(a_i)$. We still have to show that $\wedge_in(a_i)
\prec n(\wedge_ia_i)$. Consider
$\xi'(p')$ such that $\wedge_in(a_i) \in \xi'(p')$. This implies
that $n(a_j) \in \xi'(p')\ \forall\ j$. But from this follows that
$a_j \in \xi(m(p'))\ \forall\ j$ and hence $\wedge_ia_i \in
\xi(m(p'))$. As a consequence we have
$n(\wedge_ia_i) \in \xi'(p')$. But then we have shown that
$\wedge_in(a_i) \prec n(\wedge_ia_i)$. As a consequence we have
$n(\wedge_ia_i) = \wedge_in(a_i)$. We have $n(I) \prec I'$.
Consider $p' \in \Sigma'$, then $I' \in \xi'(p')$. We also have $I
\in \xi(m(p'))$, which implies that $n(I) \in \xi'(p')$. This
proves that $I' \prec n(I)$ and hence $n(I) = I'$. In an analogous
way we prove that $n(0) = 0'$. Suppose that $p' \prec q'$. We then
have $\xi'(q') \subset \xi'(p')$. From this follows that
$n^{-1}(\xi'(q'))
\subset n^{-1}(\xi'(p'))$. As a consequence we have $\xi(m(q'))
\subset \xi(m(p'))$ and this implies that $m(p') \prec m(q')$.
\begin{proposition}
Suppose that we have two state property systems $(\Sigma, {\cal L},
\xi)$ and
$(\Sigma', {\cal L}', \xi')$ connected by a morphism
$(m,n)$. Consider the Cartan maps $\kappa$ and $\kappa'$ that
connect these state property systems with their corresponding
closure systems $(\Sigma, {\cal F})$ and
$(\Sigma', {\cal F}')$. For $a \in {\cal L}$ we have:
\begin{equation}
m^{-1}(\kappa(a)) = \kappa'(n(a))
\end{equation}
\end{proposition}
Proof: We have: $p' \in m^{-1}(\kappa(a)) \Leftrightarrow m(p') \in
\kappa(a) \Leftrightarrow a \in \xi(m(p'))
\Leftrightarrow n(a) \in \xi'(p') \Leftrightarrow p'
\in \kappa'(n(a))$.
\begin{theorem} \label{cont}
Suppose that we have two state property systems $(\Sigma, {\cal L}, 
\xi)$ and
$(\Sigma', {\cal L}', \xi')$ connected by a morphism $(m,n)$
and the Cartan maps $\kappa$ and $\kappa'$ that connect these state
property systems with their corresponding closure systems $(\Sigma,
{\cal F})$ and
$(\Sigma', {\cal F}')$. The function $m$ is a continuous
function for the closure systems.
\end{theorem}
Proof: Take a closed subset $F \in {\cal F}$ and consider $m^{-1}(F)$. Since $F \in {\cal F}$ we have $a \in {\cal L}$
such that $\kappa(a) = F$. From the foregoing theorem we have $m^{-1}(F) = m^{-1}(\kappa(a)) = \kappa'(n(a)) \in {\cal
F}'$. This shows that $m$ is continuous.

\bigskip
\noindent
We are now at the point of making explicit the
powerful representation that the closure system gives for a state
property system. Let us identify the morphisms of the closure
systems that correspond to the morphisms that we have introduced in
the state property systems.
\begin{theorem} \label{morph}
Suppose that we have two closure systems $(\Sigma, {\cal F})$ and
$(\Sigma', {\cal F}')$ and a continuous function $m:
\Sigma'
\rightarrow \Sigma$. Consider the state property systems $(\Sigma,
{\cal L},  \xi)$ and
$(\Sigma', {\cal L}', \xi')$ corresponding with these two
closure systems, as proposed in theorem~\ref{closure}. If we define
the couple $(m,n)$ such that
\begin{equation}
n = m^{-1}
\end{equation}
then (m,n) is a morphism between the two state property systems.
\end{theorem}
Proof: We have to prove that the couple $(m,m^{-1})$ satisfies that
properties of a state property morphism as put forward in
definition~\ref{mor}. Since $m$ is continuous we have that $m^{-1}$
is a function from ${\cal F}$ to ${\cal F}'$. Let us show now
formula~\ref{comm} using the definition of $\xi$ and $\xi'$ as put
forward in theorem~\ref{closure}. We have
$F
\in
\xi(m(p'))
\Leftrightarrow m(p')
\in F
\Leftrightarrow p'
\in m^{-1}(F)
\Leftrightarrow m^{-1}(F) \in \xi'(p')$.

\bigskip
\noindent
Theorem~\ref{statprop} and theorem~\ref{closure} show that there is
a natural correspondence between state property systems and closure
systems. Theorem \ref{cont} and \ref{morph} show that also the
morphisms of both structures correspond. This indicates that the
correspondence may be categorical. Indeed, we will analyze the
categorical aspect of this correspondence in detail in (Aerts,
Colebunders, Van der Voorde and Van Steirteghem 1998) and show that
the category of state property systems and its morphisms and the
category of closure spaces and continuous functions are equivalent
categories.

The set of all testable properties is given by $\cup_{e \in {\cal
E}}{\cal L}(e)$. Let us remark that a priori $\cup_{e \in {\cal
E}}{\cal L}(e)$ is not a complete pre-order set. This seems to
contradict the results of earlier work. Indeed in (Piron 1976,
1989, 1990, Aerts 1981, 1982, 1983) it is shown that the set of all
testable properties, hence $\cup_{e \in {\cal E}}{\cal L}(e)$, is a
complete pre-order set. We remark that in these earlier approaches
equivalent properties are identified such that identified state
property entities are considered: the complete pre-order set is
then a complete lattice, but this is not the origin of the problem
that we want to point out here. We want to explain why in the
earlier approaches completeness was derived for the set of all
testable properties while here we can only derive it for the set of
testable properties connected to one definite experiment. First we
remark that in the earlier approaches the complete pre-order set
was constructed by introducing explicitly all the mixed
experiments. If we consider the mixed experiment
$e({\cal E})$ and the set of $e({\cal E})$-testable properties
${\cal L}(e({\cal E}))$, it can be shown that `under a certain
condition', the set of $e({\cal E})$-testable properties contains
all the other sets of $e(E)$-testable properties, where $E \subset
{\cal E}$. This means that $\cup_{e \in {\cal E}}{\cal L}(e) =
{\cal L}(e({\cal E}))$. We have shown in (Aerts 1994) that the
condition that implies this equality is a condition of
`distinguishable experiments'. This condition of `distinguishable
experiments' leading to the completeness of the set of all testable
properties was unconsciously assumed in the already mentioned
earlier approaches (Piron 1976, 1989, 1990, Aerts 1981, 1982, 1983).
There it was taken for granted that an experiment, called test,
question or experimental project in (Piron 1976, 1989, 1990, Aerts
1981, 1982, 1983), that can be distinguished from all the others,
can be associated with each property (e.g. by labeling the test by
means of the property). At first sight it seems indeed that it is
always possible to do so. But in a formalism like the one we
propose here, the properties as well as the experiments that can be
used to test these properties, are given from the start. It is
against the `rules of the game' to introduce new experiments for
the properties just with the aim of being able to distinguish them
from all the others. So we must conclude that the completeness can
a priori only be shown for the set of testable properties connected
to a definite experiment. Let us demonstrate the details of this
situation in our formalism.
\begin{proposition}
Consider an entity $S(M({\cal E}), M(\Sigma), M(X), {\cal O})$, and
suppose that $e(E) \in M({\cal E})$ is a mixed experiment and
consider $A \subset O(e(E))$. We have:
\begin{equation}
eig_{e(E)}(A) = \cap_{e \in E} eig_e(A \cap O(e))
\end{equation}
\end{proposition}
Proof:
$p \in eig_{e(E)}(A)$ $\Leftrightarrow$  $O(e(E),p) \subset
A$ $\Leftrightarrow$ $\cup_{e \in E}O(e,p) \subset A$ 
$\Leftrightarrow$ $O(e,p) \subset A \ \forall e \in
E$ $\Leftrightarrow$ $O(e,p) \subset A \cap O(e) \ \forall e \in E$
$\Leftrightarrow$ $p \in eig_e(A \cap O(e)) \ \forall e \in
E$ $\Leftrightarrow$ $p \in \cap_{e \in E}eig_e(A \cap O(e))$.
\begin{definition} {\bf (distinguishable experiment entity)}
Suppose that we have an entity $S({\cal E}, \Sigma, X, {\cal
O})$. We say that two experiments $e, f
\in {\cal E}$  are distinguishable iff $O(e) \cap O(f)
=\emptyset$. We say that the entity $S$ is a `distinguishable
experiment entity' iff $\ \forall \ e, f \in {\cal E}$ we have that
$e$ and $f$ are distinguishable.
\end{definition}
Two experiments $f$ and $g$ are distinguishable if they can
be distinguished from each other by means of their outcomes. Let us
explain intuitively in the spirit of (Piron 1976, 1989, 1990, Aerts
1981, 1982, 1983) why distinguishable experiments are necessary for
the completeness of the set of testable properties. We will use the
concept of `test', `question' or `experimental project' as it was
introduced in (Piron 1976, 1989, 1990, Aerts 1981, 1982, 1983)
without explicitly defining it again. The reader not acquainted
with this concept can better skip this section and again pick up
just before the next proposition. There the intuitive reasoning that
we will give now is repeated in the approach being developed in this
paper.

Suppose that we consider a test
$\alpha(f,A)$, consisting of performing the experiment
$f$ and giving the positive answer `yes' if the outcome is in $A$,
and a test $\alpha(g,B)$, consisting of performing the experiment
$g$ and giving a positive answer `yes' if the outcome is in $B$.
To `prove' the completeness one introduces in (Piron 1976, 1990,
Aerts 1981, 1982, 1983) the concept of `product test', and if
$\alpha(f,A)$ tests whether the property
$a(f,A)$ is actual and
$\alpha(g,B)$ tests whether the property $a(g,B)$ is actual, then
$\alpha(f,A) \cdot \alpha(g,B)$ tests whether an infimum of the
properties $a(f,A)$ and $a(g,B)$ is actual. It is by requiring that
the set of tests on the entity $S$ contains all the product tests,
that the pre-order set of testable properties becomes complete,
because an infimum exists for each subset of properties. The
product test is defined by means of the experiment
$e(\{f,g\})$, and is given by
$\alpha(e(\{f,g\}),A \cup B)$, consisting of performing the
experiment
$e(\{f,g\})$ and giving a positive answer `yes' if the outcome is
in $A \cup B$. We remark that, although the product test can always
be defined, it only tests whether the two properties $a(f,A)$ and
$a(g,B)$ are actual, if $f$ and $g$ are distinguishable
experiments. Indeed, suppose that $f$ and $g$ are not
distinguishable, then $O(f) \cap O(g) \not= \emptyset$, which means
that there is at least one outcome $x \in O(f) \cap O(g)$. Suppose
that $A$ does not contain this outcome while $B$ does, then it is
possible that the entity $S$ is in a state $p$ such that $e$ has as
possible outcomes the set $A \cup  \{x\}$, which is a state where
$a(g,A)$ is not actual, and where $g$ has as possible outcomes $B$.
Then $e(\{f,g\})$ has as possible outcomes $A \cup B$, which means
that in this state $p$ the test $\alpha(e(\{f,g\}),A \cup B)$ gives
with certainty a positive outcome. This shows that in this case of
non distinguishable experiments, $\alpha(e(\{f,g\}),A \cup B)$ does
not test the actuality of the infimum of the properties $a(f,A)$
and $a(g,B)$.
\begin{proposition} \label{clos}
Suppose that we have an entity $S(M({\cal E}), M(\Sigma), M(X),
{\cal O})$. Suppose that we denote by ${\cal F}(e({\cal E}))$ the
collection of eigenstate sets of the experiment $e({\cal E})$. If
all the experiments are distinguishable then for $E \subset {\cal
E}$ we have:
\begin{equation}
{\cal F}(e(E)) \subset
{\cal F}(e({\cal E}))
\end{equation}
\end{proposition}
Proof: Consider an arbitrary element $F \in {\cal F}(e(E))$. Then
there exists $A
\subset O(e(E))$ such that $F = eig_{e(E)}(A)$. Consider $A' = A
\cup (\cup_{e \in {{\cal E}, e \not\in E}}O(e))$, then we have
$eig_{e({\cal E})}(A') = eig_{e(E)}(A)$, which shows that $F \in
{\cal F}(e({\cal E}))$.
\begin{theorem}
Suppose that $S(M({\cal E}), M(\Sigma), M(X), {\cal O})$ is a
distinguishable experiment entity such that $e({\cal E})
\in M({\cal E})$. We then have:
\begin{equation}
\cup_{e \in M({\cal E})}{\cal F}(e) = {\cal F}(e({\cal E}))
\end{equation}
\end{theorem}
Proof: From proposition~\ref{clos} we have ${\cal F}(e) \subset
{\cal F}(e({\cal E}))$ for all $e \in M({\cal E})$, which implies
that $\cup_{e \in M({\cal E})}{\cal F}(e) \subset {\cal F}(e({\cal
E}))$. Since
$e({\cal E}) \in M({\cal E})$ we have ${\cal F}(e({\cal E}))
\subset \cup_{e \in M({\cal E})}{\cal F}(e)$.

\bigskip
\noindent
For such a distinguishable experiment entity we can also prove that
the set of all testable properties is a complete pre-order set.
\begin{theorem}
Suppose that $S(M({\cal E}), M(\Sigma), M(X), {\cal O}, {\cal L},
\xi)$ is a distinguishable experiment entity such that $e({\cal E})
\in M({\cal E})$. We then have:
\begin{equation}
\cup_{e \in M({\cal E})}{\cal L}(e) = {\cal L}(e({\cal E}))
\end{equation}
and the set of testable properties
$\cup_{e \in M({\cal E})}{\cal L}(e)$ is a complete pre-order set
with a maximal element $I$, such that $\kappa(I) = M(\Sigma)$, and
a minimal element $0$, such that $\kappa(0) = \emptyset$.
\end{theorem}
\begin{theorem}
Suppose that $S(M({\cal E}), M(\Sigma), M(X), {\cal O}, {\cal L},
\xi)$ is an identified distinguishable experiment entity such that
$e({\cal E}) \in M({\cal E})$. Then the state property system
$(\Sigma, {\cal L}(e({\cal E})), \xi_{e({\cal E})})$ describes the
state property entity $S(\Sigma, {\cal L}(e({\cal E})),
\xi_{e({\cal E})})$, and ${\cal L}(e({\cal E}))$ contains all
testable properties of the entity.
\end{theorem}
We mention that all the calculations in the earlier approaches
(Piron 1976, 1989, 1990, Aerts 1981, 1982, 1983) actually take
place in the state property system
$(\Sigma, {\cal L}(e({\cal E})), \xi_{e({\cal E})})$.
\section{The eigen-closure}
We analyze now how for a state experiment outcome entity closure
structures can be introduced in a natural way on the product set
${\cal E} \times \Sigma$ and the set of experiments ${\cal E}$.
\begin{definition} {\bf (central eigen map)}
Let us consider an entity $S({\cal E}, \Sigma, X, {\cal O})$. For $A
\subset X$ we introduce:
\begin{equation}
\begin{array}{ll}
eig : {\cal P}(X) \rightarrow {\cal P}({\cal E} \times \Sigma) &
A \mapsto eig(A)
\end{array}
\end{equation}
such that
\begin{equation}
(e,p) \in eig(A) \Leftrightarrow O(e,p) \subset A
\end{equation}
\end{definition}
\begin{definition} {\bf (eigen maps on the
experiments)} \label{eigexp} Let us consider an entity $S({\cal E},
\Sigma, X, {\cal O})$. For $p \in \Sigma$ we define a map $eig_p$,
that we call the eigen map corresponding to the state $p$: 
\begin{equation}
\begin{array}{ll}
eig_p : {\cal P}(O(p)) \rightarrow {\cal P}({\cal E}) &
A \mapsto eig_p(A) 
\end{array}
\end{equation}
\begin{equation}
e \in eig_p(A) \Leftrightarrow O(e,p) \subset A
\end{equation}
\end{definition}
\begin{proposition}
Let us consider an entity $S({\cal E}, \Sigma, X, {\cal O})$
and the central eigen map $eig : {\cal P}(X)
\rightarrow {\cal P}({\cal E} \times \Sigma), A \mapsto
eig(A)$, then for $A_i \subset X$ we have :
\begin{equation}
eig(\cap_iA_i) = \cap_ieig_(A_i)
\end{equation}
\end{proposition} 
Proof: $(e,p) \in eig(\cap_iA_i) \Leftrightarrow O(e,p) \subset
\cap_iA_i \Leftrightarrow O(e,p) \subset A_i \ \forall i
\Leftrightarrow (e,p) \in eig(A_i) \ \forall i \Leftrightarrow
(e,p) \in \cap_ieig(A_i).$
\begin{proposition}
The map $eig_p$ introduced in definition~\ref{eigexp} satisfies the
following properties:
\begin{eqnarray}
eig_p (\emptyset) &=& \emptyset \\
eig_p (O(p)) &=& {\cal E} \\
eig_p(\cap_iA_i) &=& \cap_ieig_p(A_i)
\end{eqnarray}
\end{proposition}
\begin{definition}
Let us consider an entity $S({\cal E}, \Sigma, X, {\cal O})$ and
the eigen maps $eig_p$, $p \in \Sigma$. We denote the image of an
eigen map $eig_p$ by ${\cal G}(p)$, in other words:
\begin{equation}
{\cal G}(p) = \{G \subset {\cal E} \ \vert\ \exists\ A \subset
O(p), G = eig_p(A)\}
\end{equation}
\end{definition} 
\begin{definition}
Let us consider an entity $S({\cal E}, \Sigma, X, {\cal O})$
and the central eigen map $eig$. We denote the set of all images of
$eig$ by ${\cal Y}$, hence 
\begin{equation}
{\cal Y} = \{Y \subset {\cal E} \times \Sigma \ \vert \ \exists \ A
\subset X, Y = eig(A)\}
\end{equation}
\end{definition}
We have shown that ${\cal F}(e)$ is a closure system on
$\Sigma$. In an analogous way we show that ${\cal Y}$ is a closure
system on ${\cal E} \times \Sigma$ and ${\cal G}(p)$ is a closure
system on ${\cal E}$.
\begin{theorem}
Let us consider an entity $S({\cal E}, \Sigma, X, {\cal O})$, then
${\cal Y}$ and ${\cal G}(p)$ are closure systems for every $e \in
{\cal E}, p \in
\Sigma$, respectively on ${\cal E} \times \Sigma$, $\Sigma$
and $\cal E$.
\end{theorem}
Proof: We give the proof for ${\cal Y}$. Suppose that $Y_i \in
{\cal Y}$. Then $\exists \ A_i \subset X$ such that $Y_i =
eig(A_i)$. We have $(e,p) \in eig(\cap_iA_i) \Leftrightarrow (e,p)
\in \cap_ieig(A_i) = \cap_iY_i$.
\begin{definition} {\bf (generating set)}
Suppose we have a set $Z$ and ${\cal F}$
is the set of closed subsets corresponding to a closure operator
$cl$ on $Z$. The collection $\cal B \subset {\cal F}$ is a
`generating set' for
$\cal F$ iff for each subset $F \in {\cal F}$ we have a family
$B_i \in  {\cal B}$ such that $F = \cap_iB_i$
\end{definition}
\begin{proposition}
Suppose we have a set $Z$ equipped with a closure $cl$ and $\cal B$
is a generating set for the set of closed subsets $\cal F$. Then
for an  arbitrary subset $K \subset Z$ we have : 
\begin{equation}
cl(K) = \cap_{K \subset B, B\in {\cal B}} B
\end{equation}
\end{proposition}
Proof: We know that $cl(K) = \cap_{K \subset F, F \in {\cal
F}}F$. Because $\cal B$ is a generating set for $\cal F$ we have $F
= \cap_{F
\subset B, B \in {\cal B}}B $. Hence $cl(K) = \cap_{K \subset
F}(\cap_{F
\subset B}B) = \cap_{K \subset B, B \in {\cal B}}B$.

\bigskip
\noindent
On the set of states $\Sigma$ we have a collection of closure
systems ${\cal F}(e), e \in {\cal E}$. It is easy to show that they
generate a global closure system on $\Sigma$.
\begin{theorem}
Let us consider an entity $S({\cal E}, \Sigma, X, {\cal O})$ and the
set of eigen maps $\{eig_e \vert e \in {\cal E}\}$ and
corresponding closure systems ${\cal F}(e)$. Put
${\cal A} = \cup_{e \in {\cal E}}{\cal F}(e)$, and consider: 
\begin{equation}
{\cal F} = \{\cap_iA_i \vert A_i \in {\cal A}\}
\end{equation}
Then ${\cal F}$ is a closure system on $\Sigma$ generated by ${\cal
A}$.
\end{theorem}
Proof: Consider $F_i \in {\cal F}$. Then there exist $A_{ij} \in
{\cal A}$ such that $F_i =
\cap_jA_{ij}$. We now have $\cap_iF_i = \cap_i\cap_jA_{ij}$ which
shows that $\cap_iF_i \in {\cal F}$.

\bigskip
\noindent In an analogous way the set of closure systems ${\cal
G}(p)$ on ${\cal E}$ generates a global closure system on ${\cal
E}$.
\begin{theorem}
Let us consider an entity $S({\cal E}, \Sigma, X, {\cal O})$ and
the set of eigen maps $\{eig_p \vert p \in
\Sigma\}$ and corresponding closure systems ${\cal G}(p)$. Put
${\cal C} = \cup_{p \in \Sigma}{\cal G}(p)$, and consider: 
\begin{equation}
{\cal G} = \{\cap_iC_i \vert C_i \in {\cal C}\}
\end{equation}
Then ${\cal G}$ is a closure system on ${\cal E}$ generated by ${\cal
C}$.
\end{theorem}
\begin{definition} Let us consider an entity $S({\cal E}, \Sigma, X,
{\cal O})$, we shall call ${\cal Y}$, ${\cal F}$ and ${\cal G}$
respectively the central eigen, the state eigen and the experiment
eigen closure system and denote them from now on ${\cal F}_{eig},
{\cal G}_{eig}$, and ${\cal Y}_{eig}$. To make notations not to
heavy we will denote the closure operator for each of the closure
system by
$cl_{eig}$. Hence :
\begin{equation}
\begin{array}{ll}
cl_{eig} : {\cal P}({\cal E} \times \Sigma) \rightarrow {\cal
Y}_{eig} \subset {\cal P}({\cal E} \times \Sigma) & K
\mapsto cl_{eig}(K) =
\cap_{K
\subset Y, Y \in {\cal Y}}Y \\
cl_{eig} : {\cal P}(\Sigma) \rightarrow {\cal F}_{eig}
\subset {\cal P}(\Sigma) & K \mapsto cl_{eig}(K) = \cap_{K
\subset F, F \in {\cal F}}F \\
cl_{eig} : {\cal P}({\cal E}) \rightarrow {\cal G}_{eig} \subset
{\cal P}({\cal E}) & K \mapsto cl_{eig}(K) =
\cap_{K
\subset G, G \in {\cal G}}G
\end{array}
\end{equation}
\end{definition}
We could ask ourselves now what the relation is between
the closures ${\cal Y}_{eig}$, ${\cal F}_{eig}$ and ${\cal
G}_{eig}$. Could it be that the state eigen closure and the
experiment eigen closure are in some way `traces' of the central
eigen closure? To see whether this is the case, let us introduce :
\begin{definition}
Let us consider an entity $S({\cal E}, \Sigma, X, {\cal O})$ and
the central eigen closure ${\cal Y}_{eig}$. For $Y
\in {\cal Y}_{eig}$ we introduce:
\begin{equation}
Y_{state} = \{p \in \Sigma \ \vert\ \forall\ e \in {\cal E}, (e,p)
\in Y \}
\end{equation}
We then define :
\begin{equation}
{\cal Y}_{eig}(state) = \{Y_{state}\ \vert\ Y \in {\cal Y}_{eig}\}
\end{equation}
\end{definition}
\begin{proposition}
${\cal Y}_{eig}(state)$ is a closure system on the set of states
$\Sigma$.
\end{proposition}
Proof: Consider ${\cal E} \times \Sigma \in {\cal Y}_{eig}$, then
$({\cal E} \times \Sigma)_{state} = \Sigma$, which shows that
$\Sigma \in {\cal Y}_{eig}(state)$. Obviously $\emptyset \in {\cal
Y}_{eig}(state)$. Consider now
$Z_i \in {\cal Y}_{eig}(state)$, which means that there exists $Y_i
\in {\cal Y}_{eig}$ such that $Z_i = (Y_i)_{state}$. Consider
$(\cap_iY_i)_{state}$. We have $p \in (\cap_iY_i)_{state}
\Leftrightarrow \ \forall e \in {\cal E}, (e,p) \in \cap_iY_i
\Leftrightarrow \ \forall\ e \in {\cal E}, \ \forall i, \ (e,p) \in
Y_i
\Leftrightarrow \ \forall i, \ p \in (Y_i)_{state} \Leftrightarrow
p \in \cap_i(Y_i)_{state}.$

\bigskip
\noindent In the example of section~\ref{exam} we show that in
general ${\cal F}_{eig}$ is not equal to
${\cal Y}_{eig}(state)$, but we can prove the equality for
distinguishable experiment entities.
\begin{theorem}
Let us consider a distinguishable experiment entity $S({\cal E},
\Sigma, X, {\cal O})$ with central eigen closure system
${\cal Y}_{eig}$ and state eigen closure system ${\cal F}_{eig}$.
Then we have:
\begin{equation}
{\cal F}_{eig} = {\cal Y}_{eig}(state)
\end{equation}
\end{theorem}
Proof: It is enough to show that each element of the generating set
${\cal A}$ of the state eigen closure system ${\cal F}_{eig}$ also
belongs to ${\cal Y}_{eig}(state)$. Suppose that $A \subset O(e)$
and hence $eig_e(A) \in {\cal A}$. Consider now the set $B =
\cup_{f \in {\cal E}, f \not= e}O(f) \cup A$. Remark first that
$O(e,p) \subset A \Rightarrow O(f,p) \subset B, \ \forall f \in
{\cal E}$. Let us show that because $S$ is a distinguishable
experiment entity we also have the inverse implication. Let us
remark that $B \cap O(g) = \cup_{f \in {\cal E}, f \not= e}[(O(f)
\cap O(g)) \cup (A \cap O(g))]$, which shows that $B \cap O(e) = A$. Let us now consider $p \in eig(B)_{state}
\Rightarrow \forall f \in {\cal E}, O(f,p) \subset B  \Rightarrow
O(e,p) \subset B \cap O(e) = A \Rightarrow p \in eig_e(A)$. So
$eig_e(A) = (eig(B))_{state} \in {\cal Y}(state)$. For the converse suppose $eig(A)_{state} \in {\cal Y}(state)$ where $A \subset X$, then  $eig(A)_{state} = \cap_{e \in {\cal E}}eig_e(A \cap O(e))$.

\bigskip
\noindent
To finish this section on the eigen closures we show that there is
also a very natural closure structure on the set of outcome on an
entity.
\begin{definition} \label{cloutc}
Let us consider an entity $S({\cal E}, \Sigma, X, {\cal O})$. For
$A \subset X$ we define:
\begin{equation}
cl(A) = \cap_{O(e,p) \subset A^C}O(e,p)^C
\end{equation}
\end{definition}
\begin{theorem}
Let us consider an entity $S({\cal E}, \Sigma, X, {\cal O})$ and
the map $cl$ introduced in definition~\ref{cloutc}. Then
$cl$ is a closure on $X$. All the $O(e,p)$ are open sets for this
closure and:
\begin{equation}
O(e,p) \subset cl(A)^C \Leftrightarrow O(e,p) \subset A^C
\end{equation}
\begin{equation}
eig(A) = eig(int(A))
\end{equation}
\end{theorem}
Proof: Clearly $A \subset cl(A)$. If $A \subset B$ then $cl(A)
\subset cl(B)$. We have that $cl(\emptyset) = \emptyset$. We have
to show now that $cl(cl(A)) \subset cl(A)$. Let us first prove that
$O(e,p) \subset cl(A)^C \Leftrightarrow O(e,p)
\subset A^C$ because from this follows immediately the closure
property that we are left to prove. Suppose that $O(e,p)
\subset cl(A)^C$, then $O(e,p) \subset \cup_{O(f,q) \subset
A^C}O(f,q) \subset A^C$. On the other hand, suppose that
$O(e,p) \subset A^C$, then $cl(A) \subset \cup_{O(f,q) \subset
A^C}O(f,q) \subset O(e,p)^C$, which implies that $O(e,p)
\subset cl(A)^C$. Now we have $cl(cl(A)) = \cap_{O(e,p) \subset
cl(A)^C}O(e,p)^C = \cap_{O(e,p) \subset A^C}O(e,p)^C = cl(A)$.
\section{Orthogonality and ortho-closure}
The orthogonality relations give rise to a closure in a natural
way.
\begin{proposition} Consider a set $Z$ equipped with an
orthogonality relation $\perp$, and define for $K \subset Z$ the
set $K^\perp = \{p
\vert\ p \perp q, q \in K \}$, and 
\begin{equation}
cl(K) = (K^\perp)^\perp
\end{equation}
then 
$cl$ is a
closure operator, that we shall call the ortho closure
operator connected to $\perp$.
\end{proposition}
Proof: see (Birkhoff 1948).
\begin{proposition}
Let us denote the collection of ortho closed subsets by ${\cal
Y}_{orth}$, then it can easily be shown  that this closure system
is orthocomplemented, which means that the map $^\perp : {\cal 
Y}_{orth} \to {\cal Y}_{orth}$ satisfies:
\begin{equation}
\begin{array}{lll}
K \subset L \Rightarrow L^\perp \subset K^\perp &
K^{\perp\perp} = K &
K \cap K^\perp = \emptyset
\end{array}
\end{equation}
\end{proposition}
\begin{proposition} The following formulas are satisfied in 
${\cal Y}_{orth}$, for
$Y_i \in {\cal Y}_{orth}$:
\begin{equation}
\begin{array}{l}
(\cap_iY_i)^\perp = cl(\cup_iY_i^\perp) \\
(\cup_iY_i)^\perp = \cap_iY_i^\perp \\
cl(Y \cup Y^\perp) = Z 
\end{array}
\end{equation}
\end{proposition}
 Proof: Let $Y \in {\cal Y}_{orth}$: (1)
$cl_{orth} (\cup_iY_i^\perp) \subset Y$ 
$\Leftrightarrow$
$\cup_iY_i^\perp \subset Y$ $\Leftrightarrow$ $Y_i^\perp \subset Y\ 
\forall\ i$
$\Leftrightarrow$ $Y^\perp \subset Y_i\ \ \forall\ i$
$\Leftrightarrow$ 
$Y^\perp \subset
\cap_iY_i$ $\Leftrightarrow$ $(\cap_iY_i)^\perp \subset Y$. From
this  follows that
$cl_{orth}(\cup_iY_i^\perp) = (\cap_iY_i)^\perp$. (2) $Y \subset
\cap_iY_i^\perp$
$\Leftrightarrow$ $Y_i \subset Y^\perp\ \forall\ i$
$\Leftrightarrow$ 
$\cup_iY_i \subset
Y^\perp$ $\Leftrightarrow$ $cl_{orth}(\cup_iY_i) \subset Y^\perp$ 
$\Leftrightarrow$ $ Y\subset
(\cup_iY_i)^\perp$. From this follows that $(\cup_iY_i)^\perp = 
\cap_iY_i^\perp$. (3)
$cl_{orth}(Y \cup Y^\perp) = (Y^\perp \cap Y)^\perp =
\emptyset^\perp = Z$

\bigskip
\noindent
An ortho closure system has a simple generating set of elements.
\begin{theorem} The set ${\cal B} = \{\{p\}^\perp \vert\ 
p \in Z \}$ is a
generating set for the set of ortho closure system ${\cal
Y}_{orth}$.
\end{theorem}
Proof: Consider any element $Y \in {\cal 
Y}$. We have $Y^\perp
= \cup_{p \in Y^\perp}\{p\}$, and hence $Y = Y^{\perp\perp} =
\cap_{p \in  Y^\perp}\{p\}^\perp$. 
\begin{definition}
Let us consider an entity $S({\cal E}, \Sigma, X, {\cal O})$. We
have defined orthogonality relations on ${\cal E} \times \Sigma$,
on $\Sigma$ and on ${\cal E}$. We will call the ortho closure
systems related to these orthogonality relation, the central ortho,
the state ortho and the experiment ortho closure system and denote
them respectively by
${\cal Y}_{orth}$, ${\cal F}_{orth}$ and ${\cal G}_{orth}$. 
\end{definition}
We can prove the following surprising result:
\begin{theorem}
Let us consider an entity $S({\cal E}, \Sigma, X, {\cal O})$
and the eigen closure systems ${\cal Y}_{eig}$,
${\cal F}_{eig}(e)$ and ${\cal G}_{eig}(p)$ and the ortho closure
systems ${\cal Y}_{orth}$, ${\cal F}_{orth}(e)$ and
${\cal G}_{orth}(p)$. We have :
\begin{equation}
\begin{array}{lll}
{\cal Y}_{orth} \subset {\cal Y}_{eig} &
{\cal F}_{orth}(e) \subset {\cal F}_{eig}(e) &
{\cal G}_{orth}(p) \subset {\cal G}_{eig}(p)
\end{array}
\end{equation}
\end{theorem}
Proof: Consider $Y \in {\cal Y}_{orth}$ and consider $A = \cup_{(e,p) \in Y^\perp}O(e,p)$.
Since $Y = (Y^\perp)^\perp$, we have
$(f,q)
\in Y \Leftrightarrow O(f,q) \cap O(e,p) = \emptyset \ \forall\
(e,p) \in Y^\perp \Leftrightarrow O(f,q)
\cap A = \emptyset \Leftrightarrow O(f,q) \subset A^C
\Leftrightarrow (f,q) \in eig(A^C)$. This shows that
$Y \in {\cal Y}_{eig}$. Consider now $F \in {\cal F}_{orth}(e)$ and
$B = \cup_{p \in F^{\perp_e}}O(e,p)$. Since $F =
(F^{\perp_e})^{\perp_e}$ we have $q \in F \Leftrightarrow O(e,q)
\cap O(e,p) = \emptyset, \ \forall p \in F^{\perp_e}
\Leftrightarrow O(e,q) \cap B = \emptyset \Leftrightarrow O(e,q)
\subset B^C \Leftrightarrow q \in eig_e(B^C)$. This shows that $F
\in {\cal F}_{eig}(e)$.
\section{Outcome, experiment and state determination and the first
separation axiom \label{expdet}}
In this section we will show that the traditional
topological separation axioms are connected to physically well
interpretable properties of the considered entities. Instead of
introducing these properties as axioms we choose to use them as
characterizations of types of entities.
\begin{definition}
Let us consider an entity $S({\cal E}, \Sigma, X, {\cal O})$. We
say that the entity is `outcome determined' iff $O(e,p) = O(f,q)
\Rightarrow (e,p) = (f,q)$.
\end{definition}
\begin{definition}
Consider a set $W$ with a closure operator $cl$. We say that $cl$
satisfies the $T_0$ separation axiom iff for $w,v \in W$ we have
$cl(w) = cl(v) \Rightarrow w = v$.
\end{definition}
\begin{proposition}
Let us consider an entity $S({\cal E}, \Sigma, X, {\cal O})$.
Suppose that
$cl_{eig}$ is the eigen closure operator on ${\cal E} \times
\Sigma$. We have :
\begin{equation}
cl_{eig}(\{(e,p)\}) = eig(O(e,p))
\end{equation}
\end{proposition}
Proof: Since $\{(e,p)\} \subset eig(O(e,p))$ we have
$cl_{eig}(\{(e,p)\}) \subset eig(O(e,p))$ because
$cl_{eig}(\{(e,p)\})$ is the smallest element of ${\cal Y}_{eig}$
that contains $\{(e,p)\}$. Let us prove now that $eig(O(e,p))
\subset cl_{eig}(\{(e,p)\})$. Since $cl_{eig}(\{(e,p)\}) \in {\cal
Y}_{eig}$ there exists a set $A \subset X$ such that
$cl_{eig}(\{(e,p)\}) = eig(A)$. From this follows that $(e,p) \in
eig(A)$, or $O(e,p) \subset A$. This implies that $eig(O(e,p))
\subset eig(A)$, and hence we have shown that $eig(O(e,p)) \subset
cl_{eig}(\{(e,p)\})$. As a consequence $cl_{eig}(\{(e,p)\}) =
eig(O(e,p))$.
\begin{theorem}
Let us consider an entity $S({\cal E}, \Sigma, X, {\cal O})$. The
entity $S$ is `outcome determined' iff the central eigen closure
operator satisfies the $T_0$ separation axiom.
\end{theorem}
Proof: Suppose that the entity is `outcome determined' and
consider $(e,p), (f,q) \in {\cal E} \times
\Sigma$ such that $cl_{eig}(\{(e,p)\}) = cl_{eig}(\{(f,q)\})$.
From the foregoing theorem then follows that $eig(O(e,p)) =
eig((O(f,q))$. This means that $(e,p) \in eig(O(f,q))$, or $O(e,p)
\subset O(f,q)$ and also $(f,q) \in eig(O(e,p))$ and hence $O(f,q)
\subset O(e,p)$. From this follows that
$O(e,p) = O(f,q)$ and hence, since the entity is
`outcome determined', we have $(e,p) = (f,q)$. This shows that
$cl_{eig}$ is $T_0$. Suppose now that the central eigen closure
operator is $T_0$, and consider $(e,p)$ and
$(f,q)$ such that $O(e,p) = O(f,q)$. Then $eig(O(e,p)) =
eig(O(f,q))$ and hence $cl_{eig}(\{(e,p)\}) = cl_{eig}(\{(f,q)\})$.
From this follows that $(e,p) = (f,q)$, and hence we have shown
that the entity is `outcome determined'.

\bigskip
\noindent
Let us investigate now the eigen closure on the set of states. We
can characterize the closure of singletons in the following way:
\begin{proposition}
Let us consider an entity $S({\cal E}, \Sigma, X, {\cal O})$.
Suppose that
$cl_{eig}$ is the state eigen closure operator on $\Sigma$. We have :
\begin{equation}
cl_{eig}(\{p)\}) = \cap_{e \in {\cal E}}eig_e(O(e,p))
\end{equation}
\end{proposition}
Proof: Since $\{p\} \subset eig_e(O(e,p)) \forall e \in {\cal E}$
we have $\{p\} \subset \cap_{e \in {\cal E}}eig_e(O(e,p))$. This
shows that $cl_{eig}(\{p\}) \subset \cap_{e \in {\cal
E}}eig_e(O(e,p))$. Let us now prove the inverse inclusion. Since
$cl_{eig}(\{p\}) \in {\cal F}_{eig}$ there exists for $e \in {\cal
E}, A(e) \subset O(e)$ such that $cl_{eig}(\{p\}) = \cap_{e \in
{\cal E}}eig_e(A(e))$. We have $\{p\} \subset \cap_{e \in {\cal
E}}eig_e(A(e))$ and hence $p \in eig_e(A(e))
\ \forall \ e \in {\cal E}$. But this implies that $O(e,p) \subset
A(e) \ \forall \ e \in {\cal E}$, which in turn implies that
$eig_e(O(e,p)) \subset eig_e(A(e)) \ \forall\ e \in {\cal E}$. As a
consequence
$\cap_{e \in {\cal E}}eig_e(O(e,p)) \subset \cap_{e \in {\cal
E}}eig_e(A(e))$, which shows that $\cap_{e
\in {\cal E}}eig_e(O(e,p)) \subset cl_{eig}(\{p\})$.
\begin{definition}
Let us consider an entity $S({\cal E}, \Sigma, X, {\cal O})$. We say
that the entity is `state determined' iff $O(e,p) = O(e,q) \
\forall\ e \in {\cal E} \Rightarrow p = q$.
\end{definition}
\begin{theorem}
Let us consider an entity $S({\cal E}, \Sigma, X, {\cal O})$. The
entity $S$ is `state determined' iff the state eigen closure
operator satisfies the $T_0$ separation axiom.
\end{theorem}
Proof: Suppose that the entity is `state determined'
and consider $p,q \in \Sigma$ such that
$cl_{eig}(\{p\}) = cl_{eig}(\{q\})$. This means that $\cap_{e \in
{\cal E}}eig_e(O(e,p)) = \cap_{e \in {\cal E}}eig_e(O(e,q))$. From
this follows that $p \in eig_e(O(e,q)) \ \forall \ e \in {\cal E}$.
Hence $O(e,p) \subset O(e,q) \ \forall\ e \in {\cal E}$. In an
analogous way we show that $O(e,q) \subset O(e,p) \ \forall \ e \in
{\cal E}$, which proves that $O(e,p) = O(e,q) \ \forall\ e \in
{\cal E}$. Since the entity is state determined we have as a
consequence that $p = q$. This proves that the state eigen closure
operator satisfies that $T_0$ separation axiom. Suppose now that
the state closure operator is $T_0$ and consider $p,q \in \Sigma$
such that $O(e,p) = O(e,q) \ \forall\ e \in {\cal E}$. Then
$\cap_{e \in {\cal E}}O(e,p) = \cap_{e \in {\cal E}}O(e,q)$, and
hence $cl_{eig}(\{p\}) = cl_{eig}(\{q\})$. From this follows that
$p = q$ and hence we have shown that the entity is `experiment
determined'.

\bigskip
\noindent
By symmetry we can formulate analogous properties for `experiment
determined' entities.
\begin{definition}
Let us consider an entity $S({\cal E}, \Sigma, X, {\cal O})$. We
say that the entity is `experiment determined' iff $O(e,p) = O(f,p) \
\forall\ p \in \Sigma \Rightarrow e = f$.
\end{definition}
\begin{theorem}
Let us consider an entity $S({\cal E}, \Sigma, X, {\cal O})$. The
entity $S$ is `experiment determined' iff the experiment eigen
closure operator satisfies the $T_0$ separation axiom.
\end{theorem}
\section{Atomic entities and the second separation axiom}
The second topological separation axiom is also connected to a
property that we can easily interpret physically.
\begin{definition}
Let us consider an entity $S({\cal E}, \Sigma, X, {\cal O})$. We
say that the entity is `central atomic' iff $(e,p) < (f,q)
\Rightarrow (e,p) = (f,q)$.
\end{definition}
\begin{definition}
Consider a set $W$ with a closure operator $cl$. We say that $cl$
satisfies the $T_1$ separation axiom iff for $w \in W$ we have
$cl(\{w\}) = \{w\}$.
\end{definition}
\begin{proposition}
Let us consider an entity $S({\cal E}, \Sigma, X, {\cal O})$. The
entity $S$ is `central atomic' iff the central eigen closure
operator satisfies the $T_1$ separation axiom.
\end{proposition}
Proof: Suppose that the entity is `central atomic' and consider
$(e,p) \in {\cal E} \times
\Sigma$. Suppose that $(f,q) \in cl_{eig}(\{(e,p)\})$, then $(f,q)
\in eig(O(e,p))$. This means that
$O(f,q) \subset O(e,p)$ and hence $(f,q) < (e,p)$. But from this
follows that $(f,q) = (e,p)$. So we have shown that
$cl_{eig}(\{(e,p)\})$ contains no other elements than $(e,p)$ and
hence $cl_{eig}(\{(e,p)\}) =
\{(e,p)\}$. On the other hand suppose now that the central eigen
closure operator is $T_1$, and consider $(e,p) < (f,q)$. We then
have $O(e,p) \subset O(f,q)$ and hence $eig(O(e,p)) \subset
eig(O(f,q))$ which implies that $\{(e,p)\} = cl_{eig}(\{(e,p)\})
\subset cl_{eig}(\{f,q)\}) = \{(f,q)\}$. This proves that $(e,p) =
(f,q)$.
\begin{definition}
Let us consider an entity $S({\cal E}, \Sigma, X, {\cal O})$. We
say that the entity is `state atomic' iff $p < q \Rightarrow p = q$.
\end{definition}
\begin{theorem}
Let us consider an entity $S({\cal E}, \Sigma, X, {\cal O})$. The
entity $S$ is `state atomic' iff the state eigen closure
operator satisfies the $T_1$ separation axiom.
\end{theorem}
Proof: Suppose that the entity is `state atomic' and consider $p
\in \Sigma$. Suppose that $q \in cl_{eig}(\{p\})$, then $q \in
\cap_{e \in {\cal E}}eig(O(e,p))$, and hence $q \in eig(O(e,p)) \
\forall\ e
\in {\cal E}$. This means that
$O(e,q) \subset O(e,p)\ \forall e \in {\cal E}$ and hence $q <
p$. But from this follows that
$q = p$. So we have shown that $cl_{eig}(\{p\})$ contains no other
elements than $p$ and hence
$cl_{eig}(\{p\}) =
\{p\}$. On the other hand suppose now that the state
closure operator is $T_1$, and consider $p < q$. We then have
$O(e,p) \subset O(e,q) \ \forall e \in {\cal E}$ and hence $\cap_{e
\in {\cal E}}eig(O(e,p))
\subset \cap_{e \in {\cal E}}eig(O(e,q))$ which implies that $\{p\} = cl_{eig}(\{p\}) \subset
cl_{eig}(\{q\}) =
\{q\}$. This proves that
$p = q$.

\bigskip
\noindent
Again for reasons of symmetry we have the corresponding theorem for `experiment atomic' entities.
\begin{definition}
Let us consider an entity $S({\cal E}, \Sigma, X, {\cal O})$. We
say that the entity is `experiment atomic' iff $e < f \Rightarrow e
= f$.
\end{definition}
\begin{theorem}
Let us consider an entity $S({\cal E}, \Sigma, X, {\cal O})$. The
entity $S$ is `experiment atomic' iff the experiment eigen
closure operator satisfies the $T_1$ separation axiom.
\end{theorem}
The following is now merely a reformulation of $T_1
\Rightarrow T_0$:
\begin{theorem}
Let us consider an entity $S({\cal E}, \Sigma, X, {\cal O})$. If
the entity is `central atomic' then it is `outcome determined'. If
it is `state atomic' then it is `state determined' and if it is
`experiment atomic' it is `experiment determined.
\end{theorem}
\section{D-classical entities}
We want to study entities with special properties that make
them `more classical'. Since the word `classical' is used in so
many different meanings in different approaches, we will choose to
introduce new names for these special properties.
\begin{definition} {\bf (d-classical entity)}
Let us consider an entity $S({\cal E}, \Sigma, X, {\cal O})$. We say
that $S$ is a `d-classical' entity (`d' for deterministic) iff \
$\forall e \in {\cal E}, p
\in \Sigma$ we have that $O(e,p)$ is a singleton which we
denote $O(e,p) = \{x(e,p)\}$.
\end{definition}
\begin{theorem}
Let us consider a d-classical entity $S({\cal E}, \Sigma, X, {\cal
O})$.
\begin{enumerate}
\item For $p \in \Sigma$ and $e \in {\cal E}$, we have that $p$ is
always an eigenstate for the experiment $e$ with eigen-outcome
$x(e,p)$.
\item If $p, q \in \Sigma$ and $e, f \in {\cal E}$ such that $p < q$
and $e < f$ then $q < p$ and
$f < e$ and hence $p \approx q$ and $e \approx f$.
\end{enumerate}
\end{theorem} 
Proof: follows immediately from the definitions.
\begin{theorem}
Let us consider a d-classical entity $S({\cal E}, \Sigma, X, {\cal
O})$. Suppose that $p, q \in \Sigma$ and $e, f \in {\cal E}$, then
we have that $p \approx q$ or $p \perp q$, and $e \approx f$ or $e
\perp f$.
\end{theorem}
Proof: Consider $p, q \in \Sigma$ and suppose that $p \not\approx
q$. This means that $p \not< q$ and $q
\not< p$. Then there exists at least one experiment $e \in {\cal
E}$ such that $O(e,p) \not\subset O(e,q)$. We have $O(e,p) =
\{x(e,p)\}$ and $O(e,q) = \{x(e,q)\}$. Hence
$O(e,p) \cap O(e,q) = \emptyset$, which shows that $p \perp q$.
This shows that non equivalent states are orthogonal. In
an analogous way we show that experiments are equivalent or
orthogonal for a d-classical entity.
\begin{theorem}
Let us consider a d-classical entity $S({\cal E}, \Sigma, X, {\cal
O})$.  If the entity is `outcome determined' then it is `central
atomic'. If the entity is `state determined' then it is `state
atomic', and if the entity is `experiment determined' then it is
`experiment atomic'
\end{theorem}
Proof: Suppose that the entity is outcome determined. Consider
$(e,p) < (f,g)$, then we have $O(e,p)
\subset O(f,q)$ and hence $\{x(e,p)\} \subset \{x(f,q)\}$. This
implies that $\{x(e,p)\} = \{x(f,q)\}$ and hence, since the entity
is `outcome determined' we have $(e,p) = (f,q)$. So we have proved
that the entity is central atomic. In an analogous way one proves
the two other implications.

\bigskip
\noindent Let us now study the closures for d-classical entities.
\begin{theorem}
Let us consider a d-classical entity $S({\cal E}, \Sigma, X, {\cal
O})$. We have for $A \subset X$:
\begin{equation}
eig(A) = \{(e,p)\ \vert\ x(e,p) \in A\} = x^{-1}(A)
\end{equation}
\begin{equation}
eig(A^C) = eig(A)^C = eig(A)^\perp
\end{equation}
\begin{equation}
{\cal Y}_{eig} = {\cal Y}_{orth}
\end{equation}
\end{theorem}
Proof: We have $(e,p) \in eig(A) \Leftrightarrow O(e,p) \subset A$.
Since $O(e,p) = \{x(e,p)\}$ we have $(e,p) \in eig(A)
\Leftrightarrow x(e,p) \in A$. This shows that $eig(A) = \{(e,p)\
\vert\ x(e,p) \in A\}$. Consider now
$(e,p) \in eig(A^C)$, then $x(e,p) \in A^C$, and hence $x(e,p)
\not\in A$, which implies that $(e,p) \not\in eig(A)$ or $(e,p) \in
eig(A)^C$. This shows that $eig(A^C) \subset eig(A)^C$. Consider
now $(e,p) \in eig(A)^C$, which means that $(e,p) \not\in eig(A)$
and hence $x(e,p) \not\in A$. Consider now an arbitrary $(f,q)
\in eig(A)$, i.e. $x(f,q) \in A$. This means that $O(e,p) \cap
O(f,q) = \{x(e,p)\} \cap \{x(f,q)\} =
\emptyset$. As a consequence $(e,p) \in eig(A)^\perp$. This shows
that $eig(A)^C \subset eig(A)^\perp$. Consider now
$(e,p) \in eig(A)^\perp$. This means that $(e,p) \perp (f,q)$ for
all $(f,q) \in eig(A)$. Hence $x(e,p) \in A^C$, which shows that
$(e,p) \in eig(A^C)$. Hence we have shown that $eig(A)^\perp
\subset eig(A^C)$. Let us prove now that ${\cal Y}_{eig} = {\cal
Y}_{orth}$. We already have ${\cal Y}_{orth} \subset {\cal
Y}_{eig}$ such that we only have to prove the inverse inclusion. If
we remark that $eig(A) = eig(A^C)^\perp$, it follows that $eig(A)
\in {\cal Y}_{orth}$.
\begin{theorem}
Let us consider a d-classical entity $S({\cal E}, \Sigma, X, {\cal
O})$. We have for $A \subset O(e)$:
\begin{equation}
eig_e(A) = \{p \ \vert\ x(e,p) \in A\}
\end{equation}
\begin{equation}
eig_e(A^C) = eig_e(A)^C = eig_e(A)^{\perp_e}
\end{equation}
\begin{equation}
{\cal F}_{eig}(e) = {\cal F}_{orth}(e)
\end{equation}
\end{theorem}
\begin{theorem}
Let us consider a d-classical entity $S({\cal E}, \Sigma, X, {\cal
O})$. We have for $A \subset O(p)$:
\begin{equation}
eig_p(A) = \{e \ \vert\ x(e,p) \in A\}
\end{equation}
\begin{equation}
eig_p(A^C) = eig_p(A)^C = eig_p(A)^{\perp_p}
\end{equation}
\begin{equation}
{\cal G}_{eig}(p) = {\cal G}_{orth}(p)
\end{equation}
\end{theorem}
A d-classical entity is a trivial type of probabilistic entity.
\begin{theorem}
Let us consider a d-classical entity $S({\cal E}, \Sigma, X, {\cal
O})$. It is a probabilistic entity where the probabilities are
defined as follows.
\begin{equation}
\begin{array}{ll}
\mu: {\cal E} \times \Sigma \times X \rightarrow [0,1] &
(e,p,y) \mapsto \mu(e,p,y)
\end{array}
\end{equation}
where $\mu(e,p,y) = 0$ if $y \not= x(e,p)$ and $\mu(e,p,x(e,p))
= 1$. 
\end{theorem}
\section{Sub entities and morphisms}
The concept of sub entity should be clearly defined. When will we
decide that a certain `piece' of an entity is a sub entity? Let us
consider two entities $S$ and $S'$, with sets of states $\Sigma$
and $\Sigma'$, sets of experiments ${\cal E}$ and
${\cal E}'$ and sets of outcomes $X$ and $X'$. If $S$ is to be a
part of $S'$, it is plausible to demand that if the entity $S'$ is
in a certain state $p'$, then the entity $S$, as part of $S'$, is
in a well defined state
$m(p')$. This defines a function:
\begin{equation}
\begin{array}{ll}
m : \Sigma' \rightarrow \Sigma &
p' \mapsto m(p')
\end{array}
\end{equation}
which is surjective - each state of the sub entity $S$ corresponds
to at least one state of the  entity $S'$, - but not necessarily
injective - different states of the entity $S'$ can give rise to
the same state of the sub entity $S$. This function formalizes:
``If $S$ is a sub entity of $S'$ then the mode of being of $S'$
determines that of $S$."

\smallskip
\noindent
Second, if $S$ is to be a part of $S'$, this should imply that with
each experiment $e$ that can be performed on $S$, corresponds an
experiment $n(e)$ that can be performed on $S'$. This again defines
a function 
\begin{equation}
\begin{array}{ll}
n : {\cal E} \rightarrow {\cal E}' &
e \mapsto n(e)
\end{array}
\end{equation}
which is injective - if experiments are different when they are
performed on the sub entity $S$, they are also different when they
are performed on the entity $S'$, but not necessarily surjective -
there can be experiments that can be performed on the entity $S'$
that have no counterpart on the sub entity $S$. 

\smallskip
\noindent
We have to express now that $S$ is really a sub entity of $S$ by
means of a requirement on the way experiments act on states and
outcomes occur. This is again a requirement of `covariance': the
reality does not depend of whether we represent a big piece of
it by means of the entity $S'$ or a smaller sub piece of it by
means of the sub entity $S$. More concretely we express this
requirement of covariance in the following way: if we perform an
experiment
$e$ on the entity
$S$ in a state $m(p')$ where $p'$ is a state of $S'$, then
the outcome $x(e,m(p'))$ occurs iff one specific outcome
$x'(n(e),p')$ occurs after the performance of the experiment
$n(e)$ on the entity $S'$ in state $p'$. This requirement again
defines a function
\begin{equation}
\begin{array}{ll}
l : X \rightarrow X' &
x \mapsto l(x)
\end{array}
\end{equation}
that is such that considering a state $p' \in \Sigma'$ and
an experiment $e \in {\cal E}$, each outcome $x' \in O(n(e),p')$
corresponds to an outcome $x \in O(e,m(p'))$, such that $l(x) =
x'$. The interpretation is that $x$ occurs for
$e$,
$S$ being in state $m(p')$ iff $l(x)$ occurs for $n(e)$, $S'$ being in state $p'$. Since only one outcome occurs at once,
this implies that the function $l$ is injective, and $O(e,m(p')$ is surjectively mapped onto $O(n(e),p')$ by $l$.

\smallskip
\noindent
We have now introduced all elements to present a definition:
\begin{definition} {\bf (sub entities)}
Suppose that $S({\cal E}, \Sigma, X, {\cal O})$ and $S'({\cal E}',
\Sigma', X', {\cal O}')$ are two entities. We say that $S$ is a sub
entity of $S'$ iff there exist a surjective function:
\begin{equation}
\begin{array}{ll}
m : \Sigma' \rightarrow \Sigma &
p' \mapsto m(p') 
\end{array}
\end{equation}
and an injective function:
\begin{equation}
\begin{array}{ll}
n : {\cal E} \rightarrow {\cal E}' &
e \mapsto n(e) 
\end{array}
\end{equation}
and an injective function
\begin{equation}
\begin{array}{ll}
l : X \rightarrow X' &
x \mapsto l(x) 
\end{array}
\end{equation}
that for each $p' \in \Sigma'$ and $e \in {\cal E}$ maps
$O(e,m(p')$ surjectively on $O(n(e),p')$ such that:
\par \noindent 
(i) if the entity $S'$ is in state $p'$, then the entity $S$ is in
state $m(p')$.
\par \noindent 
(ii) if the experiment $e$ is performed on the entity $S$, then the experiment $n(e)$ is performed on the entity
$S'$.
\par \noindent 
(iii) considering a state $p'$ of $S'$ and an experiment $e \in
{\cal E}$ then the outcome $x \in O(e,m(p')$ occurs for $e$ being
performed on $S$ in state $m(p')$ iff the outcome $l(x)$ occurs for
$n(e)$ being performed on $S'$ in state $p'$.
\end{definition}
\begin{proposition}
Suppose that $S({\cal E}, \Sigma, X, {\cal O})$ and $S'({\cal E}',
\Sigma', X', {\cal O}')$ are two entities such that
$S$ is a sub entity of
$S'$, and $m,n$ and $l$ are the connecting functions related to $S$
and $S'$. If $x,y \in X$, $e,f \in {\cal E}$ and $p',q' \in
\Sigma'$ we have:
\begin{eqnarray}
x \perp y &\Rightarrow& l(x) \perp l(y) \\
p' < q' &\Rightarrow& m(p') < m(q')    \\
e \perp f &\Rightarrow& n(e) \perp n(f) \\
(e,m(p')) < (f,m(q')) &\Leftrightarrow&  (n(e),p') < (n(f),q') \\ 
(e,m(p')) \perp (f,m(q')) &\Leftrightarrow&  (n(e),p') \perp
(n(f),q')
\end{eqnarray}
\end{proposition}
Proof: Suppose that $x \perp y$, then there exists $e \in {\cal E}$
and $p \in \Sigma$ such that $x, y \in O(e,p)$ and $x \not= y$.
Since $m$ is surjective we have a $p' \in \Sigma'$ such that $p =
m(p')$. This means that $x, y
\in O(e,m(p'))$ and hence $l(x), l(y) \in O(n(e),p')$. Since $l$ is injective we have $l(x) \not= l(y)$ and hence
$l(x) \perp l(y)$.
Suppose that $p' < q'$. This means that for all $e' \in {\cal
E}'$ we have $O'(e',p') \subset O'(e', q')$. Consider an arbitrary
$e \in {\cal E}$, then $l(O(e, m(p'))) = O'(n(e),p')
\subset O'(n(e),q') = l(O(e, m(q')))$. Since $l$ is injective
this shows that $O(e, m(p')) \subset O(e, m(q'))$. This proves that
$p < q$. Suppose that
$e
\perp f$, This means that there exists $p \in \Sigma$ such that
$O(e,p) \cap O(f,p) = \emptyset$. This implies that
$l(O(e,p) \cap O(f,p)) = l(O(e,p)) \cap l(O(f,p)) = \emptyset$.
Consider
$p'
\in
\Sigma'$ such that $m(p') = p$, then $O'(n(e),p') \cap
O'(n(f),p') = l(O(e,m(p'))) \cap l(O(f,m(p'))) = \emptyset$. This
shows that $n(e) \perp n(f)$.
We have $(e,m(p')) < (f,m(q'))$ $\Leftrightarrow$ $O(e,m(p'))
\subset O(f,m(q'))$ $\Leftrightarrow$ $l(O(e,m(p')))
\subset l(O(f,m(q')))$ $\Leftrightarrow$ $O'(n(e),p') \subset
O'(n(f),q'))$ $\Leftrightarrow$ $(n(e),p') < (n(f),q')$. We have:
$(e,m(p')) \perp (f,m(q'))$ $\Leftrightarrow$ $O(e,m(p')) \cap
O(f,m(q')) = \emptyset$
$\Leftrightarrow$
$l(O(e,m(p')))
\cap l(O(f,m(q'))) = \emptyset$ $\Leftrightarrow$ $O'(n(e),p') \cap O'(n(f),q')) = \emptyset$ $\Leftrightarrow$
$(n(e),p') \perp (n(f),q')$.

\bigskip
\noindent
We have to remark that the function $m$ does not necessarily
conserve the orthogonality relation. It can be that states of $S'$
that are orthogonal are mapped onto states of $S$ that are not
orthogonal. In the same way, the function $n$ does not necessarily
conserves the pre order relation. It can well be that experiments
that `imply' each other for the sub entity, do not `imply' each
other for the entity. But both functions are `continuous' for the
eigen closure system.
\begin{proposition}
Suppose that $S({\cal E}, \Sigma, X, {\cal O})$ and $S'({\cal E}',
\Sigma', X', {\cal O}')$ are two entities, such that
$S$ is a sub entity of
$S'$, and $m,n$ and $l$ are the connecting functions related to $S$ and $S'$. For $p' \in \Sigma'$, $e \in {\cal
E}$, $A \subset X$ and $A' \subset X'$ we have :
\begin{eqnarray}
m^{-1}(eig_e(A)) &=& eig_{n(e)}(l(A)) \\
n^{-1}(eig'_{p'}(A')) &=& eig_{m(p')}(l^{-1}(A'))
\end{eqnarray}
\end{proposition}
Proof: We have: $p' \in m^{-1}(eig_e(A))$ $\Leftrightarrow$ $m(p')
\in eig_e(A)$ $\Leftrightarrow$ $O(e,m(p'))
\subset A$ $\Leftrightarrow$ $l(O(e,m(p')) \subset l(A)$
$\Leftrightarrow$ $O(n(e),p') \subset l(A)$
$\Leftrightarrow$ $p' \in eig_{n(e)}(l(A))$. We also have: $e \in
n^{-1}(eig'_{p'}(A'))$ $\Leftrightarrow$ $n(e)
\in eig'_{p'}(A')$ $\Leftrightarrow$ $O'(n(e),p') \subset
A'$ $\Leftrightarrow$ $l(O(e,m(p')) \subset A'$
$\Leftrightarrow$ $O(e,m(p')) \subset l^{-1}(A')$ $\Leftrightarrow$ $e \in eig_{m(p')}(l^{-1}(A'))$.
\begin{theorem}
Suppose that $S({\cal E}, \Sigma, X, {\cal O})$ and $S'({\cal E}',
\Sigma', X', {\cal O}')$ are two entities, such that
$S$ is a sub entity of
$S'$, and $m,n$ and $l$ are the connecting functions related to $S$ and $S'$. Then $m$ and $n$ are continuous
functions for the eigen closure systems, or:  
\begin{equation}
F \in {\cal F}_{eig} \Rightarrow m^{-1}(F) \in {\cal F}'_{eig}
\end{equation}
\begin{equation}
G' \in {\cal G}'_{eig} \Rightarrow n^{-1}(G) \in {\cal G}_{eig}
\end{equation}
\end{theorem}
Proof: Suppose that $F \in {\cal F}_{eig}$. Then we have $F =
\cap_{e \in {\cal E}}F_e$ with $F_e \in {\cal F}_{eig}(e)$. From
the foregoing theorem follows that for each $F_e \in {\cal
F}_{eig}(e)$ we have $m^{-1}(F_e) \in {\cal F}'_{eig}(n(e))$ and
hence $m^{-1}(F_e) \in {\cal F}'_{eig}$. We have $m^{-1}(\cap_{e
\in {\cal E}}F_e) =
\cap_{e
\in {\cal E}}m^{-1}(F_e)$ which shows that $m^{-1}(F) \in {\cal
F}'_{eig}$.  Suppose that $G' \in {\cal G}'_{eig}$, then we have
that $G' = \cap_{p' \in \Sigma'}G_{p'}$ where $G_{p'} \in {\cal
G}'_{eig}(p')$. Hence we have $n^{-1}(G_{p'}) \in {\cal
G}_{eig}(m(p'))$ and hence also $n^{-1}(G_{p'}) \in {\cal
G}_{eig}$. Since we have $n^{-1}(\cap_{p' \in \Sigma'}G_{p'}) =
\cap_{p' \in \Sigma'}n^{-1}(G_{p'})$ we have
$n^{-1}(G) \in {\cal G}_{eig}$.

\bigskip
\noindent
Let us consider the situation of two probabilistic entities $S$ and
$S'$ such that $S$ is a sub entity of $S'$ and let
${\cal M}$ be the set of generalized probability measures of $S$ and
${\cal M}'$ the set of generalized probability measures of $S'$. We will call
$S$ a `probabilistic sub entity' of
$S'$ if the respective generalized probability measures
are connected in the way we will specify now. We remind that the
situation that we consider is the following: if we perform an
experiment
$e$ on the entity
$S$ in a state $m(p')$ where $p'$ is a state of $S'$, then the
outcome $x(e,m(p'))$ occurs iff one specific outcome
$x'(n(e),p') = l(x(e,m(p')))$ occurs after the performance of the
experiment $n(e)$ on the entity $S'$ in state
$p'$. This means that if we perform repeated experiments
on entities in identical states and calculate the relative
frequencies of outcomes
$x(e,m(p'))$ and outcomes $x'(n(e),p')$, they will be the
same. This means that also the limits of these relative frequencies,
i.e. the probabilities, will match.

Suppose that $S({\cal E}, \Sigma, X, {\cal O}, {\cal M})$
and $S'({\cal E}', \Sigma', X', {\cal O}', {\cal M}')$ are two
probabilistic entities. To each $\mu \in {\cal M}$ corresponds a
$\mu' \in {\cal M}'$ such that $\mu$ represents the relative
frequency operation on the sub entity $S$ and $\mu'$ represents the
corresponding relative frequency operation on the entity $S'$. And
we have $\mu(e,m(p'),x) = \mu'(n(e),p',l(x))$. Let us formalize
this physical idea.
\begin{definition}
Suppose that $S({\cal E}, \Sigma, X, {\cal O}, {\cal M})$ and
$S'({\cal E}', \Sigma', X', {\cal O}', {\cal M}')$ are two
probabilistic entities such that $S$ is a sub entity of $S'$. We
will say that $S$ is a probabilistic sub entity iff there exists an
injective function
\begin{equation}
\begin{array}{ll}
k: {\cal M} \rightarrow {\cal M}' &
\mu \mapsto k(\mu)
\end{array}
\end{equation}
such that for $e \in {\cal E}$, $p' \in
\Sigma'$ and $x \in X$ we have:
\begin{equation}
\mu(e,m(p'),x) = k(\mu)(n(e),p',l(x))
\end{equation}
\end{definition}
\section{A finite example \label{exam}} 
The first example that we discuss is a finite example.
Let us consider an entity $S$ with the following set of states
$\Sigma$, set of experiments ${\cal E}$, and sets of outcomes:
\begin{equation}
\Sigma = \{p,q,r\}, \quad {\cal E} = \{e,f,g\} 
\end{equation}
\begin{equation} \begin{array}{lll}
O(e,p) = \{x_1,x_2\} &  O(e,q) = \{x_1,x_3\} & O(e,r) =
\{x_2,x_3\} \\ 
O(f,p) = \{y_1,y_2\} & O(f,q) =
\{x_2,y_2\} & O(f,r) = \{x_3,y_1, y_2\}, \\ 
O(g,p) =
\{x_1,y_1\} & O(g,q) = \{x_2 \} & O(g,r) = \{x_1,
x_2, y_1\} 
\end{array}
\end{equation}
Then we have :
\[ 
\begin{array}{lll}
O(e) = \{x_1,x_2,x_3\} & O(f) = \{x_2,x_3,y_1,y_2\} & O(g)
=\{x_1,x_2,y_1\} \\
O(p) = \{x_1, x_2, y_1, y_2\} & O(q) = \{x_1, x_2, x_3\} & O(r) =
\{x_1, x_2, x_3, y_1, y_2\}
\end{array}
\]
\begin{equation}
X = \{x_1,x_2,x_3,y_1,y_2\}
\end{equation}

\subsection{Pre-order and orthogonality:}

We introduce a shorter notation for the
nine elements of ${\cal E}
\times \Sigma$. Let us denote $(e,p) = \lambda_{11}, (e,q)
= \lambda_{12}, (e,r) = \lambda_{13}, (f,p) = \lambda_{21}, (f,q) =
\lambda_{22}, (f,r) = \lambda_{23}, (g,p) = \lambda_{31}, (g,q) =
\lambda_{32}, (g,r) = \lambda_{33}$. We have:
\begin{equation}
\begin{array}{ccccc}
\lambda_{11} < \lambda_{33} & \lambda_{21} < \lambda_{23} &
\lambda_{32} < \lambda_{11} &
\lambda_{32} < \lambda_{13} & \lambda_{32} < \lambda_{22} \\
\lambda_{32} < \lambda_{33} & \lambda_{11} \perp \lambda_{21} &
\lambda_{11} \perp \lambda_{23} & \lambda_{12} \perp
\lambda_{21} & \lambda_{12} \perp
\lambda_{22} \\
\lambda_{12}
\perp \lambda_{32} & \lambda_{13} \perp \lambda_{21} & \lambda_{13}
\perp \lambda_{31} & 
\lambda_{21} \perp \lambda_{11} & \lambda_{21}
\perp \lambda_{12} \\
\lambda_{21} \perp
\lambda_{13} & \lambda_{21} \perp \lambda_{32} & \lambda_{22} \perp
\lambda_{12} &
\lambda_{22}
\perp \lambda_{31} & \lambda_{23} \perp \lambda_{11} \\
\lambda_{23} \perp \lambda_{32} & \lambda_{31}
\perp \lambda_{13} & \lambda_{31} \perp
\lambda_{22} & \lambda_{31} \perp \lambda_{32} & \lambda_{32}
\perp
\lambda_{12} \\
\lambda_{32} \perp \lambda_{21} & \lambda_{32} \perp \lambda_{23} 
& \lambda_{32} \perp
\lambda_{31} & \mbox{} & \mbox{}
\end{array}
\end{equation}
Let us now calculate the pre-order and orthogonality relations
on the set of states $\Sigma$ and on the set of experiments ${\cal
E}$ for this example. We have: 
\begin{equation}
\begin{array}{ccccccc}
p \not< q & q \not< p & p \not< r & r \not< p & q \not< r & r \not< q
& p \perp_g q \\
p \not\perp r &  q \not\perp r & e \not< f & f \not< e & e
\not< g & g \not< e & f \not< g \\
g \not< f & e \perp_p f & e \perp_q f & e \perp_q g & f
\not\perp g & \mbox{} & \mbox{}
\end{array}
\end{equation}
We have $q$ that is an eigenstate of $g$ with eigen-outcome $x_2$,
and hence $(g,q)$ is an eigen couple with eigen outcome $x_2$.

\subsection{The eigen closures:}

Let us now study the closure structures and let us construct the
eigen-map $eig$ and the closure system ${\cal Y}$ for our finite
example.

\begin{equation}
\begin{array}{l}
\begin{array}{ll}
eig(\{x_2\}) = \{\lambda_{32}\} & eig(\{x_1,x_2\}) =
\{\lambda_{11}, \lambda_{32}\}  \\
 eig(\{x_2,x_3\}) = \{\lambda_{13},
\lambda_{32}\} & eig(\{x_1,y_1\}) = \{\lambda_{31}\} \\
eig(\{x_1,x_3\}) = \{\lambda_{12}\} & eig(\{x_1,x_3,y_2\}) =
\{\lambda_{12}\} \\
eig(\{x_2,y_1\}) = \{\lambda_{32}\} & eig(\{x_2,y_2\}) =
\{\lambda_{22},\lambda_{32}\} \\
eig(\{y_1,y_2\}) = \{\lambda_{21}\} & eig(\{x_1,x_2,x_3\})
= \{\lambda_{11},\lambda_{12}, \lambda_{13},
\lambda_{32} \} \\
eig(\{x_1,x_3,y_1\}) = \{\lambda_{12},\lambda_{31}\}
& eig(\{x_1,x_2,y_1\}) = \{\lambda_{11},\lambda_{31}, \lambda_{32},
\lambda_{33}\} \\
eig(\{x_1,y_1,y_2\}) = \{\lambda_{21},\lambda_{31}\}
& eig(\{x_1,x_2,y_2\}) = \{\lambda_{11},\lambda_{22}, \lambda_{32}
\} \\
eig(\{x_2,x_3,y_1\}) =
\{\lambda_{13},\lambda_{32}\} & eig(\{x_2,x_3,y_2\}) =
\{\lambda_{13},\lambda_{32}, \lambda_{22}\} \\
eig(\{x_3,y_1,y_2\}) = \{\lambda_{21},\lambda_{23}\} &
eig(\{x_2,y_1,y_2\}) = \{\lambda_{21}, \lambda_{22},
\lambda_{32}\}
\end{array} \\
\ eig(\{x_1,x_3,y_1,y_2\}) = \{\lambda_{12},
\lambda_{21},\lambda_{23},\lambda_{31}\} \\
\  eig(\{x_1,x_2,x_3,y_2\}) =
\{\lambda_{11},\lambda_{12},\lambda_{13},\lambda_{22},
\lambda_{32} \} \\
\ eig(\{x_2,x_3,y_1,y_2\}) = \{\lambda_{13},
\lambda_{21},\lambda_{22},\lambda_{23}, \lambda_{32} \} \\
\ eig(\{x_1,x_2,y_1,y_2\}) = \{\lambda_{11},\lambda_{21},
\lambda_{22},\lambda_{31},\lambda_{32},\lambda_{33}\} \\
\ eig(\{x_1,x_2,x_3,y_1\}) = \{\lambda_{11},\lambda_{12},
\lambda_{13},
\lambda_{31},\lambda_{32},\lambda_{33}\}
\end{array}
\end{equation}
The images of all other subsets of $X$ are $\emptyset$ or ${\cal E}
\times \Sigma$ or already contained in the ones presented here.
Hence, if ${\cal Y}_{eig}$ is the set of eigen-closed subsets, we
have:
\begin{eqnarray}
{\cal Y}_{eig} &=& \{ \emptyset, \{\lambda_{12}\}, 
\{\lambda_{31}\}, \{\lambda_{32}\},
\{\lambda_{21}\}, \{\lambda_{11},\lambda_{32}\},
\{\lambda_{13},\lambda_{32}\}, \{\lambda_{21},\lambda_{31}\}, 
\nonumber \\  & & \{\lambda_{32},\lambda_{22}\},
 \{\lambda_{22},\lambda_{21}, \lambda_{32}\},\{\lambda_{21},
\lambda_{23}\},
\{\lambda_{11},\lambda_{31},\lambda_{32},\lambda_{33}\},
\nonumber \\
& &  \{\lambda_{11},\lambda_{22},\lambda_{32}\}, \{\lambda_{12},
\lambda_{31}\},
\{\lambda_{11},\lambda_{12},\lambda_{13},\lambda_{32}\}, 
\{\lambda_{12},\lambda_{21},\lambda_{23},
\lambda_{31}\},
\nonumber \\
& &  \{\lambda_{13}, \lambda_{32}, \lambda_{22}\},
\{\lambda_{11},\lambda_{12},\lambda_{13},\lambda_{22} \},
\{\lambda_{13},\lambda_{21},\lambda_{22},\lambda_{23},
\lambda_{32}\}, \nonumber \\ &
&\{\lambda_{11},\lambda_{21},\lambda_{22},\lambda_{31},
\lambda_{32},\lambda_{33}\},
\{\lambda_{11},\lambda_{12},\lambda_{13},
\lambda_{31},\lambda_{32},\lambda_{33}\} \} 
\end{eqnarray}
Let us now construct the closure system on the set of states.  We
have:
\begin{equation}
\begin{array}{lll}
eig_e(\{x_1,x_2\}) = \{p\} & eig_e(\{x_1,x_3\}) = \{q\} &
eig_e(\{x_2,x_3\}) = \{r\}
\end{array}
\end{equation}
and all the other images of $eig_e$ are $\emptyset$ or $\Sigma$. 
This shows that:
\begin{equation}
{\cal F}(e) = \{\emptyset, \{p\}, \{q\}, \{r\}, \Sigma\}
\end{equation}
We also have:
\begin{equation}
\begin{array}{ll}
eig_f(\{y_1,y_2\}) = \{p\} & eig_f(\{y_2,x_2\}) = \{q\} \\
eig_f(\{y_1,y_2,x_3\}) = \{p, r\} & eig_f(\{y_1,y_2,x_2\}) = 
\{p,q\}
\end{array}
\end{equation}
and the other images that are $\emptyset$ or $\Sigma$. This shows
that:
\begin{equation}
{\cal F}(f) = \{\emptyset, \{p\}, \{q\}, \{p,q\}, \{p,r\}, 
\Sigma\}
\end{equation}
And finally we have:
\begin{equation}
\begin{array}{ll}
eig_g(\{x_2\}) = \{q\} & eig_g(\{x_1,y_1\}) = \{p\}
\end{array}
\end{equation}
which shows that
\begin{equation}
{\cal F}(g) = \{\emptyset, \{p\}, \{q\}, \Sigma\}
\end{equation}
The state eigen closure system is given by:
\begin{equation}
{\cal F} = {\cal A} = \{\emptyset, \{p\}, \{q\}, \{r\}, \{p,q\}, 
\{p,r\}, \Sigma\}
\end{equation}
Let us now construct the closure system on the set of experiments. 
We have :
\begin{equation}
\begin{array}{ll}
eig_p(\{x_1,x_2\}) = \{e\} & eig_p(\{y_1,y_2\}) = \{f\} \\ 
eig_p(\{x_1,x_2,y_1\}) = \{e,g\} & eig_p(\{x_1,y_1,y_2\}) = \{f,g\}
\end{array}
\end{equation}
We have then :
\begin{equation}
{\cal G}(p) = \{ \emptyset,\{e\},\{f\},\{g\},\{e,g\},\{f,g\}, 
{\cal E}\}
\end{equation}
For the state $q$ we have:
\begin{equation}
\begin{array}{ll}
eig_q(\{x_1,x_3\}) = \{e\} & eig_q(\{x_2\}) = \{g\} \\
 eig_q(\{x_2,y_2\}) = \{f,g\} & eig_q(\{x_1,x_2,x_3\}) = 
\{e,g\}
\end{array}
\end{equation}
We have then :
\begin{equation}
{\cal G}(q) = \{ \emptyset,\{e\},\{g\},\{e,g\},\{f,g\}, {\cal E}\}
\end{equation}
Finally for $r$ we have :
\begin{equation}
\begin{array}{ll}
eig_r(\{x_2,x_3\}) = \{e\} & eig_r(\{y_1,y_2,x_3\}) = \{f\} \\
eig_r(\{x_1,x_2,y_1\}) = \{g\} & eig_r(\{y_1,y_2,x_2,x_3\}) = \{e,f\}
\\
eig_r(\{x_1,x_2,x_3,y_1\}) =
\{e,g\} & \mbox{}
\end{array}
\end{equation}
We have then :
\begin{equation}
{\cal G}(r) = \{\emptyset,\{e\},\{f\},\{g\},\{e,g\},\{e,f\}, 
{\cal E}\}
\end{equation}
This means that
\begin{equation}
{\cal G} = \{ \emptyset, \{e\},\{f\},\{g\},\{e,f\},\{e,g\},\{f,g\},
{\cal E}\} = {\cal P}({\cal E})
\end{equation}
We can easily see that in general ${\cal Y}(state)$ is different 
from ${\cal F}$ by considering our example. Indeed we have:
\begin{equation}
eig(\{x_1,x_2,y_1,y_2\})(state) = \{p\}
\end{equation}
And all the other traces from elements of ${\cal Y}$ are 
$\emptyset$ or $\Sigma$, which shows that:
\begin{equation}
{\cal Y}(state) = \{\emptyset, \{p\}, \Sigma\}
\end{equation}
This shows that for example $\{q\}$ is not contained in  ${\cal
Y}(state)$ while it is contained in ${\cal F}$.

\subsection{The ortho closures:}

Let us construct now the ortho closure systems  ${\cal Y}_{orth}$,
${\cal F}_{orth}(e)$, ${\cal F}_{orth}$.  To do this, we first
construct the generating set of elements consisting of the
orthogonal's of singletons. First we construct ${\cal Y}_{orth}$ :
\begin{equation}
\begin{array}{ll}
\{\lambda_{11}\}^\perp = \{\lambda_{21}, \lambda_{23}\} & 
\{\lambda_{12}\}^\perp = \{\lambda_{21},
\lambda_{22}, \lambda_{32}\} \\
\{\lambda_{13}\}^\perp = 
\{\lambda_{21}, \lambda_{31}\} &
\{\lambda_{21}\}^\perp = \{\lambda_{11} , \lambda_{12},
\lambda_{13}, \lambda_{32}\} \\
\{\lambda_{22}\}^\perp =
\{\lambda_{12},
\lambda_{31}\} & \{\lambda_{23}\}^\perp = \{\lambda_{11},
\lambda_{32}\} \\
\{\lambda_{31}\}^\perp = \{\lambda_{13}, \lambda_{22}, 
\lambda_{32}\} &
\{\lambda_{32}\}^\perp = \{\lambda_{12},\lambda_{21},
\lambda_{23},\lambda_{31}\} \\
\{\lambda_{33}\}^\perp = \emptyset & \mbox{}
\end{array}
\end{equation}
If we consider this collection as generating set of  elements we
find :
\begin{eqnarray}
{\cal Y}_{orth} &=& \{\emptyset, \{\lambda_{12}\}, 
\{\lambda_{21}\},\{\lambda_{31}\}, \{\lambda_{32}\},
\{\lambda_{11},\lambda_{32}\},\{\lambda_{13},\lambda_{32}\},
\{\lambda_{22},\lambda_{32}\}, \nonumber \\ & & \{\lambda_{23},
\lambda_{21}\},
\{\lambda_{21},\lambda_{31}\},\{\lambda_{12},\lambda_{31}\},
\{\lambda_{21}, \lambda_{22},\lambda_{32}\}, \{\lambda_{13},
\lambda_{22},\lambda_{32}\}, \nonumber \\ & &
\{\lambda_{11},\lambda_{12},
\lambda_{13},\lambda_{32}\},\{\lambda_{12},\lambda_{21},
\lambda_{23},\lambda_{31}\}, {\cal E} \times \Sigma \}
\end{eqnarray}
Let us now construct ${\cal F}_{orth}(e)$, ${\cal F}_{orth}(f)$ 
and ${\cal F}_{orth}(g)$ . We have :
\begin{equation}
\{p\}^{\perp_e} = \emptyset, \quad \{q\}^{\perp_e} = \emptyset, 
\quad \{r\}^{\perp_e} = \emptyset
\end{equation}
From this follows that:
\begin{equation}
{\cal F}_{orth}(e) = \{\emptyset, \Sigma\}
\end{equation}
In an analogues way we have:
\begin{equation}
{\cal F}_{orth}(f) = \{\emptyset, \Sigma\}
\end{equation}
Let us now construct ${\cal F}_{orth}(g)$. We have :
\begin{equation}
\{p\}^{\perp_g} = \{q\}, \quad \{q\}^{\perp_g} = \{p\}, \quad 
\{r\}^{\perp_g} = \emptyset
\end{equation}
From this follows that:
\begin{equation}
{\cal F}_{orth}(g) = \{\emptyset, \{p\}, \{q\}, \Sigma\}
\end{equation}
It also follows that
\begin{equation}
{\cal F}_{orth} = \{\emptyset, \{p\}, \{q\}, \Sigma\}
\end{equation}
Again we can see that the trace of the ortho closure system is 
not equal to the state ortho closure system in general. Indeed we
have:
\begin{equation}
{\cal Y}_{orth}(state) = \{\emptyset, \{p\}, \Sigma\}
\end{equation}
The example shows us that the eigen closures are in general different
from the ortho closures.

\subsection{Special properties:}

We can easily check that our entity is `outcome determined'.    
Let us calculate the eigen closures of the singletons. We have :
\begin{equation}
\begin{array}{l}
cl_{eig}(\{\lambda_{11}\}) = \{\lambda_{11}, \lambda_{32}\} = 
eig(O(\lambda_{11})) \\
cl_{eig}(\{\lambda_{12}\}) = \{\lambda_{12}\} = 
eig(O(\lambda_{12})) \\
cl_{eig}(\{\lambda_{13}\}) = \{\lambda_{13}, \lambda_{32}\} = 
eig(O(\lambda_{13})) \\
cl_{eig}(\{\lambda_{21}\}) = \{\lambda_{21}\} = 
eig(O(\lambda_{21})) \\
cl_{eig}(\{\lambda_{22}\}) = \{\lambda_{22}, \lambda_{32}\} = 
eig(O(\lambda_{22})) \\
cl_{eig}(\{\lambda_{23}\}) = \{\lambda_{22}, \lambda_{23}\} = 
eig(O(\lambda_{23})) \\
cl_{eig}(\{\lambda_{31}\}) = \{\lambda_{31}\} = 
eig(O(\lambda_{31})) \\
cl_{eig}(\{\lambda_{32}\}) = \{\lambda_{32}\} = 
eig(O(\lambda_{32})) \\
cl_{eig}(\{\lambda_{33}\}) = \{\lambda_{11}, \lambda_{31}, 
\lambda_{32}, \lambda_{33}\} = eig(O(\lambda_{33})) 
\end{array}
\end{equation}
In this example we can also see that the ortho closure of  the
singletons is not necessarily equal to the eigen closure,  even in
the case of an `outcome determined' entity. Indeed, for example :
\begin{equation}
cl_{orth}(\{\lambda_{33}\}) = {\cal E} \times \Sigma  \not=
cl_{eig}(\{\lambda_{33}\})
\end{equation}
\section{Standard quantum mechanics} 
We describe now the way in which our formalism is
related to the complex Hilbert space model of standard
quantum mechanics. We will introduce the concepts of our approach and
illustrate what they are for standard quantum mechanics. We will see
that everything works very well except when we arrive at the
description of sub entities. There something peculiar happens, that has
been remarked early in quantum mechanics, and has been studied in
detail in (Aerts and Daubechies 1978, Aerts 1981, 1982, 1984a). We will
come back to the problem of the description of sub entities in the next
section and a proposal for its solution will lead us to the formulation
of an alternative quantum mechanics in Hilbert space where additional
`pure' states are introduced in a very natural way. Let us first
describe the non-problematic aspects of standard quantum mechanics.

For sake of simplicity of notations we consider a finite dimensional
complex Hilbert space, but it is easy to see that an analogous scheme
can be formulated for the case of an separable infinite dimensional
complex Hilbert space. Hence consider the $n$ dimensional complex
Hilbert space ${\cal H}$. Let us first introduce some concepts of the
Hilbert space that we will use in the following.
\begin{definition}
Consider a separable complex Hilbert space ${\cal H}$. We
introduce the set of unit vectors, the set of
rays, the set of orthogonal projections and the set of spectral
families of the Hilbert space:
\begin{equation}
\begin{array}{l}
{\cal U}({\cal H}) = \{c \ \vert\ c \in {\cal H}, \ \|c\|=1\} \\
{\cal R}({\cal H}) =
\{{\bar c}\ \vert\ \ {\bar c} \ {\rm is\ the \ ray\ of}\ {\cal H} \
{\rm generated\ by\ } c \in {\cal U}({\cal H}) \} \\ 
{\cal P}({\cal H}) =
\{ E_k \ \vert\ \ E_k \ {\rm is\ an \ orthogonal \ projection\ of}\
{\cal H} \} \\
{\cal S}({\cal H}) = \{ E \ \vert\ \ E \ {\rm is\ a \ spectral
\ family \ of}\ {\cal H} \}
\end{array}
\end{equation}
We will denote unit vectors by $c,d, ...$, rays by ${\bar c}, {\bar
d}, ... $, orthogonal projections by $E_k, E_l, ...$, and spectral
families by $E, D, ...$.
\end{definition}
For an entity that is described by this
Hilbert space in standard quantum theory a state
$p_{\bar c}$ is represented by a ray ${\bar c} \in {\cal R}({\cal H})$
of the Hilbert space (this will not be the case anymore in the
alternative completed quantum mechanics that we present in the next
section).

Traditionally it is said that an experiment is described by a
self adjoint operator. However, if we want to remain closer to the
physical meaning, it is well known that we can better represent the
experiment by means of the spectral family of orthogonal projections of
this self adjoint operator. Let's first mention the spectral theorem
that makes both representations equivalent.
\begin{proposition}
If $H$ is a self adjoint operator of an n dimensional complex
Hilbert space ${\cal H}$, then there exist distinct real numbers
$\lambda_1, ..., \lambda_r (1 \leq r \leq n)$ and a pairwise orthogonal
set of nonzero projections $\{E_1,...,E_r\}$ such that
\begin{equation}
\begin{array}{ll}
\sum_{k=1}^r E_k = 1 & H = \sum_{k=1}^r\lambda_kE_k
\end{array}
\end{equation}
which will be called a `spectral family' of the Hilbert
space ${\cal H}$. Conversely, if
$\{\lambda_1,...,\lambda_r\}$ is a set of distinct real numbers, and
$\{E_1,...,E_r\}$ is a pairwise orthogonal set of nonzero projections
and if the two above mentioned conditions are satisfied, and hence we
have a spectral family, then
$\{\lambda_1,...,\lambda_r\}$ is the set of distinct eigenvalues of
$H$, and for each
$k$, $E_k$ is the projection onto the eigen space corresponding to
$\lambda_k$.
\end{proposition}
That is the reason that we shall represent an experiment by the
spectral family
$E =
\{E_1,...,E_r\}$ of pairwise orthogonal nonzero projections
that satisfies the first of the two conditions mentioned in the
spectral theorem. We will not use the $\lambda_i$ to indicate the
outcomes, although we could do so, but it would show less the
underlying structure of the outcomes. Instead of this we identify an
outcome $x_{E_k}$ in the quantum model with the eigen space $E_k$ of the
Hilbert space (or with the orthogonal projector $E_k$  on this eigen
space, we will not make a distinction). The set of all outcomes
$X_{sq}$ for the standard quantum model corresponds to the set of
all orthogonal projections or equivalently the set of all closed
subspaces of the Hilbert space ${\cal P}({\cal H})$, which is a
complete atomic orthocomplemented lattice.  For an experiment $e_E$ we
have $O(e_E) = \{x_{E_1},...,x_{E_r}\}$. Suppose that the entity is in
state $p_{\bar c}$ and we consider an experiment
$e_E$, then the set of   outcomes $O(e_E, p_{\bar c})$ is determined in
the following way, for $x_{E_j} \in O(e_E)$ we have $x_{E_j} \in
O(e_E,p_{\bar c}) \Leftrightarrow E_j(c) \not= 0$.

Let us now identify the probabilities as they appear in the case of  a
quantum entity described by the standard quantum mechanical formalism.
A quantum entity is a probabilistic entity where the probabilities 
are defined as follows. Suppose that we have an experiment $e_E$, a
state $p_{\bar c}$, and an outcome $x_{E_k} \in O(e_E, p_{\bar c})$ 
then $\mu(e_E,p_{\bar c}, x_{E_k}) = <c, E_k(c)>$, where $<\ ,\ >$ is
the inproduct of the Hilbert space, is the probability that the outcome
$x_{E_k}$ occurs if the experiment $e_E$ is performed the entity being
in state $p_{\bar c}$. It is interesting to remark that the quantum
probabilities only  depend on the state and the outcome and not on the
experiment.  This is one of the essential features of standard quantum
mechanics. We have now introduced all the necessary correspondences to
present a formal definition.
\begin{definition}
Consider a probabilistic entity $S({\cal E}_{sq}, \Sigma_{sq}, X_{sq},
{\cal O}_{sq}, {\cal M}_{qs})$ and a separable complex Hilbert space
${\cal H}$, with set of unit vectors ${\cal U}({\cal H})$, set of rays
${\cal R}({\cal H})$, set of orthogonal projections ${\cal P}({\cal
H})$ and set of spectral families ${\cal S}({\cal H})$. We say that the
entity is a `standard quantum entity' iff we have:
\begin{equation}
\begin{array}{l}
{\cal E}_{sq} = \{e_E\ \vert\ E \in {\cal S}({\cal H}) \} \\
\Sigma_{sq} = \{ p_{\bar c}\ \vert\ {\bar c} \in {\cal R}({\cal H})
\} \\ 
X_{sq} = \{ x_{E_k}\ \vert\ E_k \in {\cal P}({\cal H})\} \\
{\cal O}_{sq} = \{O(e_E,p_{\bar c})\ \vert\  E \in {\cal S}({\cal
H}), {\bar c} \in {\cal R}({\cal H}) \} \\
{\cal M}_{sq} = \{ \mu\ \vert\ \mu: {\cal E}_{sq} \times \Sigma_{sq}
\times X_{sq}
\rightarrow [0,1] \ {\rm is\ a\  generalized \ probability }\}
\end{array}
\end{equation}
such that
\begin{equation}
\begin{array}{l}
O(e_E,p_{\bar c}) = \{x_{E_k}\ \vert\  E_k \in {\cal P}({\cal H}),
E_k(c) \not= 0\} \\ 
\mu(e_E,
p_{\bar c}, E_k) = <c, E_k c> {\rm\ if}\ E_k \in E \\
\mu(e_E,
p_{\bar c}, E_k) = 0 {\rm\ if}\ E_k \not\in E 
\end{array}
\end{equation}
\end{definition}

\subsection{Pre-order and orthogonality:}

Let us investigate the orthogonality relation
and show that it coincides with the orthogonality of the
Hilbert space.
\begin{proposition}
Consider a standard quantum entity $S({\cal E}_{sq}, \Sigma_{sq},
X_{sq}, {\cal O}_{sq}, {\cal M}_{qs})$. If $x_{E_1}, x_{E_2} \in
X_{sq}$ then:
\begin{equation}
 x_{E_1} \perp x_{E_2} \Leftrightarrow E_1 \perp E_2
\end{equation}
\end{proposition}
Proof: Suppose that $x_{E_1} \perp x_{E_2}$, then there exists $e_E \in
{\cal E}_{sq}$ and $p_{\bar c} \in
\Sigma_{sq}$ such that
$x_{E_1} \not= x_{E_2} \in O(e_E,p_{\bar c})$. By definition of
$e_E$ it follows that $E_1, E_2 \in E$ and hence $E_1
\perp E_2$. If on the other hand $E_1 \perp E_2$ it is always
possible to consider a spectral family $E$ such that $E_1, E_2
\in E$. Further we can choose easily a vector $c$ such that
$E_1(c) \not= 0$ and $E_2(c) \not= 0$. Then we have that
$x_{E_1}, x_{E_2} \in O(e_E,p_{\bar c})$, which proves that
$x_{E_1} \perp x_{E_2}$.

\bigskip
\noindent
It is
important to show that the orthogonality relation on the set of states
coincides with the original orthogonality relation in the Hilbert space.
\begin{proposition}
Consider a standard quantum entity $S({\cal E}_{sq}, \Sigma_{sq},
X_{sq}, {\cal O}_{sq}, {\cal M}_{qs})$. For $p_{\bar c}, p_{\bar d} \in
\Sigma_{sq}$, we have:
\begin{equation}
p_{\bar c} \perp p_{\bar d} \Leftrightarrow c \perp d
\end{equation}
\end{proposition}
Proof: Suppose that $p_{\bar c} \perp p_{\bar d}$, then there exists 
an experiment $e_E$, with $E = \{E_1, ..., E_r\}$, such that
$O(e_E,p_{\bar c}) \cap O(e_E,p_{\bar d}) = \emptyset$. This means 
that we have two subsets $K \subset
\{1,...,r\}$ and $L \subset \{1,...,r\}$ such that $K \cap L = 
\emptyset$ and $O(e_E,p_{\bar c}) = \{x_{E_i} \
\vert\ i \in K\}$ while $O(e_E,p_{\bar d}) = \{x_{E_i} \  \vert\  i \in
L\}$. We have $E_i(c) \not= 0$ for
$i \in K$ and $E_i(d) \not= 0$ for $i \in L$. This implies that 
$E_i(c) = 0$ for $i
\notin K$ and $E_i(d) = 0$ for $i \notin L$, which shows that 
$\sum_{i \notin K}E_i(c) = 0$ and $\sum_{i
\notin L}E_i(d) = 0$. And since $E_i,i \in \{1,...,r\}$ is a spectral 
family we have $\sum_{i \in K}E_i(c) = c$ and $\sum_{i \in L}E_i(d) =
d$, which shows that $c \perp d$. The other implication is
straightforward. 
\begin{proposition}
Consider a standard quantum entity $S({\cal E}_{sq}, \Sigma_{sq},
X_{sq}, {\cal O}_{sq}, {\cal M}_{qs})$. For $p_{\bar c}, p_{\bar d} \in
\Sigma_{sq}$ we have:
\begin{equation}
p_{\bar c} < p_{\bar d} \Leftrightarrow {\bar c} = {\bar d} 
\Leftrightarrow p_{\bar c} = p_{\bar d}
\end{equation}
\end{proposition}
Proof: Suppose that ${\bar c} \not= {\bar d}$. We do not have  to
consider the situation where ${\bar c}
\perp {\bar d}$ since then certainly $p_{\bar c} \not< p_{\bar d}$. 
Hence suppose that ${\bar c}
\not\perp {\bar d}$. Let us construct an experiment by means of a set 
of spectral projections
$\{E_1, ..., E_r\}$ where $E_k$ is a one dimensional projector that 
is orthogonal to ${\bar d}$ but not orthogonal to ${\bar c}$. This is
always possible if the Hilbert space has dimension $\ge 2$. For this
experiment $e_E$ we have that $O(e_E,p_{\bar c})$ contains the outcome
$x_{E_k}$, while
$O(e_E,p_{\bar d})$ does not contain it. This shows that $p_{\bar c} 
\not< p_{\bar d}$. If the Hilbert space has dimension 1, the
proposition is trivially satisfied. 
\begin{theorem} \label{sqatom}
A standard quantum entity $S({\cal E}_{sq}, \Sigma_{sq},
X_{sq}, {\cal O}_{sq}, {\cal M}_{qs})$ is state atomic.
\end{theorem}
\begin{proposition}
Consider a standard quantum entity $S({\cal E}_{sq}, \Sigma_{sq},
X_{sq}, {\cal O}_{sq}, {\cal M}_{qs})$. For $e_E, e_F \in {\cal
E}_{sq}$ we have :
\begin{equation}
e_E = e_F \quad {\rm or} \quad e_E \perp e_F
\end{equation}
\end{proposition}
Proof: For a Hilbert space of dimension 1 the proposition is 
trivially satisfied. Hence consider a Hilbert space of at least
dimension 2. Consider two experiments $e_E \not= e_F$. This situation
is of the following nature. We have $E =
\{E_1,...,E_s,E_{s+1},...E_r\}$ and $F =
\{E_1,...,E_s,F_{s+1},...,F_t\}$ where $s$ is the number of spectral
projections that are equal, hence $F_i \not= E_j$. Let us take now a
vector $c \in (\sum_{i=1}^s E_i)^\perp$, which is always possible since
$e_E \not= e_F$, i.e. $E \not= F$. We then have $O(e_E,p_{\bar c}) \cap
O(e_F,p_{\bar c}) = \emptyset$, which proves that $e_E \perp e_F$.

\bigskip
\noindent
For the orthogonality and pre order relation on ${\cal E}  \times
\Sigma$ different situations are possible.
\begin{proposition}
Consider a standard quantum entity $S({\cal E}_{sq}, \Sigma_{sq},
X_{sq}, {\cal O}_{sq}, {\cal M}_{qs})$. For $(e_E,p_{\bar
c}), (e_E,p_{\bar d}) \in {\cal E}_{sq} \times \Sigma_{sq}$ we have:
\begin{equation}
(e_E,p_{\bar c}) < (e_E,p_{\bar d}) \Leftrightarrow R(c) = c
\end{equation}
where 
\begin{equation}
\begin{array}{l}
R = \sum_{x_{E_k} \in O(e_E,p_{\bar d})}E_k
\end{array}
\end{equation}
\end{proposition}
Proof: We have: $R(c) = c$ $\Leftrightarrow$ $E_k(c) = 0$ for 
$x_{E_k} \not\in O(e_E,p_{\bar d})$
$\Leftrightarrow$ $O(e_E,p_{\bar c}) \subset O(e_E,p_{\bar d})$ 
$\Leftrightarrow$ $(e_E,p_{\bar c}) < (e_E,p_{\bar d})$.
\begin{proposition}
Consider a standard quantum entity $S({\cal E}_{sq}, \Sigma_{sq},
X_{sq}, {\cal O}_{sq}, {\cal M}_{qs})$. For $(e_E,p_{\bar
c}), (e_F,p_{\bar d}) \in {\cal E}_{sq} \times \Sigma_{sq}$ we have:
\begin{equation}
(e_E,p_{\bar c}) < (e_F,p_{\bar d}) \Leftrightarrow R(c) = c  \ {\rm
and} \ T(c) = c
\end{equation}
where \begin{equation}
\begin{array}{ll}
R = \sum_{x_{E_k} \in O(e_E,p_{\bar d})}E_k & {\rm and}  \quad T =
\sum_{E_k \in E \cap F}E_k
\end{array}
\end{equation}
\end{proposition}
\begin{proposition}
Consider a standard quantum entity $S({\cal E}_{sq}, \Sigma_{sq},
X_{sq}, {\cal O}_{sq}, {\cal M}_{qs})$. For $(e_E,p_{\bar
c}), (e_F,p_{\bar d}) \in {\cal E}_{sq} \times \Sigma_{sq}$ we have:
\begin{equation}
(e_E,p_{\bar c}) \perp (e_F,p_{\bar d}) \Leftrightarrow T(c) =  c \
{\rm or} \ T(d) = d
\end{equation}
where
\begin{equation}
\begin{array}{l}
T = \sum_{E_k \not\in E \cap F}E_k
\end{array}
\end{equation}
\end{proposition}
The concept of eigen states coincides with the
traditional one.

\subsection{The eigen closures:}

Let us construct the eigen closures for the standard Hilbert
space model. We can prove the following proposition:
\begin{proposition}
Consider a standard quantum entity $S({\cal E}_{sq}, \Sigma_{sq},
X_{sq}, {\cal O}_{sq}, {\cal M}_{qs})$. For an experiment
$e_E$, with $E = \{E_1, ..., E_r\}$ and $A
\subset O(e_E)$, we have:
\begin{equation} 
p_{\bar c} \in eig_{e_E}(A) \Leftrightarrow c \in R(A) ({\cal H})
\end{equation}
where
\begin{equation}
\begin{array}{l}
R(A) = \sum_{x_{E_k} \in A}E_k
\end{array}
\end{equation}
\end{proposition}
Proof: $p_{\bar c} \in eig_{e_E}(A)$ $\Leftrightarrow$ $O(e_E,p_{\bar
c}) \subset A$ $\Leftrightarrow$
$E_k(c) = 0$ for
$x_{E_k} \not\in A$ $\Leftrightarrow$ $R(A)(c) = c$
$\Leftrightarrow$ $c \in R(A)({\cal H})$.

\bigskip
\noindent
This proposition shows that the $eig_{e_E}(A)$ correspond to
the orthogonal projections or closed subspaces of the Hilbert space.
\begin{proposition}
Consider a standard quantum entity $S({\cal E}_{sq}, \Sigma_{sq},
X_{sq}, {\cal O}_{sq}, {\cal M}_{qs})$. For an arbitrary $R$,
orthogonal projection of ${\cal H}$, and the spectral set $E =
\{R,1-R\}$ we have: 
\begin{equation}
c \in R({\cal H}) \Leftrightarrow p_{\bar c} \in eig_{e_E}(\{R\}) 
\end{equation}
\end{proposition}
Proof: $c \in R({\cal H})$ $\Leftrightarrow$ $R(c) = c$
$\Leftrightarrow$ $(1-R)(c) = 0$ $\Leftrightarrow$
$O(e_E,p_{\bar c}) = \{R\}$ $\Leftrightarrow$ $p_{\bar c} \in
eig_{e_E}(\{R\})$.

\bigskip
\noindent
Form these propositions follows that the state eigen closure system for
the standard quantum mechanical model is isomorphic with the closure
structure of the Hilbert space.

\subsection{The ortho closures:}

Let us investigate the ortho closure system of standard
quantum mechanics and prove that the state ortho closure system
coincides completely with the state eigen closure system.
\begin{theorem}
Consider a standard quantum entity $S({\cal E}_{sq}, \Sigma_{sq},
X_{sq}, {\cal O}_{sq}, {\cal M}_{qs})$. For $K \subset
\Sigma_{sq}$ and $L =  \{ c \ \vert\ p_{\bar c}
\in K \}$ we have:  
\begin{eqnarray}
K^\perp &=& \{p_{\bar c} \ \vert\ c \in L^\perp\} \\
cl_{orth}(K) &=& \{p_{\bar c} \ \vert\ c \in cl(L)\}
\end{eqnarray}
where $cl$ is the closure operator in the Hilbert space.
Suppose that $L$ is a closed subspace of ${\cal H}$, and $F  =
\{p_{\bar c} \ \vert\ c \in L\}$, then we have $F \in {\cal F}_{orth}$.
For the standard quantum mechanical model we have:
\begin{equation}
{\cal F}_{eig} = {\cal F}_{orth}
\end{equation}
\end{theorem}
Proof: We have $K^\perp = \{p_{\bar c} \ \vert\ p_{\bar c}  \perp
p_{\bar d}, p_{\bar d} \in K\}$ $= \{p_{\bar c} \
\vert\  c \perp d, d \in L\}$ $= \{p_{\bar c} \ \vert\  c \in 
L^\perp\}$. From this follows that $cl_{orth}(K) = (K^\perp)^\perp =
\{p_{\bar c} \ \vert\ c \in (L^\perp)^\perp\} = \{p_{\bar c} \ \vert\ c
\in cl(L)\}$. Consider now $L$ to be a closed subspace of the Hilbert
space and $F = \{p_{\bar c} \ \vert\ c \in L\}$. Then $cl_{orth}(F) =
\{p_{\bar c} \ \vert\ c \in cl(L)\} = \{p_{\bar c} \ \vert\ c  \in L\}
= F$, which shows that $F \in {\cal F}_{orth}$.

\bigskip
\noindent So for the standard quantum mechanical formalism the  eigen
closure system and the ortho closure system coincide. As a consequence
the eigen closure system is orthocomplemented.
\section{Completed quantum mechanics: a possible
solution of the sub entity problem} \label{comquant}
For standard quantum mechanics a sub entity is described by means  of
the tensor product procedure of the Hilbert spaces. Let us explain
shortly how this procedure works. Let
$S$ and $S'$ be described in complex Hilbert spaces ${\cal H}$ and 
${\cal H}'$, such that
${\cal H}' = {\cal H} \otimes {\cal G}$ where ${\cal G}$ is another 
complex Hilbert space. In this situation `standard quantum mechanics
says that' the entity $S'$ consists of two sub entities, one described
by the Hilbert space ${\cal H}$ (this is
$S$), and one described by the Hilbert space ${\cal G}$ (let us call 
this entity $T$). We have studied this situation in detail in earlier
work (Aerts and Daubechies 1978, Aerts 1984a), and will here only show
how this scheme fits (and does not fit - and this will be the reason to
`change' standard quantum mechanics and formulate a new
`completed' quantum mechanics within Hilbert space) into the general
description of a sub entity that we have developed in this new
approach. 

Let us consider the entity $S({\cal E}_{sq},
\Sigma_{sq}, X_{sq}, {\cal O}_{sq}, {\cal M}_{qs})$ described in the
Hilbert space ${\cal H}$ and the entity
$S'({\cal E}'_{sq},
\Sigma'_{sq}, X'_{sq}, {\cal O}'_{sq}, {\cal M}'_{qs})$
described in the Hilbert space ${\cal
H}'$ and suppose that $S$ is a sub entity of $S'$. Let us identify the
connection functions
$m, n, l$ and
$k$. Let us first do this for the functions $n$ and $l$,
because we will see that we will hit upon a strange situation for
the functions $m$ and $k$.  
$m$. We have:
\begin{equation}
\begin{array}{ll}
n: {\cal E}_{sq} \rightarrow {\cal E}'_{sq} &
e_E \mapsto e'_{E'} = n(e_E) 
\end{array}
\end{equation}
\begin{equation}
E' = \{E_1 \otimes I_{\cal G}, E_2 \otimes I_{\cal G}, ... , E_k 
\otimes I_{\cal G} \}  
\end{equation}
\begin{equation}
\begin{array}{ll}
l: X_{sq} \rightarrow X'_{sq} &
x_{E_k} \mapsto x'_{E'_{k'}} = l(x_{E_k})
\end{array}
\end{equation}
\begin{equation}
E'_{k'} = E_k \otimes I_{\cal G}
\end{equation}
These two functions show that for the standard tensor product 
procedure of standard quantum mechanics we can make correspond with
each experiment $e_E$ on the sub entity $S$ a unique experiment
$e'_{E'}$ on the big entity
$S'$, and also with each outcome $x_{E_k}$ of the sub entity $S$ 
corresponds a unique outcome $x'_{E'_{k'}}$ of the big entity
$S'$. 

The requirement that with each state $p'_{\bar c'}$ of the big  entity
$S'$ corresponds a unique state of the sub entity $S$ is not satisfied
in this tensor product procedure within standard quantum mechanics. It
is only met for some of the states of the big entity $S'$, namely for
the product states. Indeed if we consider a state
$p'_{\bar c'}$ where $c' = c \otimes d$, the function $m$ can be 
defined as follows: $m(p'_{\bar c'}) = p_{\bar c}$. But for a general
state of $S'$, and especially a non product state, i.e. $p'_{\bar c'}$,
where $c' =
\sum_ic_i \otimes d_i$, this can not be done.

Let us consider the
natural correspondence between the probabilities of the sub entity and
the big entity, that defines that function $k$, and see that also here
we have a correspondence only in the case that the big entity is in a
product state. Consider a probability measure $\mu$ for the sub
entity, the big entity being in a product state $p_{\bar c'}$ with
$c' = c \otimes d$. Hence we have
$\mu(e_E,m(p'_{\bar c'}), E_k) = <c, E_kc>$. The corresponding
probability measure $\mu'$ for the big entity should be such that
$\mu'(n(e_E), p_{\bar c'}, l(x_{E_k})) = \mu(e_E,m(p'_{\bar c'}),
E_k)$. If we put $\mu'(n(e_E), p_{\bar c'}, l(x_{E_k})) = <\bar c',
(E_k \otimes I_{\cal G}) \bar c'>$ then this is satisfied. Indeed we
have $<\bar c', (E_k \otimes I_{\cal G}) \bar c'> = <c \otimes d, (E_k
\otimes I_{\cal G}) c \otimes d> = <c, E_k c> <d, I_{\cal G} d> = <c,
E_k c> <d, d> = <c, E_k c>$. So we can define $\mu' = k(\mu)$.

The analysis we come to make means that the tensor product
procedure of standard quantum  mechanics cannot be used to describe sub
entities of the new approach. In (Aerts 1984b) we show that some of the
traditional axioms that lead to standard quantum mechanics are at the
origin of this problem. More specifically these are the axioms of
orthocomplementation, the covering law and the axiom of atomicity
(Aerts 1984b). The problem of the description of compound entities and
quantum axiomatics (which includes the problem of the description of sub
entities) has also been identified in other axiomatic approaches
(Randall and Foulis 1981, Pulmannova 1983, 1984, 1985, Aerts and
Valckenborgh 1998), and possibilities to replace the failing axioms are
under investigation (Aerts and Van Steirteghem 1998).

\subsection{The sub entity problem in standard quantum mechanics}

What we will come up here now is a completely new possibility to 
solve this problem. If  the `solution' that we propose here is correct
this will automatically lead to the formulation of a new
`completed' quantum mechanics in Hilbert space. Let us explain how we
came to this possible solution.

The main problem is that if the big entity is in a non product state 
represented by a ray of the tensor product Hilbert space ${\cal H}
\otimes {\cal G}$, the sub entities are not in a state represented by a
ray of one of the Hilbert spaces ${\cal H}$ or ${\cal G}$. This seems
to indicate that the sub entities `are not in a state' even if the big
entity `is in a state'. This is of course very difficult to imagine.
Indeed, if a piece of reality (the big entity) is in a certain state,
then also a `piece' of this `piece' of reality (in this case the sub
entities) should be in a state. It is hard to conceive of a reality
that would not satisfy such an elementary and fundamental property. Let
us indicate from now on the deep conceptual problem that we just come to
state by `the sub entity problem of standard quantum mechanics'.

 We have to remark that this problem was known from the early days  of
quantum mechanics but concealed more or less by the confusion that
often exists between pure states and mixtures. Let us explain this
first. The reality of a quantum entity in standard quantum mechanics is
represented by a pure state, namely a ray of the corresponding Hilbert
space. So what we have called `states' in this article are often called
`pure states'. Mixed states (what we also have called mixed states in
this article - see section~\ref{mixedstat}) are represented in standard
quantum mechanics by density matrices (positive self adjoint operators
with trace equal to 1). But although a mixed state is also called a
state it does not represent the reality of the entity under
consideration, but a lack of knowledge about this reality. This means
that if the entity is in a mixed state, it is actually in a pure state,
and the mixed state just describes the lack of knowledge that we have
about the pure state it is in. We have remarked that the deep conceptual
problem that we indicate here was noticed already in the early days of
quantum mechanics, but disguised by the existence of the two types of
states, pure states and mixed states. Indeed in most books on quantum
mechanics it is mentioned that for the description of sub entities by
means of the tensor product procedure it is so that the big entity can
be in a pure state (and a non product state is meant here) such that
the sub entities will be in mixed states and not in pure states (see
for example Jauch 1968 section 11-8 and Cohen-Tannoudji 1973, p 306).
The fact that the sub entities, although they are not in a pure state,
are at least in a mixed state, seems at first sight to be some kind of
a solution to the `sub entity problem in standard quantum mechanics'.
Although a little further reflection shows that it is not: indeed, if a
sub entity is in a mixed state, it should anyhow be in a pure state,
and this mixed state should just describe our lack of knowledge about
this pure state. So the `sub entity problem of standard quantum
mechanics' is not solved at all. Probably because quantum mechanics is
anyhow entailed with a lot of paradoxes and mysteries, the deep
problem of the sub entity description was unconsciously just added to
the list by the majority of physicists.

Way back, published in a paper in 1984, we have shown already that in
a more general approach we can  define pure states for the sub
entities, but they will not be `atoms' of the lattice of properties
(Aerts 1984b). Now it can easily be shown that within the general
lattice approach (very similar to the approach that we have exposed in
this paper in section~\ref{statpropent}) standard quantum mechanics
gives rise to an atomic property lattice, the rays of the Hilbert space
representing the atoms of the lattice (see also theorem~\ref{sqatom}
of this paper which proves the `state atomicity'). This
means that the non atomic pure states that we had identified in (Aerts
1984b) can anyhow not been represented within the standard quantum
mechanical formalism. We must admit that the finding of the existence
of non atomic pure states in the 1984 paper, even from the point of
view of generalized quantum formalisms, seemed also to us very far
reaching and difficult to interpret physically. Indeed
intuitively it seems to be so that only atomic states should
represent pure states. We know now that this is a wrong intuition. But
to explain why we have to present first the other pieces of the puzzle.

A second piece of the puzzle appeared when in 1990 we built a model of
a mechanistic classical laboratory situation violating Bell inequalities
with $\sqrt{2}$, exactly `in the same way' as its violations by the EPR 
experiments (Aerts 1990). With this model we tried to show that Bell
inequalities can be violated in the macroscopic world with the same
numerical value as the quantum violation. What is interesting for
the problem of the description of sub entities is that new `pure'
states were introduced in this model. We will see in a moment that the
possibility of existence of these new states lead to a solution of the
problem of the description of sub entities within a Hilbert space
setting but different from standard quantum mechanics.

More pieces of the puzzle appeared steadily during the elaboration of 
the general formalism presented in the present paper. We started to
work on this formalism during the first half of the
eighties, reformulating and elaborating some of the concepts during
the years. Then it became clear that the new states introduced in
(Aerts 1990), although they are `pure' states in the model, they appear
as non-atomic states in the general formalism. This made us understand
that the first intuition that classified non-atomic states as no good
candidates for pure states was a wrong intuition. Let us present now the
total scheme of our solution.

\subsection{The quantum machine: a macroscopic spin model}

We have introduced this example on earlier
occasions (Aerts, 1986, Aerts 1990, Aerts and Durt 1994, Aerts 1995),
and will  use it here to illustrate the solution of the sub entity
problem of standard quantum mechanics that we want to present and we
will show how all the pieces of the puzzle fit together. The quantum
machine is in fact a model for the spin of a spin ${1 \over 2}$ quantum
entity. Let us present it in some detail such that this section is self
contained. 
The entity $S_{qm}$ that we consider is a point
particle $P$ that can move on the surface of a sphere denoted by 
$surf$ with center $0$ (the origin of a three dimensional real space)
and radius
$1$. The unit vector $v$ giving the location of the particle on the 
surface of the sphere represents the state
$p_v$ of the particle (see Fig. 1,a), when it is at the surface of 
the sphere. Hence the collection of all possible states of the entity
$S_{qm}$ that we consider is given by :
\begin{equation}
\Sigma_{qm} = \{p_v\
\vert\ v \in surf\}
\end{equation}

\vskip 0.7 cm

\hskip 2 cm \includegraphics{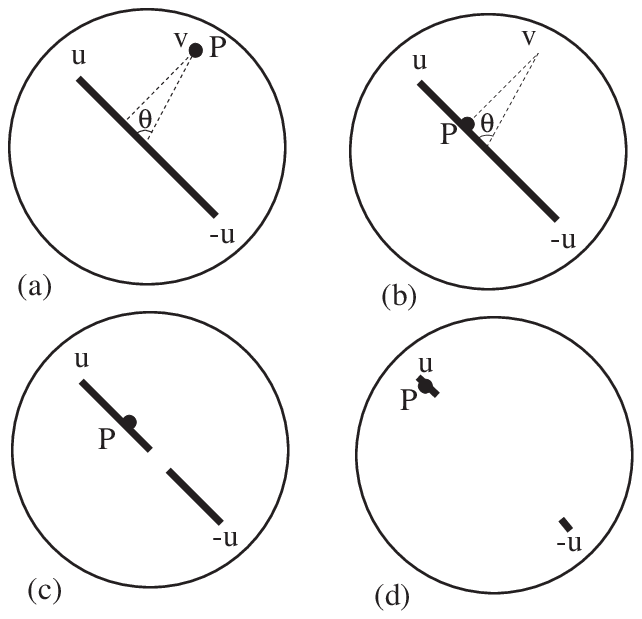}

\begin{quotation}
\noindent \baselineskip= 7pt \smallroman Fig. 1 : A representation of the 
quantum machine. In (a) the physical entity $\scriptstyle P$ is in
state $\scriptstyle p_v$ in the point $\scriptstyle v$, and the elastic
corresponding to the experiment $\scriptstyle e_{u}$ is installed
between the two diametrically opposed points $\scriptstyle u$ and 
$\scriptstyle -u$. In (b) the particle $\scriptstyle P$ falls
orthogonally onto the elastic  and sticks to it. In (c) the elastic
breaks and the particle $\scriptstyle P$ is pulled  towards the point
$\scriptstyle u$, such that (d) it arrives at the point  $\scriptstyle
u$, and the experiment $\scriptstyle e_u$ gets the outcome
$\scriptstyle  o^u_1$.
\end{quotation}
We define the following experiments. For each point  $u \in surf$,
we introduce the experiment $e_u$. We consider the diametrically
opposite point $-u$, and install an elastic band of length 2, such that
it is fixed with one of its end-points in $u$ and the other end-point
in $-u$. Once the elastic is installed, the particle
$P$ falls from its original place $v$ orthogonally onto the elastic, 
and sticks on it (Fig 1,b). Then the elastic breaks and the particle
$P$, attached to one of the two pieces of the elastic (Fig 1,c), moves
to one of the two end-points $u$ or $-u$  (Fig 1,d). Depending on
whether the particle $P$ arrives in $u$ (as in  Fig 1) or in $-u$, we
give the outcome $o^u_1$ or $o^u_2$ to $e_u$. Hence for the quantum
machine we have :
\begin{equation}
{\cal E}_{qm} = \{e_u\ \vert\ u \in
surf\}
\end{equation}
If we consider the two unit vectors $v, u \in surf$ we can have the 
following possibilities.

\smallskip
\noindent
(1) If we have $v = u$ then $O(e_u, p_v) = \{o^u_1\}$, (2) if we  have
$v = -u$ then $O(e_u,p_v) = \{o^u_2\}$, (3) if we have $v \not=u$ and
$v \not= -u$ then $O(e_u, p_v) = \{o^u_1, o^u_2\}$. This shows that:
\begin{equation}
X_{qm} = \{o^u_1, o^u_2 \ \vert\ u \in surf\}
\end{equation}
The probabilities are easily calculated. The probability,  $\mu(e_u,
p_v, o^u_1)$, that the particle $P$ ends up in point $u$ and hence
experiment $e_u$ gives outcome $o^u_1$ is given by the length of the
piece of elastic
$L_1$ divided by the total length of the elastic. 
\vskip 0.7 cm

\hskip 3.5 cm \includegraphics{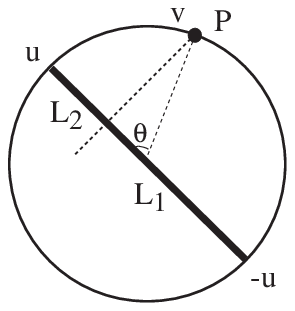}

\begin{quotation}
\noindent \baselineskip= 7pt \smallroman Fig. 2 : A representation of the 
experimental process in the plane where it takes place. The elastic of
length 2, corresponding to the experiment
$\scriptstyle e_u$, is installed between $\scriptstyle u$ and 
$\scriptstyle -u$. The probability, $\scriptstyle \mu(e_u, p_v,
o^u_1)$, that the particle  $\scriptstyle P$ ends up in point
$\scriptstyle u$ under influence of the experiment $\scriptstyle e_u$
is given by the length of the piece of elastic $\scriptstyle L_1$
divided by the total length of the elastic. The probability,
$\scriptstyle
\mu(e_u, p_v, o_2^u)$, that the particle $\scriptstyle P$ ends up  in
point $\scriptstyle -u$ is given by the length of the piece of elastic
$\scriptstyle L_2$ divided by the total length of the elastic.
\end{quotation}
The probability, $\mu(e_u, p_v, o_2^u)$, that the particle
$P$ ends up in point
$-u$, and hence experiment $e_u$ gives outcome $o^u_2$ is given by the length of the piece of elastic $L_2$
divided by the total length of the elastic. This gives us:
\begin{eqnarray}
\mu(e_u, p_v, o^u_1) &=& {L_1\over 2} = {1 \over 2}(1 + cos\theta) = cos^2{\theta\over 2} \\
\mu(e_u, p_v, o^u_2) &=& {L_2\over 2} = {1 \over 2}(1 - cos\theta) = sin^2{\theta\over 2}
\end{eqnarray}
These are exactly the standard quantum mechanical probabilities 
connected to the spin of a spin${1 \over 2}$ quantum particle described
in a 2 dimensional complex Hilbert space.

Let us present shortly also
the quantum description. The state $p_v$ is represented by $p_{\bar c^v}$ where 
\begin{equation}
c^v = (cos{\theta \over 2}  e^{i{\phi
\over 2}}, sin{\theta \over 2}  e^{-i{\phi \over 2}})  \label{raystate}
\end{equation}
and the experiment $e_u$ is represented by $e_{E^u}$ where  $E^u =
\{E^u_1, E^u_2\}$ is the spectral family with spectral projections:
\begin{equation}
E^u_1 = \left( \begin{array}{ccc} 
1 & 0  \\
0 & 0 
\end{array} \right)
\quad 
E^u_2 = \left( \begin{array}{ccc} 
0 & 0  \\
0 & 1 
\end{array} \right)
\end{equation}
We remark that we have chosen the basis of the two dimensional 
complex Hilbert space that describes our spin to coincide with the
eigenvectors of $e_u$, hence $c^u = (1,0)$ and $c^{-u} = (0,1)$, but
this does not endanger the generality of our description. Let us verify
that the quantum mechanical calculation recovers the probabilities of
our model. Indeed we have:
\begin{equation}
\begin{array}{l}
\mu_q(e_{E^u}, p_{\bar c^v}, o_1) = <c^v, E^u_1c^v> =  cos^2{\theta
\over 2} = \mu(e_u, p_v, o^u_1) \\
\mu_q(e_{E^u}, p_{\bar c^v}, o_2) = <c^v, E^u_2c^v> =  sin^2{\theta
\over 2} = \mu(e_u, p_v, o^u_2)
\end{array}
\end{equation}
This completes our model for the spin of a spin${1 \over 2}$  quantum
entity in standard quantum mechanics.

\subsection{The new state space: the completed quantum machine}

In the example that we proposed in (Aerts 1990) we used two  spin
models as the one presented here and introduced new states on both
models with the aim of presenting a situation that violates the Bell
inequalities exactly as in the case of the singlet spin state of two
coupled spin${1 \over 2}$ particles do. We indeed introduced a state
for both spin models that corresponds to the point in the center of
each sphere, and connecting these two states by a rigid rod we could
generate a violation of Bell's inequalities. Let us introduce this
state corresponding to the center $0$ of the sphere now explicitly and
call it $p_0$. We clearly see that if we apply one of the experiments
$e_u$ to the point now being in the state $p_0$, hence being located in
the center of the sphere, the probability corresponding to the
respective outcomes is ${1 \over 2}$, and hence the set of
possible outcomes is $\{o^u_1, o^u_2\}$ for any $u$. So we have:
\begin{equation}
\begin{array}{lll}
\mu(e_u, p_0, o^u_1) = {1 \over 2} & \mu(e_u, p_0, o^u_2) =  {1
\over 2} & \forall\ u \in surf
\end{array}
\end{equation}
\begin{equation}
\begin{array}{ll}
O(e_u,p_0) = \{o^u_1,o^u_2\} & \forall\ u \in surf
\end{array}
\end{equation}
If we consider the general definition of the `state implication'  that
we have introduced in definition~\ref{defprestat} then we can see that
\begin{equation}
\begin{array}{lll}
p_v < p_0  &
p_0 \not< p_v & \forall\ v \in surf
\end{array}
\end{equation}
which shows that $p_0$ is `not an atom' of the pre-ordered set  of
states. This means that we have `identified' a possible `non-mixture'
state (meaning with `non-mixture' that it really represents the reality
of the entity and not a lack of knowledge about this reality) that is
not an atom of the pre-ordered set of states. Is this a candidate for
the `non-mixed' states that we identified in (Aerts 1984b) and that
were non-atoms? We will see that it is. Let us proceed now and
explicitly define all the new states that we want to introduce in our
example. Since it will not be the same example anymore we will call
this new quantum machine the `completed' quantum machine.

The entity $S_{cqm}$ (completed quantum machine) that we consider  is
again a point particle $P$ that can move inside and on the surface of a
sphere denoted by $ball = \{w \ \vert\  \|w\| \leq 1\}$ with center
$0$ (the origin of a three dimensional real space) and radius
$1$. The vector $w$ giving the location of the particle inside  the
sphere represents the state
$p_w$ of the particle (see Fig. 3). The experiments that we consider
for this completed quantum machine are the same as the one we
considered for the quantum machine. This means that the set of
outcomes and the set of experiments are given by:
\begin{equation}
\begin{array}{ll}
\Sigma_{cqm} = \{p_w\ \vert\ w \in ball\} &
{\cal E}_{cqm} = \{e_u\ \vert\ u \in
surf\}
\end{array}
\end{equation}
Before we start to calculate the probabilities for the completed 
quantum entity we remark the following. Because the sphere is a convex
set, each vector $w \in ball$ can be written as a convex linear
combination of two vectors $v$ and $-v$ on the surface of the sphere
(see Fig.3). 
\vskip 0.7 cm

\hskip 3.5 cm \includegraphics{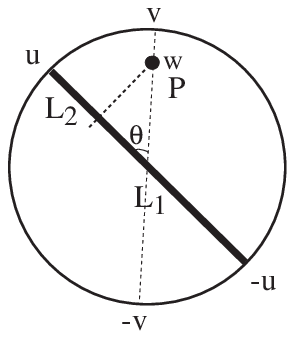}

\begin{quotation}
\noindent \baselineskip= 7pt \smallroman Fig. 3 : A representation of the 
experimental process in the case of the `completed' quantum machine.
The elastic of length 2, corresponding to the experiment
$\scriptstyle e_u$, is installed between $\scriptstyle u$ and 
$\scriptstyle -u$. The probability, $\scriptstyle \mu(e_u, p_w,
o^u_1)$, that the particle  $\scriptstyle P$ ends up in point
$\scriptstyle u$ under influence of the experiment $\scriptstyle e_u$
is given by the length of the piece of elastic $\scriptstyle L_1$
divided by the total length of the elastic. The probability,
$\scriptstyle
\mu(e_u, p_w, o_2^u)$, that the particle $\scriptstyle P$ ends up  in
point $\scriptstyle -u$ is given by the length of the piece of elastic
$\scriptstyle L_2$ divided by the total length of the elastic.
\end{quotation}
More concretely this means that we can write (referring to
the $w$ and $v$ and $-v$ in the Figure 3):
\begin{equation}
w = a \cdot v - b \cdot v,\ \  a, b \leq 1, \ \  a + b = 1
\end{equation} 
Hence, if we introduce these convex combination coefficients $a, b$ 
we have $w = (a-b) \cdot v$. Let us calculate now the transition
probabilities for a completed quantum machine entity being in a general
state
$p_w$ with $w
\in ball$ and hence
$\|w\| \leq 1$ (see Fig.3). Again the probability
$\mu(e_u, p_w, o^u_1)$, that the particle $P$ ends up in point $u$ 
and hence experiment $e_u$ gives outcome
$o^u_1$ is given by the length of the piece of elastic
$L_1$ divided by the total length of the elastic. The probability,  
$-u$, and hence experiment $e_u$ gives outcome $o^u_2$ is given by 
the length of the piece of elastic $L_2$ divided by the total length of
the elastic. This means that we have:
\begin{eqnarray} \label{probform}
\mu(e_u, p_w, o^u_1) &=& {L_1\over 2} = {1 \over 2}(1 + (a-b) 
cos\theta) = a cos^2{\theta\over 2} + b sin^2{\theta\over 2} \\
\mu(e_u, p_w, o^u_2) &=& {L_2\over 2} = {1 \over 2}(1 - (a-b) 
cos\theta) = a sin^2{\theta\over 2} + b cos^2{\theta\over 2}
\end{eqnarray}
These are new probabilities that will never be obtained if we  limit
the set of states to the rays of the two dimensional complex Hilbert
space as it is the case for the (non completed) quantum machine. The
question is now the following: can be find another mathematical entity,
connected in some way or another to the Hilbert space, that would allow
us, with a new quantum rule for calculating probabilities, to find back
these probabilities? The answer is yes, but now we have to proceed very
carefully not to get into too much confusion. We will show that these
new `pure' states of the interior of the sphere can be represented
by using density matrices, the same matrices that are used within the
standard quantum formalism to represent mixed states. And the standard
quantum mechanical formula that is used to calculate the probabilities
connected to mixed states, represented by density matrices, can also be
used to calculate the probabilities that we have identified here. But
of course the meaning will be different: in our case this standard
formula will represent a transition probability from one pure state to
another and not the probability connected to the change of a mixed
state. Let us show all this explicitly and to do this construct the
density matrices in question.

The well known quantum formula for the calculation of the
probabilities  for an outcome $x_{E_k}$ if an experiment $e_E$ is
performed, where $E = \{E_1,...,E_k,...,E_n\}$ is the spectral
decomposition corresponding to the experiment, and where the quantum
entity is in a mixed state
$p$ represented by the density matrix
$W$, is the following:
\begin{equation}
\mu(e_E, p, E_k) = tr(W \cdot E_k) \label{probcal}
\end{equation}
where $tr$ is the trace of the matrix.

A standard quantum mechanical calculation shows that the density 
matrix representing the ray state $c_v = (cos{\theta
\over 2}  e^{i{\phi
\over 2}}, sin{\theta \over 2}  e^{-i{\phi \over 2}})$ (see 
\ref{raystate}) is given by:
\begin{equation}
W(v)  = \left( \begin{array}{ccc} 
cos^2{\theta \over 2} & sin{\theta \over 2}cos{\theta \over 2}e^{-i\phi}  \\
sin{\theta \over 2}cos{\theta \over 2}e^{i\phi} & sin^2{\theta \over 2} 
\end{array} \right)
\end{equation}
and the density matrix representing the diametrically opposed ray 
state $c_{-v}$ is given by:
\begin{equation}
W(-v)  = \left( \begin{array}{ccc} 
sin^2{\theta \over 2}  & -sin{\theta \over 2}cos{\theta \over
2}e^{-i\phi}  \\ -sin{\theta \over 2}cos{\theta \over 2}e^{i\phi} &
cos^2{\theta \over 2} 
\end{array} \right)
\end{equation}
We will show now that the convex linear combination of these two 
density matrices with convex weights $a$ and
$b$ represents the state $p_w$ if we use the standard quantum 
mechanical formula (formula~\ref{probcal}) to calculate the transition
probabilities. If, for $w = av + b(-v)$, we put :
\begin{equation}
W(w) = aW(v) + b W(-v)
\end{equation}
we have:
\begin{equation}
W(w) = \left( \begin{array}{ccc} 
acos^2{\theta \over 2} + bsin^2{\theta \over 2} &  (a-b)sin{\theta
\over 2}cos{\theta \over 2}e^{-i\phi}  \\ (a-b)sin{\theta \over
2}cos{\theta \over 2}e^{i\phi} & asin^2{\theta \over 2} + bcos^2{\theta
\over 2} 
\end{array} \right)
\end{equation}
and it is easy to calculate now the transition probabilities  using
formula~\ref{probcal}. We have:
\begin{equation}
W(w) \cdot E_1 = \left( \begin{array}{ccc} 
acos^2{\theta \over 2} + bsin^2{\theta \over 2} & 0  \\
(a-b)sin{\theta \over 2}cos{\theta \over 2}e^{i\phi} & 0 
\end{array} \right)
\end{equation}
and hence, comparing with formula~\ref{probform}, we find:
\begin{equation}
tr(W(w) \cdot E_1) = acos^2{\theta \over 2} + bsin^2{\theta  \over 2} =
\mu(e_u, p_w, o^u_1)
\end{equation}
In an analogous way we find that:
\begin{equation}
tr(W(w) \cdot E_2) = asin^2{\theta \over 2} + bcos^2{\theta  \over 2} =
\mu(e_u, p_w, o^u_2)
\end{equation}
So we have shown that we can represent each one of the new states
$p_w$ by the density matrix $W(w)$ if we use formula~\ref{probcal} for
the calculation of the transition probabilities.

Let us also prove that each density operator
represents one of the new states $p_w$. We
can show this easily by using the general properties of density
matrices. Since a density operator is a self-adjoint operator, we can
find an orthonormal base of the two dimensional Hilbert space were it
is diagonal. Since it is a positive operator with trace equal to
$1$ it will have two real numbers $a,b$ such that $0 \leq a \leq 1$ 
and $0 \leq b \leq 1$ and $a + b = 1$ on its diagonal. Suppose that $v$
and $-v$ are the diametrically opposed points of the
sphere representing the base vectors. Then the density operator
represents the state corresponding to the point $(a - b)v$.

Although we have done all the calculations here only explicitly  for
the case of a two dimensional complex Hilbert space representing the
spin of a spin${1 \over 2}$ quantum entity, it can be shown easily that
this procedure is generally valid for an arbitrary quantum entity with
a arbitrary dimensional Hilbert space. The new non-product (hence pure)
states that we need to introduce to solve the `sub entity problem' of
standard quantum mechanics can be represented in a similar way by
density operators. We show in much more detail the new aspect of this
new approach to Hilbert space quantum mechanics in a forthcoming paper
(Aerts 1989).

We have not yet properly defined for the general case what is a 
density operator. Let us do this now such that we can prove that the
step that we want to propose, namely interpreting the density operators
as `also' representing `pure' states, within a new `completed'
Hilbert space formalism, solves our original `sub entity problem'.

\subsection{Completed quantum mechanics}

A density operator $W$ in the case of a general complex Hilbert space
${\cal H}$ is a positive self-adjoint operator with trace equal to $1$.
Only if $W^2 = W$ it represents a projection operator on a ray of the
Hilbert space and hence a `ray state'. If $W^2 \not= W$ the density
operator represents one of the new states that is not a ray state, but
is still a pure state. The same density operator of course also still
represents a mixed state like in the standard quantum mechanics. We
remark that even in the standard quantum mechanics several distinct
mixed states are represented by the same density operator, such that
the `double' representation that we introduce for this mathematical
object does not lead to additional conceptual problems. We just have to
be aware for which type of state we use the specific representation of
a specific density operator.

To get out of the confusion with the different types of states and
their representations we will now introduce some new concepts. 
\begin{definition}
Consider a separable complex Hilbert space ${\cal H}$. We introduce
the set of density operators ${\cal W}({\cal H})$. A density operator
is a positive self-adjoint operator with trace equal to $1$. We will
denote density operators by $W, V,...$.
\end{definition}
The set of all density operators ${\cal W}({\cal H})$ is a convex set,
subspace of the vector space of all bounded operators. This means that
if we consider a set $(W_i)_i$ of density operators and a set $(a_i)_i$
of real numbers such that $\sum_ia_i =1$, then $\sum_ia_iW_i$ is also
a density operator. It can be shown that the $W \in {\cal W}({\cal
H})$ we have $W^2 = W$ iff $W$ is an orthogonal projection on a one
dimensional subspace of ${\cal H}$. The density operators that equal
their product are the extremal points of the convex set ${\cal
W}({\cal H})$ and they represent the `ray' states. This also means
that every density operator can be written as the convex sum of such
ray state density operators. We have now all the necessary material to
present a formal definition of a completed quantum entity.
\begin{definition}
Consider a probabilistic entity $S({\cal E}_{cq}, \Sigma_{cq}, X_{cq},
{\cal O}_{cq}, {\cal M}_{cs})$ and a separable complex Hilbert space
${\cal H}$, with set of density operators ${\cal W}({\cal H})$, set of
orthogonal projections
${\cal P}({\cal H})$ and set of spectral families ${\cal S}({\cal H})$.
We say that the entity is a `completed quantum entity' iff we have:
\begin{equation}
\begin{array}{l}
{\cal E}_{cq} = \{e_E\ \vert\ E \in {\cal S}({\cal H}) \} \\
\Sigma_{cq} = \{ p_W \ \vert\ W \in {\cal W}({\cal H}) \} \\ 
X_{cq} = \{ x_{E_k}\ \vert\ E_k \in {\cal P}({\cal H})\} \\
{\cal O}_{cq} = \{O(e_E,p_W)\ \vert\  E \in {\cal S}({\cal
H}), W \in {\cal W}({\cal H}) \} \\
{\cal M}_{cq} = \{ \mu\ \vert\ \mu: {\cal E}_{cq} \times \Sigma_{cq}
\times X_{cq} \rightarrow [0,1] \ {\rm is\ a\  generalized \ probability
}\}
\end{array}
\end{equation}
such that
\begin{equation}
\begin{array}{l}
O(e_E,p_W) = \{x_{E_k}\ \vert\  E_k \in {\cal P}({\cal H}), E_k \in E,
tr(W  E_k) \not= 0 \} \\ 
\mu(e_E, p_W, E_k) = tr(W  E_k) {\rm\ if}\ E_k \in E \\
\mu(e_E, p_W, E_k) = 0 {\rm\ if}\ E_k \not\in E 
\end{array}
\end{equation}
\end{definition}
For a completed quantum entity we can solve the problem of the
description of the sub entity. Let us consider again the situation of
a completed quantum entity $S({\cal E}_{cq}, \Sigma_{cq}, X_{cq},
{\cal O}_{cq}, {\cal M}_{cs})$, described in a Hilbert space ${\cal
H}$ that is a sub entity of a completed quantum entity $S'({\cal
E}'_{cq}, \Sigma'_{cq}, X'_{cq}, {\cal O}'_{cq}, {\cal M}'_{cs})$
described in a Hilbert space ${\cal H}'$.

The functions $n$ and $l$ are defined as in the case of standard
quantum mechanics, namely:
\begin{equation}
\begin{array}{ll}
n: {\cal E}_{sq} \rightarrow {\cal E}'_{sq} &
e_E \mapsto e'_{E'} = n(e_E) 
\end{array}
\end{equation}
\begin{equation}
E' = \{E_1 \otimes I_{\cal G}, E_2 \otimes I_{\cal G}, ... , E_k 
\otimes I_{\cal G} \}  
\end{equation}
\begin{equation}
\begin{array}{ll}
l: X_{sq} \rightarrow X'_{sq} &
x_{E_k} \mapsto x'_{E'_{k'}} = l(x_{E_k})
\end{array}
\end{equation}
\begin{equation}
E'_{k'} = E_k \otimes I_{\cal G}
\end{equation}
Let us now consider a state $p'_{W'}$ of the big entity $S'$. Let us
show that there is one unique state $m(p'_{W'}) = p_W$ of the entity
$S$ such that
$tr(W'E'_{k'}) = tr(W E_k)$ and hence $\mu(e_E, m(p'_{W'}), E_k) =
k(\mu)(n(e_E), p'_{W'}, i(E_k))$. 
\begin{proposition} \label{uniqdens}
Let us suppose that we have three Hilbert spaces ${\cal H}, {\cal G}$
and ${\cal H}'$ such that ${\cal H}' = {\cal H} \otimes {\cal G}$.
For a density operator $W' \in {\cal W}'({\cal H}')$ there exists a
unique density operator $W \in {\cal W}({\cal H})$ such that for
an arbitrary $E_k \in {\cal P}({\cal H})$ we have $tr(W'E_k \otimes
I_{\cal G}) = tr(WE_k)$. We will denote $W = \hat{m}(W')$.
\end{proposition}
Proof: We first proof that $W$ is unique if it exists. Suppose that we
would have two density operators $W, V \in {\cal W}({\cal H})$ such
that $tr(W'E_k \otimes I_{\cal G}) = tr(WE_k) = tr(VE_k)\ \forall\ E_k
\in {\cal P}({\cal H})$. If we consider especially the projection
operator $E_c$ on an arbitrary ray ${\bar c}$ of the Hilbert space
${\cal H}$, then we have $tr(WE_c) = <c,Wc> = <c,Vc> = tr(VE_c)$. This
shows that $<c,Wc> = <c,Vc>\ \forall\ c \in {\cal H}$ and as a
consequence $W = V$.

Suppose that $W$ is a solution for an arbitrary $W'$. We know that
$W'$ can be written as the convex sum $\sum_{c'}a(c')W_{c'}$ of
density operators $W'_{c'}$ corresponding to projections on the
rays ${\bar c}'$, and hence with $\sum_{c'}a(c') = 1$. Since the
linearity of the trace we have $tr(W'E_k \otimes I_{\cal G}) =
\sum_{c'}a(c')tr(W'_{c'}E_k)$, which shows that if we construct the
density operator $W$ for the case where $W' = W'_{c'}$ is a density
operator corresponding to a ray ${\bar c}'$ we have a solution for the
general situation. 

This means that we have only to construct a solution for the case of a
density operator $W'_{c'}$ corresponding to a ray ${\bar c}'$ of the
big entity $S'$. Let us first show that $W$ has trace equal to 1.
Suppose that we consider an orthonormal base $(c_i)_i$ of ${\cal H}$ and
an orthonormal base
$(d_j)_j$ of ${\cal G}$, then $(c_i \otimes d_j)_{ij}$ is an
orthonormal base of ${\cal H}'$. This means that we can write $c' =
\sum_{ij}a_{ij}c_i \otimes d_j$. We have:
\begin{equation}
\begin{array}{l}
1 = <c',c'> = \sum_{ijkl}a_{ij}a_{kl}<c_i \otimes d_j, c_k \otimes
d_l> \\
= \sum_{ijkl}a_{ij}a^*_{kl}<c_i,c_k><d_j, d_l>
= \sum_{ijkl}a_{ij}a^*_{kl}\delta_{ik}\delta_{jl}
= \sum_{ij}\|a\|^2_{ij}
\end{array}
\end{equation}
Let us now use the correspondence of the
probabilities as required by the sub entity relation. We have $tr(WE_k)
= tr(W'E_k \otimes I_{\cal G}$ for all $E_k \in {\cal P}({\cal H})$.
Take especially $E_k$ to be the projector on $c_m$, and lets notate
this projector by $E_m$. Then we have: 
\begin{equation} \label{caldens}
\begin{array}{l}
<c_m, Wc_m> = tr(WE_m) = tr(W'E_m \otimes I_{\cal G}) = <c', E_m \otimes
I_{\cal G} c'> \\
= <\sum_{ij}a_{ij} c_i \otimes d_j, E_m \otimes I_{\cal
G} \sum_{kl} a_{kl} c_k \otimes d_l> \\
 = <\sum_{ij}a_{ij} c_i \otimes
d_j, \sum_{kl} a_{kl} (E_mc_k) \otimes d_l> \\
= <\sum_{ij}a_{ij} c_i
\otimes d_j,
\sum_l a_{ml} c_m \otimes d_l> = \sum_{ijl} a_{ij} a^*_{ml}
\delta_{im}\delta_{jl} \\
 = \sum_j \|a_{mj}\|^2
\end{array}
\end{equation}
This shows that:
\begin{equation}
\begin{array}{l}
tr(W) = \sum_m<c_m,Wc_m> = \sum_{mj}\|a_{mj}\|^2 = 1 
\end{array}
\end{equation}
We can easily calculate, using \ref{caldens}, the matrix elements of $W$
in a base were
$W$ is diagonal (this always exists since $W$ is a self-adjoint
operator).

\bigskip
\noindent
The result of theorem~\ref{uniqdens} makes it now possible for us to
define unambiguously the functions $m$ and $k$. Indeed:
\begin{equation}
\begin{array}{ll}
m: \Sigma'_{cq} \rightarrow \Sigma_{cq} &
p_{W'} \mapsto p_W = m(p_{W'}) 
\end{array}
\end{equation}
\begin{equation}
W = \hat{m}(W') 
\end{equation}
\begin{equation}
\begin{array}{ll}
 k: {\cal M}_{cq} \rightarrow {\cal M}'_{cq} &
\mu \mapsto k(\mu) 
\end{array}
\end{equation}
\begin{equation}
\begin{array}{l}
tr(WE_k) = tr(\hat{m}(W')E_k) = \mu(e_{E}, m(p_{W'}),x_{E_k}) \\
= k(\mu)(n(e_{E}), p_{W'}, l(x_{E_k})) 
= tr(W'E_k \otimes I_{\cal G}) \ {\rm \ for}\ E_k \in E
\end{array}
\end{equation}
\section{Conclusion} We have announced in the introduction that we
would elaborate as essential components of a general operational and
realistic formalism the structures of the states, the experiments, the
outcomes, the probabilities and the symmetries. We have treated the
structures of the states, experiments and outcomes in some detail and
point out now the aspects that are still missing and will be presented
in forthcoming work. If we think of Piron's representation theorem
(Piron 1976) that is formulated within the category of a state property
systems (see Aerts, Colebunders, Vervoort and Van steirteghem 1998), it
takes (1) completeness (2) atomicity, (3) orthocomplementation, (4) weak
modularity and (5) the covering law to arrive at a structure that is
isomorphic with a generalized Hilbert space. For an updated version of
the axioms necessary for this representation theorem, also
incorporating the resent result of Sol\`er, we refer to (Aerts and Van
Steirteghem 1998). We have treated the completeness and the atomicity
in the formalism presented here. We have shown that the completeness of
the whole set of properties can only be derived for the case of
distinguishable experiments entities and we have proven that the
atomicity is equivalent to the $T_1$ separation property for the eigen
closure structure. We have introduced the ortho closure structure and
this closure structure gives rise in a natural way to an
orthocomplementation. This is the reason that it is possible to
introduce the orthocomplementation by postulating that the eigen
closure structure has to coincide with the ortho closure structure, as
we have proposed in (Aerts 1994). We have not made this step in this
article because we want to study the problem of the introduction of an
orthocomplementation in a more detailed way in forthcoming work. We
want to mention that because we have made the choice to treat the
states and the properties of an entity as independent concepts, what
was not the case in the earlier approaches, we have identified a new
axiom, that we have called 'state determination (see also Aerts 1994).
We have not touched weak modularity and the covering law: this will be
done in future work. We have also only shortly introduced the concept
of probability and left the elaboration of it for future investigation.
We have not spoken at all of the symmetries and want to mention shortly
how we will analyze this aspect. Considering the group of automorphism
of our basic mathematical structure, we want to introduce the
symmetries as group representations of the different physical groups
that are connected to the different symmetries.

\section{Acknowledgments}

Diederik Aerts is Senior Research Associate of the Fund for 
Scientific Research and thanks Sven Aerts and Bart Van Steirteghem for
discussions about the content of this paper.

\section{References}

\begin{description}

\item Aerts, D., 1981, {\it The one and the many,\/} Doctoral Thesis, Free
University of Brussels, Brussels, 1981. 

\item Aerts, D., 1982, ``Description of many physical entities  without the
paradoxes encountered in quantum mechanics", {\it Found. Phys.\/}, {\bf
12}, 1131.

\item Aerts, D., 1983a, ``Classical theories and Non Classical Theories as a
Special Case of a More General Theory", {\it J. Math.  Phys.\/} {\bf
24}, 2441.

\item Aerts, D., 1983b, ``The description of one and many physical systems",
in {\it Foundations of Quantum Mechanics\/}, eds. C. Gruber, A.V.C.P.
Lausanne, 63.

\item Aerts, D., 1984a, ``Construction of a structure which makes it possible to describe the joint system
of a classical and a quantum system", {\it Rep. Math. Phys.\/}, {\bf 20}, 421.

\item Aerts, D., 1984b, ``Construction of the tensor product for lattices of
properties of physical entities", {\it J. Math. Phys.\/}, {\bf 25},
1434.

\item Aerts, D., 1986, ``A Possible Explanation for the Probabilities of
Quantum Mechanics", {\it J. Math. Phys.\/} {\bf 27}, 202.

\item Aerts, D., 1991, ``A mechanistic classical laboratory situation
violating the Bell inequalities with
$\sqrt{2}$, exactly `in the same way' as its violations by the EPR
experiments", {\it Helv. Phys. Acta,} {\bf 64}, 1.

\item Aerts, D., 1994, ``Quantum Structures, Separated Physical Entities and
Probability", {\it Found. Phys.\/} {\bf 24}, 1227.

\item Aerts, D., 1995, ``Quantum structures : an attempt to explain their
appearance in nature", {\it Int. J. Theor. Phys.\/} {\bf 34}, 1165.

\item Aerts, D., 1998, ``A possible solution of the sub entity problem of
standard quantum mechanics leading to a new type of Hilbert space
quantum mechanics", preprint FUND-CLEA, Free University of Brussels.

\item Aerts, D., Coecke, B.,  Durt, T. and Valckenborgh, F., 1997,
``Quantum, Classical and Intermediate; a Model on the Poincar\'e
Sphere, {\it Tatra Mountains Math. Publ.\/} {\bf 10}, 225.

\item Aerts, D., Coecke, B.,  Durt, T. and Valckenborgh, F., 1997,
``Quantum, Classical and Intermediate; the Vanishing Vector Space
Structure", {\it Tatra Mountains Math. Publ.\/} {\bf 10}, 241.

\item Aerts, D., Colebunders, E., Van der Voorde, A. and Van Steirteghem,
B., 1998, ``State property systems and closure spaces: a study of
categorical equivalence'', International Journal of Theoretical
Physics, this issue.

\item Aerts, D. and Daubechies, I., 1978, ``Physical Justification for using
the Tensor Product to describe two Quantum Systems as one Joint
System", {\it Helv. Phys. Acta} {\bf 51} , 661.

\item Aerts, D. and Durt, T., 1994, ``Quantum. Classical and Intermediate, an
illustrative example", {\it Found. Phys.\/} {\bf 24}, 1353.

\item Aerts, D. and Valckenborgh, F., 1998, ``Lattice
extensions and the description of compound entities", FUND, Brussels
Free University, preprint.

\item Aerts, D. and Van Steirteghem, B., 1998, ``Quantum Axiomatics and a
theorem of M.P. Sol\`er", submitted to International Journal of
Theoretical Physics.

\item Birkhoff, G. and Von Neumann, J., 1936, ``The logic of
quantum mechanics", {\it Annals of Mathematics}, {\bf 37}, 823.

\item Cohen-Tannoudji C, Diu B and Lalo{\"e}, F., 1973, {\it M{\'e}canique
Quantique, Tome I}, Hermann, Paris.

\item Foulis, D., Piron C. and Randall, C., 1983, ``Realism, operationalism,
and quantum mechanics", {\it Found. Phys.}, {\bf 13}, 813.

\item Foulis, D. and Randall, C., 1981 ``What are quantum logics and what
ought they to be?" in {\it Current Issues in Quantum Logic}, E.
Beltrametti and B. van Fraassen, eds., Plenum Press, New York, NY, 35.

\item Jauch, J., 1968, {\it Foundations of Quantum Mechanics},
Addison-Wesley, Reading, Mass.

\item Mackey,G.W., 1963, {\it Mathematical foundations of quantum
mechanics}, Benjamin, Reading Massachusetts.

\item Piron, C., 1964, ``Axiomatique quantique", {\it Helv. Phys. Acta}, {\bf
37}, 439.

\item Piron, C., 1976, {\it Foundations of Quantum Physics,\/}
Reading, Mass., W. A. Benjamin.

\item Piron, C., 1989, ``Recent Developments in Quantum
Mechanics", Helv. Phys. Acta, {\bf 62}, 82.

\item Piron, C., 1990, {\it M\`ecanique Quantique: bases et applications,},
Press Polytechnique de Lausanne.

\item Pulmannova, S., 1983, ``Coupling of quantum logics", {\it Int. J.
Theor. Phys.}, {\bf 22}, 837.

\item Pulmannova, S., 1984, ``On the product of quantum logics", {\it Suppl.
Circulo Mat. Palermo, Ser. II}, 231.

\item Pulmannova, S., 1985, ``Tensor products of quantum logics", {\it
J. Math. Phys.}, {\bf 26}, 1.

\item Randall, C. and Foulis, D., 1976 ``A mathematical setting for inductive
reasoning", in    {\it Foundations of Probability Theory, Statistical
Inference, and Statistical Theories of Science III}, C. Hooker, ed.,
Reidel , Dordrecht, 169.

\item Randall, C. and Foulis, D., 1978, ``The operational approach to
quantum mechanics", in {\it Physical Theories as Logico-Operational
structures,} Hooker, C.A. (ed.), Reidel, Dordrecht, Holland, 167.

\item Randall, C. and Foulis, D., 1981 ``Operational statistics and tensor
products", in   {\it Interpretations and Foundations of Quantum Theory}, H.
Neumann, ed., B.I. Wissenschaftsverslag, Bibliographisches Institut,
Mannheim, 21.

\item Randall, C. and Foulis, D.,1983, `` Properties and
operational propositions in quantum mechanics", {\it  Found. Phys.},
{\bf 13}, 835.

\item Von Neumann, J. 1932, {\it Mathematische Grundlagen
der Quanten-Mechanik}, Springer-Verlag, Berlin.

\item Varadarajan, V., 1968, {\it Geometry of quantum theory I\&II}, von
Nostrand, Princeton, New Jersey.

\item Zierler, N., 1961, ``Axioms for non-relativistic quantum mechanics",
{\it Pacific Journal of Mathematics}, {\bf 11}, 1151.

\end{description}

\end{document}